\documentclass[11pt]{iopart}
\pdfoutput=1

\usepackage{etoolbox}
\makeatletter
\newrobustcmd{\fixappendix}{%
  \patchcmd{\l@section}{1.5em}{7em}{}{}%
  \patchcmd{\l@subsection}{2.3em}{7em}{}{}%
}
\makeatother

\expandafter\let\csname equation*\endcsname=\relax
\expandafter\let\csname endequation*\endcsname=\relax

\usepackage[utf8]{inputenc}
\usepackage{amsmath}
\usepackage{amsfonts}
\usepackage{amssymb}
\usepackage{slashed}
\usepackage{graphicx}
\usepackage{cuted}
\usepackage{lipsum,comment}
\usepackage{color}
\usepackage{aas_macros}
\definecolor{RedWine}{rgb}{0.743,0,0}
\definecolor{RoyalBlue}{rgb}{0.25,.41,.88}
\definecolor{ForestGreen}{rgb}{.13,.54,.13}
\usepackage{cite} 
\usepackage[backref,breaklinks,colorlinks,citecolor=ForestGreen]{hyperref}
\usepackage{wrapfig}

\newcommand{\arctanh}{\textrm{arctanh}\ \!}
\newcommand{\pd}{\partial}
\newcommand{\scri}{\mathcal{I}}

\usepackage[normalem]{ulem}

\makeatletter
\newcommand{\mainmatter}{%
  \setcounter{footnote}{0}%
  \patchcmd{\@makefntext}{\fnsymbol}{\arabic}{}{}%
  \patchcmd{\@thefnmark}{\fnsymbol}{\arabic}{}{}%
  \def\@makefnmark{\textsuperscript{\arabic{footnote}}}%
}
\makeatother

\usepackage{caption}
\DeclareCaptionFormat{myformat}{\small \textbf{#1}#2#3\par}
\begin{document}

\title[Gravitational Scattering]{\hspace{-10pt}An asymptotic framework for  gravitational scattering}
\author{Geoffrey Comp\`ere$^1$, Samuel E Gralla$^2$,
Hongji Wei$^2$\footnote{hongjiwei@arizona.edu}}
\address{$^1$ Universit\'{e} Libre de Bruxelles, International Solvay Institutes, CP 231, B-1050 Brussels, Belgium}
\address{$^2$ Department of Physics, University of Arizona, Tucson, Arizona 85721, USA}

\begin{abstract} 
Asymptotically flat spacetimes have been studied in five separate regions: future/past timelike infinity $i^\pm$, future/past null infinity $\scri^\pm$, and spatial infinity $i^0$.  We formulate assumptions and definitions such that the five infinities share a single Bondi-Metzner-Sachs (BMS) group of asymptotic symmetries and associated charges.  We show how individual ingoing/outgoing massive bodies may be ascribed initial/final BMS charges and derive global conservation laws stating that the change in total charge is balanced by the corresponding radiative flux.  This framework provides a foundation for the study of asymptotically flat spacetimes containing ingoing and outgoing massive bodies, i.e., for generalized gravitational scattering.  Among the new implications are rigorous definitions for quantities like initial/final spin, scattering angle, and impact parameter in multi-body spacetimes, without the use of any preferred background structure.
\end{abstract}

\maketitle

\makeatletter
\def\@oddhead{\textit{Gravitational Scattering} \hfill \thepage}
\makeatother

\tableofcontents

\mainmatter
\clearpage
\section{Introduction}

Scattering experiments provide an effective way to understand and test physical law.  While gravitational scattering cannot at present be executed in the lab, it can be observed in nature and studied in theory.  The final stages of classical $2 \to 1$ scattering are now routinely observed \cite{LIGO}, and the existence (or not) of a quantum S-matrix lies at the heart of puzzles in quantum gravity (e.g., \cite{HOOFT198761,AMATI198781,Strominger2014,Damour:2017zjx,DiVecchia:2020ymx,Prabhu2022}).

In this paper we will study classical $m \to n$ gravitational scattering of macroscopic massive bodies in four spacetime dimensions.  We will  develop a general formalism that characterizes the initial/final state of the bodies as well as the incoming/outgoing gravitational radiation, organized in terms of symmetries and globally conserved charges.  Such a framework should be directly useful for classical calculations, for example in understanding puzzles regarding angular momentum \cite{Ashtekar:2019rpv,Compere:2021inq,Chen:2022fbu,DiVecchia2022,Veneziano2022,Javadinezhad:2022ldc,Gralla2022a,Gralla2022b,Manohar2022,Bini2022,Riva:2023xxm} or comparing gravitational waveforms obtained from different numerical or analytical methods \cite{Mitman:2020bjf,Mitman:2021xkq,Mitman:2022kwt}. 
We also hope for some insight into quantum scattering, although none will be given in this paper.

It is perhaps surprising that a fully general framework for macroscopic gravitational scattering does not already exist, given the tremendous effort devoted to the study of compact object interactions.  Indeed, a framework \textit{almost} exists, in that various asymptotic formalisms \cite{bms1,bms2,1962PhRv..128.2851S,1972JMP....13..956G,1977asst.conf....1G,Ashtekar:1978zz,beig_einsteins_1982,beig1984integration,1992GReGr..24.1185H,compere_relaxing_2011,troessaert_bms4_2018,Henneaux:2018cst,Prabhu:2021cgk,chakraborty_supertranslations_2022,capone_charge_2022} need only be tweaked and stitched together. However, the necessary tweaking and stitching turns out to be subtle and intricate, and we report here a rather significant sartorial effort.  The basic elements are the five separate asymptotic ``locations'':
\begin{itemize}
\item Future timelike infinity $i^+$, where massive bodies end up
\item Future null infinity $\scri^+$, where radiation ends up
\item Spatial infinity $i^0$, where spacelike slices end
\item Past null infinity $\scri^-$, where incoming radiation comes from
\item Past timelike infinity $i^-$, where massive bodies come from
\end{itemize}
In a gravitational scattering experiment, $i^\pm$ should encode the final/initial state of the bodies, $\scri^\pm$ should encode the outgoing/incoming radiation, and $i^0$ should be a trivial spectator.  Our contribution will be to flesh out how $i^\pm$ can encode the state of massive bodies and provide a way to unify the five infinities, such that all share a single group of symmetries and conserved charges.  This allows the entire collection of asymptotic regions to be regarded as a single asymptotic \textit{frame} in which the experiment takes place.
 
\begin{wrapfigure}[]{r}{.33\textwidth}
\includegraphics[width=\linewidth]{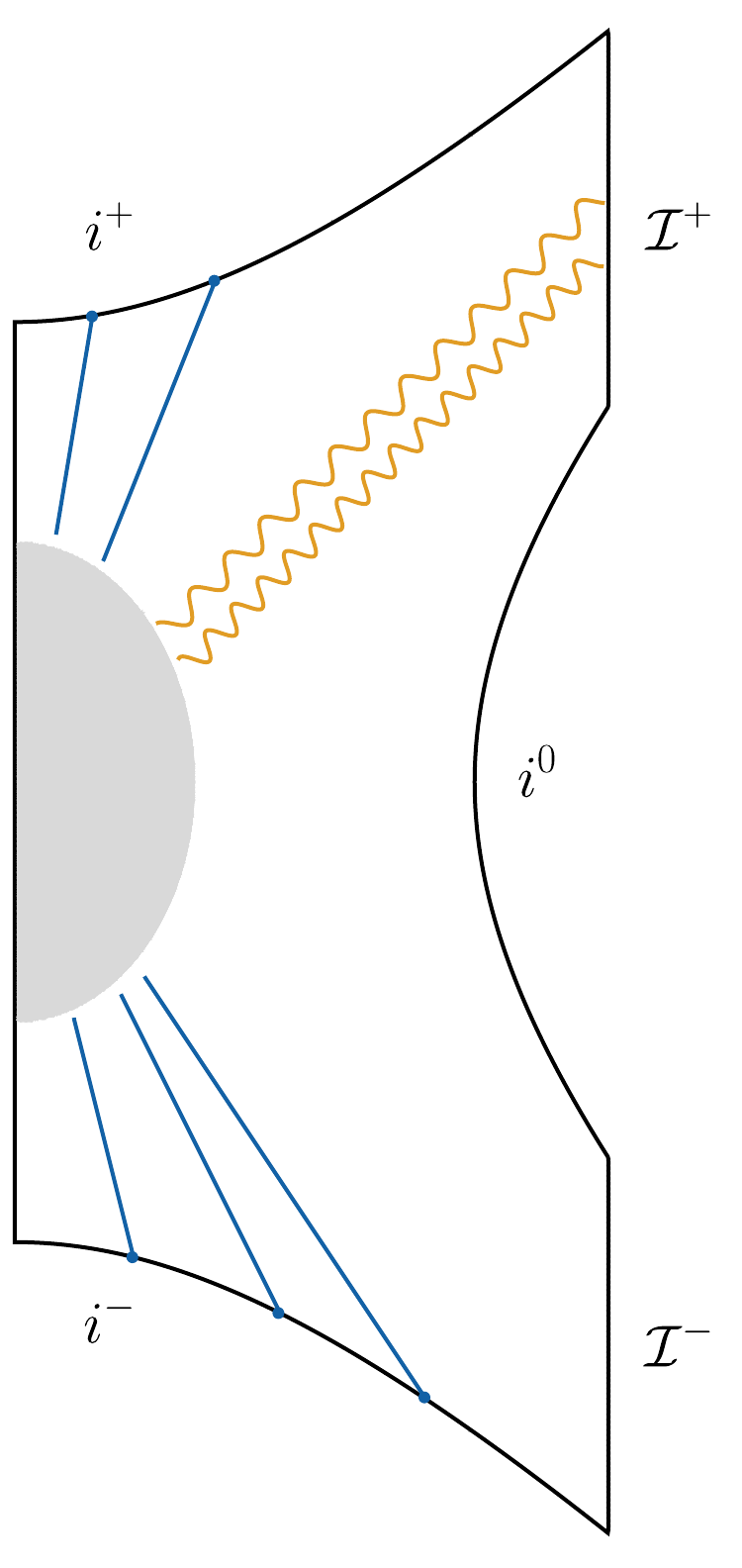}
\caption{$3\to2$ scattering}
\label{fig:cool}
\end{wrapfigure}

One notable feature of gravitational scattering is that the freedom of frame is not just Poincar\'e transformations, but the larger ``BMS group'' named after Bondi, van der Burg, Metzner, and Sachs \cite{bms1,bms2,1962PhRv..128.2851S}.  The (infinitely many) additional transformations are called \textit{supertranslations} and reflect additional freedom in setting up distant detectors of radiation using asymptotic data only.  The associated charges mix with the Poincar\'e charges  under valid changes of BMS frame and hence cannot be ignored.\footnote{$\,\,$ In an asymptotic spacetime region where the Bondi news asymptotically vanishes, one could require as a supplementary boundary condition that the electric part of the shear asymptotically vanishes, which selects a particular Poincar\'e subgroup of BMS; alternatively, one can use the electric part of the shear to construct supertranslation-invariant Poincar\'e charges in this region \cite{Javadinezhad:2018urv,Compere:2019gft,Chen:2021szm,Compere:2021inq,Chen:2021kug,Javadinezhad:2022hhl}, as reviewed in \ref{sec:invBMS}. However, after the passage of radiation the final and initial frames in general will differ by a supertranslation, encoding the displacement memory \cite{Strominger:2014pwa}.} For example, angular momentum is needed to define the impact parameter, but angular momentum changes under a supertranslation.  To fully describe a gravitational scattering experiment, the full BMS frame must be specified, and the dynamics can be interpreted by following all the BMS charges as they are exchanged between bodies and lost/gained due to outgoing/incoming radiation. Our main result is a corresponding framework for gravitational scattering featuring a single BMS group with global conservation laws for all BMS charges.  The charges provide definitions of all the usual quantities employed to describe a scattering result (e.g., impact parameter, spin, scattering angle). The general approach is illustrated in Fig.~\ref{fig:cool}.  

While we view this framework as a significant step forward, it certainly has limitations.  First, we simply assume the spacetime properties that we need, rather than proving that they arise from some specific kind of initial data.  While our assumptions appear to be compatible with all known exact and approximate solutions that could reasonably be called classical gravitational scattering, future discoveries could easily invalidate one or more of our assumptions.  However, we expect in this case the framework could be suitably modified to accommodate more general situations: the basic approach is sound.

A second limitation is the lack of a practical method for calculating the conserved charges.  The application of this framework requires the construction of special coordinate systems at each infinity of interest, which must satisfy a long list of regularity conditions.  While this approach makes an important conceptual point about the presence of a single BMS group governing gravitational scattering, it is not convenient in practice.  Further work is necessary to find practical methods of employing the results found herein.

This paper is organized as follows.  We begin in Sec.~\ref{sec:flat} with a broad overview of the five infinities in the example of flat spacetime.  We next consider $i^+$ in curved spacetime in Sec.~\ref{sec:i+}, establishing the BMS group and its conserved charges.  We then review $\scri^+$ in Sec.~\ref{sec:scri+} and match to $i^+$ in Sec.~\ref{sec:i+scri+}.  The past infinities ($i^-$, $\scri^-$, and their matching) are discussed in Sec.~\ref{sec:past-infinities}.  In Sec.~\ref{sec:i0} we obtain the BMS group at $i^0$ and we match to $\scri^\pm$ in Sec.~\ref{sec:i0scripm}.  In Sec.~\ref{sec:flux-balance} we put everything together to derive the flux balance laws and discuss their physical use in scattering problems. Finally, we conclude in Sec.~\ref{sec:conclusion} with a summary of the novel aspects of this work.  \ref{sec:equation} provides mode solutions to certain differential equations that arise on $i^\pm$ and $i^0$; \ref{sec:schwarzschild}, \ref{sec:translated-schwarzschild}, and \ref{sec:Kerr} analyze the Schwarzschild and Kerr metrics in various asymptotic regions; and \ref{sec:invBMS} reviews the construction of supertranslation-invariant charges in non-radiative regions.  We use units where $G=c=1$ and our signature is $-+++$.

\section{Overview in flat spacetime}\label{sec:flat}
 
 Before delving into technical details, it will be helpful to provide an overview in the special case of flat spacetime.  Newcomers to the field of asymptotics may appreciate the pedagogical introduction, while seasoned hands may find a new perspective.  In particular, we regard $\scri^\pm$ as timelike (rather than null) and use a ``puzzle piece'' diagram (instead of a conformal diagram) to represent the five infinities.  Our approach is to introduce special coordinates for each infinity, building on the seminal work of BMS \cite{bms1,bms2,1962PhRv..128.2851S} for $\mathcal{I}^\pm$, the analogous approach for $i^0$ due to Beig and Schmidt \cite{beig_einsteins_1982}, and the latter's generalization to $i^\pm$ \cite{chakraborty_supertranslations_2022}.  In this section we simply introduce the relevant coordinates in flat spacetime and discuss our general notation and philosophy.  Later sections review the curved spacetime frameworks and derive our results.

\subsection{Flat spacetime: \texorpdfstring{$i^+$}{i}}
\label{sec:flat-future-timelike-infinity}

We begin with future timelike infinity $i^+$, where the final state of the bodies will be recorded.  A first guess for $i^+$ would be some kind of limit $t \to \infty$ in Minkowski-like coordinates $(t,x,y,z)$.  However, this limit is inconvenient for describing outgoing particles, since  particles with non-zero velocity end up at infinite spatial distance from the origin---they do not have a limiting position in the limiting metric.  Furthermore, the limit does not naturally distinguish between timelike and null trajectories, meaning that special care would be required to exclude radiation.  A natural solution is to use velocity instead of position for the spacelike coordinates, i.e., to pick spatial coordinates $\phi^a$ such that curves of constant $\phi^a$ represent outgoing particle trajectories at late times.  Such particles then reach a finite position $\phi^a$ in the limiting spatial metric by construction, and we can restrict to subluminal velocities by carefully choosing the range of $\phi^a$.

To see how to implement this idea, consider a family of particles emanating from the origin $t=r=0$ of Minkowski spacetime, parameterized by their direction of motion $(\theta,\phi)$ and their (constant) velocity $v=r/t$.  If we choose the rapidity $\rho=\arctanh (r/t)$ as a coordinate, then the standard coordinate ranges of $(\rho,\theta,\phi)$ as spherical coordinates guarantee that we include only massive particles (and all massive particles).  For the timelike coordinate, we choose the proper 
 time $\tau=\sqrt{t^2-r^2}$, since it does not prefer any particular velocity.  Beginning with Minkowski space and making the coordinate transformation
\begin{align}
    t = \tau \cosh \rho\,, \qquad 
    r = \tau \sinh \rho\,,\label{transi+}
\end{align}
we find
\begin{align}
    ds^2 & = -d\tau^2 + \tau^2 \left(d\rho^2 + \sinh^2 \!\rho \ \! \gamma_{AB} dx^A dx^B \right) \label{milne} \\
    & = -d\tau^2 + \tau^2h^+_{ab}d\phi^ad\phi^b\,,\label{h+}
\end{align}
where $\gamma_{AB}$ is the metric of the unit sphere $S_2$ and $h^+_{ab}$ is the metric of (one sheet of) the two-sheet unit hyperboloid, otherwise known as Euclidean AdS${}_3$.  

We may regard the coordinates $\phi^a=(\rho, \theta, \phi)$ as spanning a unit hyperboloid $\mathcal{H}^+$  whose points represent velocities of massive particles.  As $\tau \to \infty$, all massive free particles (not just those emanating from the origin) have asymptotically constant $\phi^a$ and are naturally assigned to a point of $\mathcal{H}^+$.  The same can be said for outgoing particles in any spacetime whose metric agrees asymptotically with Eq.~\eqref{milne} as $\tau \to \infty$.  This is the key idea: a spacetime is considered asymptotically flat near future timelike infinity if there exist coordinates where its metric agrees with Eq.~\eqref{milne} at large $\tau$, and timelike infinity $i^+ \cong \mathcal{H}^+$ is the hyperboloid of outgoing velocities spanned by $\phi^a$.  A precise definition will be given in Sec.~\ref{sec:i+}.

\begin{figure}
\includegraphics[width=\textwidth]{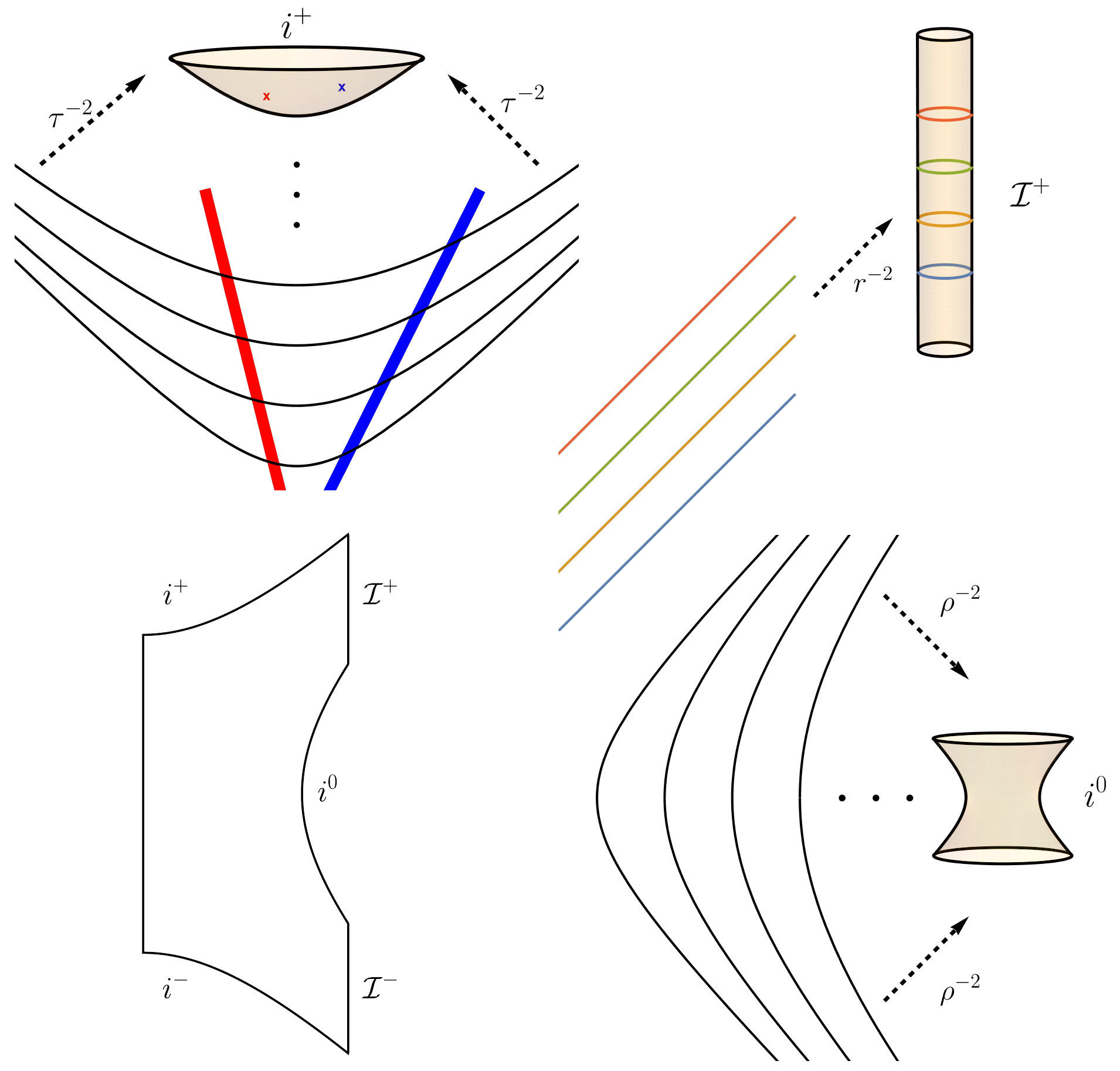}
\caption{\noindent Diagrams for asymptotically flat spacetime.  Future timelike infinity $i^+$ (top left) is a unit hyperboloid obtained from rescaling the induced metric on late-time hyperboloids.  Finite-sized massive bodies (red, blue) approach points of $i^+$ corresponding to their asymptotic velocities.  Future null infinity $\scri^+$ (top right) is a cylinder, the celestial sphere cross time, whose cross-sections (colors) represent the arrival times of light rays (correspondingly colored).  In our treatment, only the angular metric is rescaled; $\scri^+$ does not arise from the induced metric on any bulk surface.  Spatial infinity $i^0$ (bottom right) is defined analogously to $i^+$ using suitably rescaled Lorentzian hyperboloids at large radius $\rho$.  Past null infinity $\scri^-$ and past timelike infinity $i^-$ are the time-reverses of their future counterparts.  The five infinities are naturally represented on a ``puzzle piece'' diagram (bottom left).
 }\label{fig:iplus}
\end{figure}

In the above discussion we contemplate point particles that ``live'' in the spacetime but do not affect it.  To discuss gravitational scattering in full generality, we must include finite-sized self-gravitating bodies, such as black holes, that strongly affect the metric for all $\tau \to \infty$.  However, since distances on $i^+$ are infinitely rescaled,
\begin{align}\label{hplus}
h^+_{ab} = \lim_{\tau \to \infty}\frac{g_{ab}}{\tau^2} \qquad \textrm{(fixing $\rho,\theta,\phi$)}\,,
\end{align}
we expect all finite-sized bodies to become effectively pointlike.  Indeed, we shall show that this happens in a beautiful way (see also \cite{chakraborty_supertranslations_2022}).  We will see that the finite-sized bodies shrink to zero size and disappear entirely in the limit to $i^+$, but leave behind an effective point particle description at higher orders in $\tau^{-1}$, in terms of fields defined on $i^+$ that have poles at the points $\phi^a$ that the bodies ``disappeared to''  (Fig.~\ref{fig:iplus}). In particular, these fields can be used to assign a full set of BMS charges to each asymptotic body.  In this way, even spacetimes with fully self-gravitating bodies can be asymptotically flat in the sense of agreeing with flat spacetime \eqref{milne} at large $\tau$, and the asymptotic state is nicely encoded in subleading corrections that define fields on $i^+$.\footnote{$\,\,$ At the order in $\tau$ to which we work, information about the multipolar structure of each body is not present (beyond the mass and spin).  We expect that higher multipoles can be recovered at higher order in $\tau$.  However, even the multipole expansion does not fully specify the body.  It should be possible to recover full information about the body by introducing an alternative ``near-zone'' limit where one remains at fixed distance to the body at late times, and considering both limits at higher order in $\tau$.  (See, e.g., \cite{Gralla2008} for analogous limits in the context of the motion of small bodies.)  At present we will not formalize this procedure, simply assuming that in any given case, the type of bodies considered can be relayed along with the BMS charges as part of the overall description of the calculation or experiment.}

\subsection{Flat spacetime: $\scri^+$}

We turn now to null infinity $\scri^+$, where the state of outgoing radiation should be recorded.  Since radiation moves at a fixed speed (the speed of light), the space of velocities is reduced to the two-dimensional space of angles $(\theta,\phi)$.  While it would be mathematically legitimate (and arguably more mathematically natural) to pursue a correspondingly two-dimensional $\scri^+$, we will be guided by the physical idea that $\scri^+$ should represent an array of distant detectors that record gravitational-wave signals as a function of time.  We will denote the time coordinate(s) of the detectors by $u$.  If we return to a swarm of particles emanating from $t=r=0$ in Minkowski spacetime (this time moving only at light speed), then it is clear that for $u$ to be a faithful ``distant detector'' time coordinate (referring to the same radiation as $r\to\infty$), we must take it to be the retarded time,
\begin{align}\label{udef}
u = t-r\, .
\end{align}
We are therefore led to consider retarded coordinates $(u,r,\theta,\phi)$, where the Minkowski metric takes the form
\begin{align}\label{retarded}
ds^2 = -du^2 - 2 du dr + r^2 \gamma_{AB} dx^A dx^B\,.
\end{align}
We can then adopt a similar definition of asymptotic flatness: a metric is asymptotically flat at future null infinity if there exist coordinates where the metric agrees with \eqref{retarded} as $r \to \infty$.  The subleading corrections encode the effects of gravitational radiation, as worked out by many authors over many years.  Our precise assumptions are given in Sec.~\ref{sec:scri+}.

In analogy with our treatment of $i^+$, we will take $\scri^+$ to be the manifold spanned by $(u,x^A)$, thought of as living ``at large $r$''.  To find a metric for this manifold, the analogous approach would be to rescale the induced metric on constant-$r$ surfaces by a factor of $r^{-2}$ as $r\to\infty$, similarly to \eqref{hplus}.  However, it is easily checked that this results in a degenerate metric.  The physical difficulty is that by rescaling time, we force experimenters at successively larger radii to use successively slower clocks, such that time stops entirely in the limit.  By contrast, it is quite natural to rescale distances tangent to the sphere, since tangent vectors then have fixed length in the limit and can represent ``arms'' of a gravitational-wave detector.  For this reason we will only consider a rescaled angular metric, which is just the unit sphere metric,
\begin{align}\label{scriplusmetric}
\gamma_{AB} = \lim_{r \to \infty} \frac{1}{r^2} g_{AB} \qquad  \textrm{(fixing $u,\theta,\phi$)}\,.
\end{align}
We call the space spanned by $(\theta,\phi)$ the celestial sphere and assign it the sphere metric $\gamma_{AB}$.  We do not define a spacetime metric on $\scri^+$; it has the structure of the celestial sphere crossed with time.  The lack of spacetime structure on $\scri^+$ results from the physical asymmetry in the meaning of its points as the \textit{time} of receipt of radiation moving in a particular \textit{direction}.\footnote{$\,\,$ In the conformal approach to asymptotics \cite{Penrose1962}, $\scri^+$ is a null surface, which in flat spacetime is equal to the $v \to \infty$ limit of a surface of constant $v=t+r$.  This construction has the counter-intuitive feature that $\scri^+$ records the times at which radially \textit{ingoing} radiation intersects the outgoing radiation.  It is impossible to station detectors near the standard $\scri^+$, since they would have to be moving towards the source at the speed of light.}  

An alternative approach to describe $\scri^+$ (which we do not use in this paper) consists in extending the two-dimensional metric $\gamma_{AB}$ onto a non-invertible metric $\gamma_{ab}$ of signature $(0,+,+)$ with coordinates $x^a=(u,x^A)$ with the property that the vector $n^a \partial_a = \partial_u$ is a degenerate direction: $n^a \gamma_{ab}=0$. The couple 
$(\gamma_{ab}, n^a)$ then forms a so-called Carrollian structure at $\scri^+$ which is left invariant under the transformations $\mathcal L_\xi \gamma_{ab}=2\alpha \gamma_{ab}$, $\mathcal L_\xi n^a=-\alpha n^a$ with $\alpha=\alpha(u,x^A)$. Such transformations form the three-dimensional conformal Carrollian algebra which is isomorphic to the BMS algebra \cite{Duval:2014uva,Duval:2014lpa,Donnay:2022wvx} and therefore adequately represents the global symmetries of $\scri^+$.

An interesting contrast between $\scri^+$ and $i^+$ is that massless particles can arrive at $\scri^+$ with the same momentum, but at different times, whereas each point of $i^+$ counts \textit{all} mass that approaches with the corresponding momentum.  In other words, a train of massless particles registers as a worldline of $\scri^+$, but it would seem that a similar train of massive particles registers as a point of $i^+$.  However, a train of massive particles with same momentum cannot be maintained all the way to $i^+$ due to their mutual interaction. (In particular, after boosting into a frame where the particles are stationary, it is clear that they would attract and collapse.)  No such restriction exists for massless particles, helping to explain the distinction between $i^+$ and $\scri^+$.

\subsection{Flat spacetime: $i^0$}\label{sec:flati0}

Next consider spatial infinity $i^0$, which will represent the asymptotic region at spacelike separation.  Since by definition this region is causally disconnected from the entire scattering experiment, it will not encode the state of any physical objects, but it will be mathematically important for relating $\scri^+$ and $\scri^-$.  Just as $i^+$ is the manifold of subluminal asymptotic velocities, we want $i^0$ to be the manifold of superluminal asymptotic velocities.  So instead of using the rapidity $\rho=\arctanh(r/t)$ (which would now be imaginary), we define the analogous quantity $\hat{\tau}=\arctanh(t/r)$.  We let $\hat{\tau}$ range over $(-\infty,\infty)$ so that both incoming and outgoing velocities are included, allowing $i^0$ to link $\scri^-$ and $\scri^+$ (which describe only ingoing and outgoing velocities, respectively).  The coordinates $\phi^{\hat{a}}=(\hat{\tau},x^A)$ then encode all relevant superluminal trajectories.   (We use the notation that a hat on the index indicates that hatted coordinates are considered.)  For the fourth coordinate, analogously to $\tau=\sqrt{t^2-\rho^2}$ used for $i^+$, we use $\hat{\rho}=\sqrt{r^2-t^2}$ for $i^0$.  Thus our transformation is
\begin{align}\label{transi0}
t = \hat{\rho} \sinh \hat{\tau}\,, \qquad r = \hat{\rho} \cosh \hat{\tau}\,.
\end{align}
Making this transformation, the Minkowski metric becomes
\begin{align}
ds^2 & = d\hat{\rho}^2 + \hat{\rho}^2\left(-d\hat{\tau}^2 + \cosh^2 \!\hat{\tau} \gamma_{AB} dx^A dx^B\right) \\
& = d\hat{\rho}^2 + \hat{\rho}^2 h^0_{\hat{a}\hat{b}} d\phi^{\hat{a}} d\phi^{\hat{b}}\,,\label{h0}
\end{align}
where now $h^0_{\hat{a}\hat{b}}$ is the metric on the one-sheet unit hyperboloid $\mathcal{H}_0$ spanned by $\phi^{{\hat{a}}}=(\hat{\tau},x^A)$, otherwise known as Lorentzian $dS_3$.  A spacetime is called asymptotically flat at spatial infinity if there exist coordinates such that it asymptotically agrees with this form as $\hat{\rho}\to\infty$, and $i^0$ is the hyperboloid $\mathcal{H}^0$.  Precise assumptions will be given in Sec.~\ref{sec:i0}.  Analogously to \eqref{hplus}, we may write
\begin{align}
h_{\hat{a}\hat{b}}^0 = \lim_{\hat{\rho} \to \infty} \frac{1}{\hat{\rho}^2} g_{\hat{a}\hat{b}} \qquad \textrm{(fixing $\hat{\tau},\theta,\phi$)}\,.
\end{align}

\subsection{Analytic continuation between $i^+$ and $i^0$}\label{sec:analytic}

The definition of $i^+$ involved coordinates $(\tau,\phi^a)$ valid in the causal future $t>r$ of the origin of coordinates, while the definition of $i^0$ involved coordinates $(\hat{\rho},\phi^{\hat{a}})$ valid in the spacelike region $|t|<r$.  Comparing Eqs.~\eqref{transi+}--\eqref{h+} and \eqref{transi0}--\eqref{h0}, we see that the two approaches are related by an analytic continuation and field redefinition,
\begin{align}\label{magic}
\hat{\rho} = i \tau, \qquad \hat{\tau} = \rho - \frac{i \pi}{2}, \qquad  h^0 = - h^+\,.
\end{align}
The first two formulas (without the third) define an invertible map between the regions $t>r$ (relevant for $i^+$) and $|t|<r$ (relevant to $i^0$).\footnote{$\,\,$ Formally speaking, we first complexify and maximally extend Minkowski space, then apply the complex diffeomorphism, and finally restrict to the real submanifold where the coordinates take their desired ranges.}  The last two formulas (without the first) provide a map from $i^0$ to $i^+$.  The first and last formula (without the second) provide a map from $i^+$ to $i^0$.  The three together will be useful in relating results between $i^0$ and $i^+$.  In fact, we shall import most of the needed formulas at $i^+$ from existing analysis at $i^0$.

\subsection{Flat spacetime: $\scri^-$}\label{sec:flatscri-}

Just as $\scri^+$ records the state of outgoing radiation, $\scri^-$ will record the state of incoming radiation.  We  imagine emitters stationed at large radius $r$ at each angle $(\theta,\phi)$ in the distant past, sending radiation inward at each time $v$ on their clocks.  The formulas analogous to \eqref{udef}, \eqref{retarded}, and \eqref{scriplusmetric} are
\begin{align}
    v & = t+r \label{vdef}\,, \\
    ds^2 & = -dv^2 + 2 dv dr + r^2 \gamma_{AB} dx^A dx^B  \label{advanced} \,,\\
    \gamma_{AB} & = \lim_{r \to \infty} \frac{1}{r^2} g_{AB} \qquad \textrm{(fixing $v,\theta,\phi$)}\,,\label{scriminusmetric}
\end{align}
and $\scri^-$ is the manifold spanned by $(v,\theta,\phi)$, thought of as living at large $r$.  It has the structure of the celestial sphere (metric $\gamma_{AB}$) cross time.  Notice that radiation leaving $\scri^-$ from an angle $(\theta,\phi)$ has momentum along the antipodal direction $(\pi-\theta,\phi+\pi)$.  This is different from $\scri^+$, where the momentum direction and asymptotic angle are coincident.\footnote{$\,\,$ If we instead labeled points on $\scri^-$ by their momentum direction (as we do for $\scri^+$), we would find that our angular coordinates on $\scri^-$ are antipodally related to our angular coordinates on $\scri^+$, as in \cite{Strominger2014}.}

Two examples are helpful to keep in mind.  A null geodesic (straight line) arrives at $\scri^+$ with the same momentum it had at $\scri^-$, but at an antipodal point on the celestial sphere.  A uniformly accelerated particle instead arrives at $\scri^+$ with antipodally related momentum, but at the same point on the celestial sphere.  Noting that the orbits of a boost Killing field are uniformly accelerating particles, we see that the angular components of a boost Killing field will be antipodally related between $\scri^\pm$.  This is a first hint of the fact that charges defined intrinsically on $\scri^+$ and $\scri^-$ will match antipodally on the celestial sphere \cite{Strominger2014}.

Finally, notice that the formulas \eqref{retarded} and \eqref{scriplusmetric} for $\scri^+$ are related to corresponding formulas \eqref{advanced} and \eqref{scriminusmetric} by $u \leftrightarrow -v$ together with the stipulation that both $u$ and $v$ increase towards the future.  Mathematically, this corresponds a combination of time-reversal (viewed actively) and a coordinate transformation.  Time-reversal $\mathcal{T}$ acts on the retarded coordinates as $\mathcal{T}[u,r,\theta,\phi]=[-u -2r,r,\theta,\phi]$ and maps points near $\scri^+$ to points near $\scri^-$, leaving the metric invariant and reversing the time orientation.  A (passive) coordinate transformation $v=-\mathcal{T}[u]$ then expresses the metric near $\scri^-$ as \eqref{scriminusmetric}, with $v$ increasing toward the future. Since $\mathcal{T}[t]=-t$ and $\mathcal{T}[r]=r$, applying $\mathcal{T}$ to \eqref{udef} reproduces \eqref{vdef}.  This reasoning will allow us to relate results between $\scri^\pm$ by sending $u \leftrightarrow -v$.

\subsection{Flat spacetime: $i^-$}\label{sec:flati-}

Just as $i^+$ records the final state of massive bodies, $i^-$ will record the initial state of massive bodies.  We again imagine a swarm of particles of all different subluminal velocities, this time entering from the distant past and converging on the origin of coordinates.  We introduce coordinates $\phi^{\bar{a}}=(\bar{\rho},\theta,\phi)$ with $\bar\rho > 0$ that are constant on the particle worldlines, this time with $(\theta,\phi)$ representing the angle of origin, which is antipodally related to the momentum direction. Introducing the time coordinate $\bar{\tau} = -\sqrt{t^2-r^2}$, the analogs of Eqs.~\eqref{transi+}--\eqref{hplus} are
\begin{align}
    t = \bar{\tau} \cosh \bar{\rho}\,, \qquad 
    r = -\bar{\tau} \sinh \bar{\rho}\,,\label{transi-}
\end{align}
and 
\begin{align}
    ds^2 & = -d\bar{\tau}^2 + \bar{\tau}^2 \left(d\bar{\rho}^2 + \sinh^2 \!\bar{\rho} \ \! \gamma_{AB} dx^A dx^B \right) \label{milne2} \\
    & = -d\bar{\tau}^2 + \bar{\tau}^2 h^-_{\bar{a}\bar{b}}d\phi^{\bar{a}}d\phi^{\bar{b}}\,,\label{h-}
\end{align}
with
\begin{align}\label{hminus}
h^-_{\bar{a}\bar{b}} = \lim_{\bar{\tau} \to -\infty}\frac{g_{\bar{a}\bar{b}}}{\bar{\tau}^2} \qquad \textrm{(fixing $\bar{\rho},\theta,\phi$)}\,,
\end{align}
and $i^-$ is the hyperboloid $\mathcal{H}^-$ spanned by $\phi^{\bar{a}}=(\bar{\rho},\theta,\phi)$ with the ranges of spherical coordinates, thought of as living at large negative $\tau$.  The metric components $h_{\bar{a}\bar{b}}$ of $i^-$ are equal to the metric components $h_{ab}$ of $i^+$ when compared in the natural way, but we keep these metrics conceptually distinct.

The relationship between $i^\pm$ is analogous to the relationship between $\scri^\pm$ discussed at the conclusion of section \ref{sec:flatscri-}.  The formulas \eqref{milne} and \eqref{hplus} for $i^+$ are related to corresponding formulas \eqref{milne2} and \eqref{hminus} for $i^-$ by $\tau \leftrightarrow -\bar{\tau}$, together with the stipulation that both $\tau$ and $\bar{\tau}$ increase towards the future.  
\begin{wrapfigure}[]{r}{.3\textwidth}
\includegraphics[width=\linewidth]{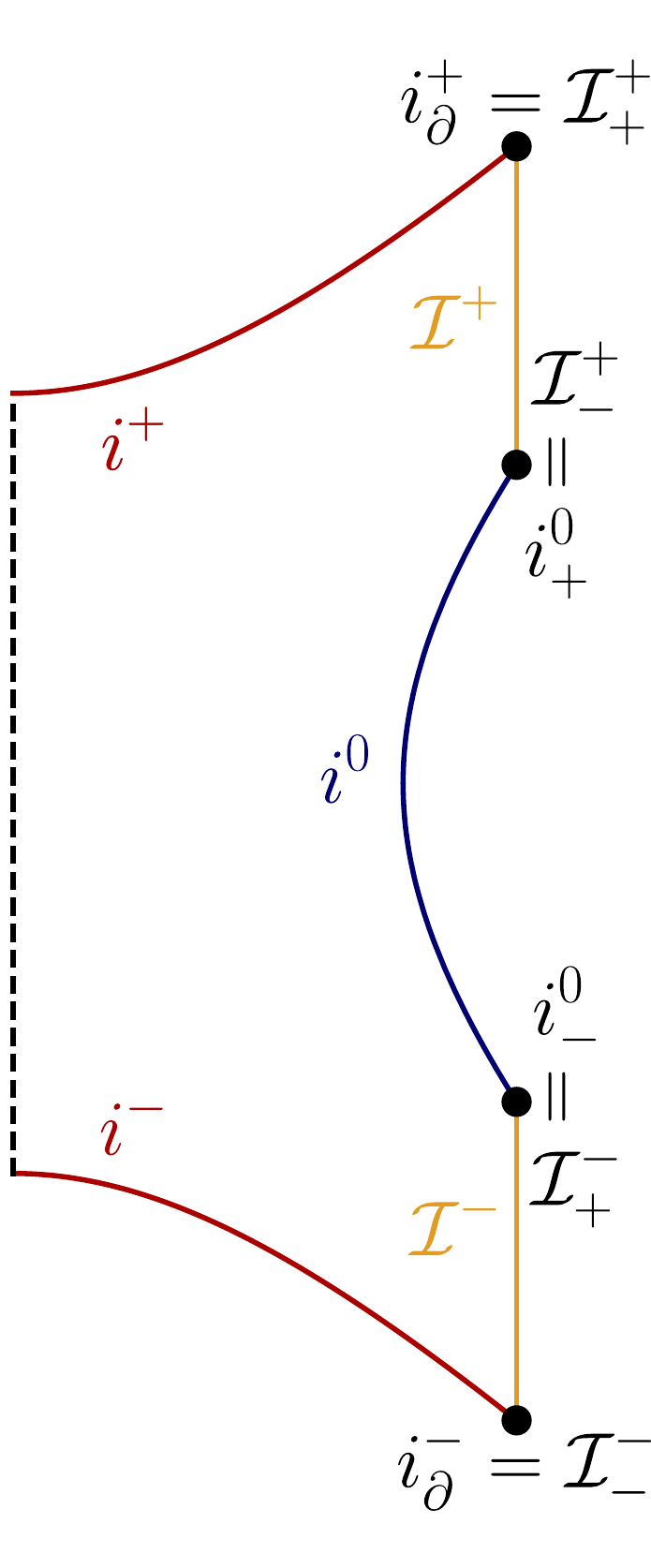}
\caption{Overlap regions}
\label{fig:overlaps}
\end{wrapfigure}
This corresponds mathematically to time-reversal (viewed actively) together with a coordinate transformation. 
Time-reversal $\mathcal{T}$ acts as $\mathcal{T}(\tau,\rho,\theta,\phi) = (-\tau,-\rho,\theta,\phi)$ and maps points near $i^+$ to points near $i^-$, leaving the metric invariant and reversing the time orientation.  A (passive) coordinate transformation $\bar{\tau}=-\mathcal{T}[\tau], \bar{\rho}=\mathcal{T}[\rho]$ then expresses the metric near $i^-$ as \eqref{h-}, with $\bar{\tau}$ increasing towards the future.  Since $\mathcal{T}[t]=-t$ and $\mathcal{T}[r]=r$, applying $\mathcal{T}$ to \eqref{transi+} reproduces \eqref{transi-}. This reasoning will allow us to relate results between $i^\pm$ by sending $\tau \leftrightarrow -\tau$.

\subsection{Overlap regions}\label{sec:overlaps}

We have introduced five asymptotic boundaries $i^+$, $\scri^+$, $i^0$, $\scri^-$ and $i^-$.  These boundaries meet at four regions, which we denote as $i^{+}_\partial=\scri^+_+$, $\scri^+_-=i^0_+$, $i^0_-=\scri^-_+$, $\scri^-_-=i^{-}_\partial$, as indicated in Fig.~\ref{fig:overlaps}.  In the main body, we define charges separately at each of the five infinities and construct coordinate transformations that relate them via matched asymptotic expansions at the four overlap regions.

\begin{figure}[!hptb]
\begin{center}
\includegraphics[width=0.85\textwidth]{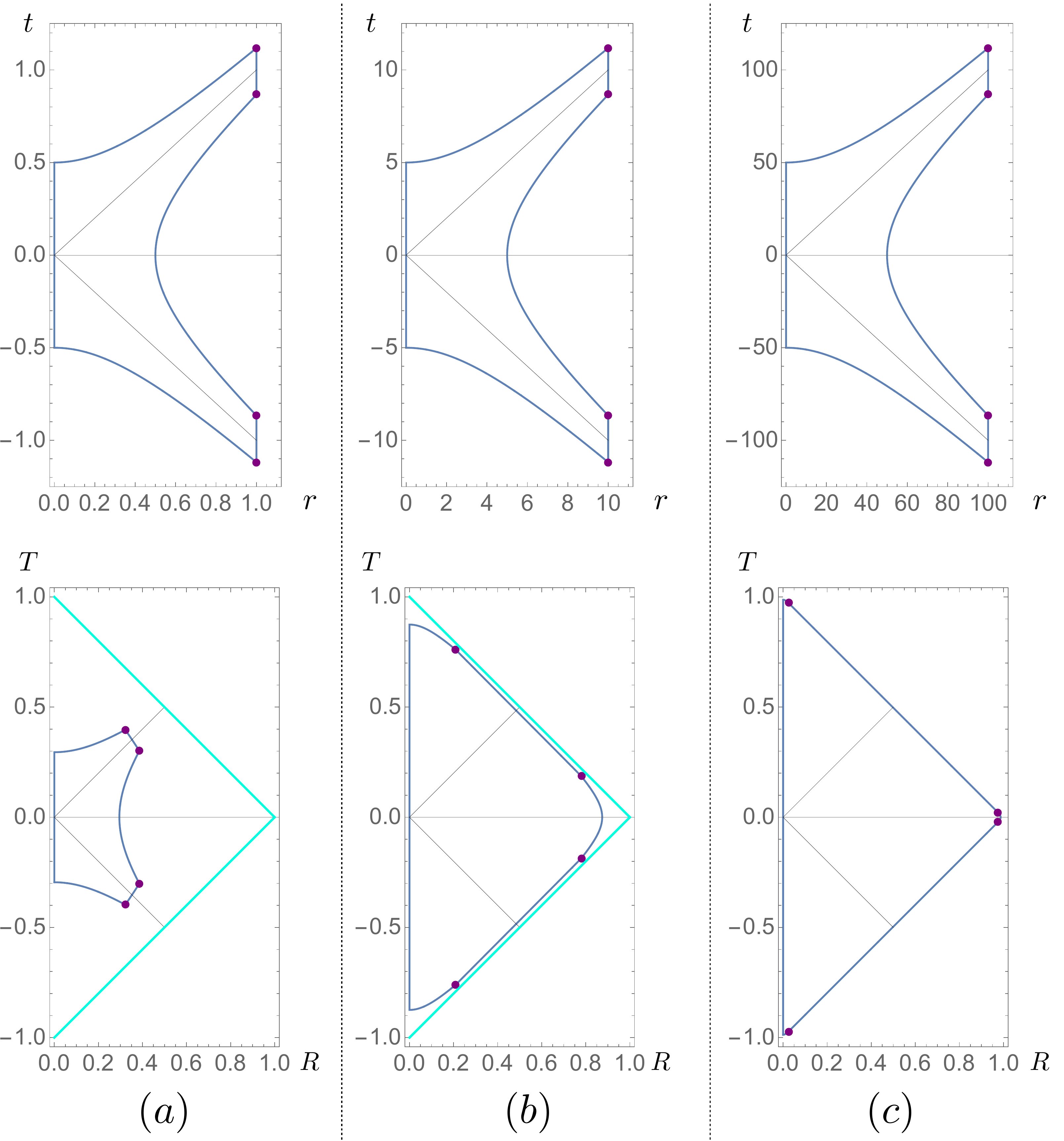}
\end{center}
\caption{\noindent The five infinities can be represented as the boundaries of the region defined by $\vert t^2-r^2 \vert < \mathcal{T}$ and $r<\mathcal{R}$ for large constants $\mathcal{T}$ and $\mathcal{R}$.  Here we plot this boundary for a sequence (a) $\mathcal{T}=0.25$, $\mathcal{R}=1$; (b) $\mathcal{T}=25$ and $\mathcal{R}=10$; and (c) $\mathcal{T}=2500$ and $\mathcal{R}=100$, showing the 4 overlapping boundary regions as bullets.  The ``puzzle piece'' shape is invariant on a Minkowski spacetime diagram (top), but converges to a diamond on a conformal diagram (bottom).  (Here $T=U+V$ and $R=V-U$ are the standard conformal coordinates constructed from $\tan U = t-r$ and $\tan V=t+r$.)  The puzzle piece can adequately represent all asymptotic regions, while the conformal diagram does not resolve $i^+$, $i^-$ and $i^0$, nor any of the overlap regions.}\label{fig:PenrosePuzzle}
\end{figure}

In our approach, the natural representation of the five infinities is as a ``puzzle piece'', already displayed in Figs.~\ref{fig:cool}--\ref{fig:overlaps}.  In Fig.~\ref{fig:PenrosePuzzle} we compare with the alternative conformal description of asymptotics.

\section{Future timelike infinity $i^+$}\label{sec:i+}
Asymptotics at timelike infinity were studied previously in Refs.~\cite{1998JMP....39.6573G,1989CQGra...6.1075C, 1982RSPSA.381..323P, Campiglia:2015qka, chakraborty_supertranslations_2022}.  Our approach is most similar to Ref.~\cite{chakraborty_supertranslations_2022}, except that rather than working with Einstein's equation near $i^+$, we will import results from Ref.~\cite{compere_relaxing_2011} at $i^0$ via the analytic continuation discussed in Sec.~\ref{sec:analytic}.  This produces a more general form of the expansion (including log terms that are necessary for scattering spacetimes) and also provides definitions of all charges corresponding to the asymptotic symmetries.  We also impose an additional coordinate condition that reduces the symmetry group to BMS.

An asymptotically flat spacetime at timelike infinity is defined as a spacetime that admits the following expansion, 
\begin{align}
    ds^2 & = \left(-1 - \frac{2\sigma}{\tau}-\frac{\sigma^2}{\tau^2}+ o(\tau^{-2})\right)d\tau^2 + o(\tau^{-2})\tau d\tau d\phi^a \nonumber \\
    &+\tau^2\biggr(h_{ab}+\tau^{-1}(k_{ab}-2\sigma h_{ab}) + \frac{\log \tau}{\tau^2}i_{ab} + \tau^{-2}j_{ab} + o(\tau^{-2})\biggr)d\phi^ad\phi^b, \label{Beig-Schmidt}
\end{align}
where $h_{ab}$ is the hyperboloid metric denoted $h^+$ in \eqref{h+}.  The expansion defines $i^+$ as the manifold spanned by $\phi^a$ with metric $h_{ab}$, together with tensor fields $\sigma$, $k_{ab}$, $i_{ab}$, and $j_{ab}$ on $i^+$. Eq.~\eqref{Beig-Schmidt} will be called the Beig-Schmidt expansion after the original work at spatial infinity \cite{beig_einsteins_1982}.  The fields $\sigma$ and $k_{ab}$ will be called ``first-order'' and the fields $i_{ab}$ and $j_{ab}$ will be called ``second order''.  These names correspond to the \textit{relative} order in $\tau$, where log terms are counted as order unity.  For example, the first-order field $\sigma$ appears at $O(\tau^{-1})$ in $g_{\tau \tau}$ but at $O(\tau)$ in $g_{ab}$. 
 These terms would appear at the same order in $\tau$ if the expansion were done in an orthonormal basis.

This expansion at $i^+$ is obtained from the known expansion at $i^0$ by an analytic continuation as follows. 
Eq.~(3.14) of Ref.~\cite{compere_relaxing_2011} (repeated as Eq.~\eqref{Beig-Schmidt-spatial} below) provides an asymptotic expansion of the metric near $i^0$.  Performing the analytic continuation described in Sec.~\ref{sec:analytic}, we find that all displayed quantities become real after additional field redefinitions. In total we make the substitutions 
\begin{align}
    & \rho \mapsto i \tau\,, \qquad \tau \mapsto \rho - \frac{i \pi}{2}\,, \qquad \partial_\rho \mapsto -i \partial_\tau\,,\qquad \partial_\tau \mapsto \partial_\rho\,, \qquad  h_{ab} \mapsto -h_{ab}\,, \nonumber \\
    &\sigma \mapsto i \sigma\,, \qquad k_{ab} \mapsto -i k_{ab}\,, \qquad i_{ab} \mapsto i_{ab}\,, \qquad j_{ab}\mapsto j_{ab}-i\frac{\pi}{2}i_{ab}\, ,\label{continuation}
\end{align}
where $j_{ab}$ was called $h^{(2)}_{ab}$ in \cite{compere_relaxing_2011}.

The full Einstein's equation up to this order in the asymptotic expansion will be presented in Section \ref{sec:EOM} after additional assumptions are imposed.  At present, we require only the equations for the first-order fields.  Outside of sources, the vacuum Einstein equations imply
\begin{align}\label{First Einstein}
    &(D^2 - 3) \sigma = 0\,, \qquad 
    D^b k_{ab}=D_a k\,, \qquad (D^2+3)k_{ab}=(D_aD_b+h_{ab})k\,.
\end{align}
Here $k=k^a_{\;\; a}$ is the trace of $k_{ab}$ (with indices raised by $h^{ab}$) and $D_a$ is the derivative on $i^+$ (compatible with $h_{ab}$).

\subsection{Original coordinate freedom}\label{sec:original-coordinate-freedom}

We will consider spacetimes where there exist coordinates such that the metric takes the form \eqref{Beig-Schmidt}, and given such a spacetime, a coordinate system is \textit{admissible} if the metric indeed takes that form.  We insist that $h_{ab}$ is always the hyperboloid metric (equal, component-wise, to $h^+_{ab}$ in \eqref{h+})\footnote{$\,\,$ We require that the components of $h_{ab}$ are exactly those of $h^+_{ab}$ in \eqref{h+}, but this choice is arbitrary.  A choice must be made for the purposes of defining the asymptotic symmetry group, but resulting group is independent of the choice, and our results will be expressed covariantly on $i^+$, such that they can easily be applied in other coordinate systems.}, but the remaining $i^+$ tensors $\sigma,k_{ab},i_{ab},j_{ab}$ will have different values in different admissible coordinate systems. 
 We can define the variation $\delta_\xi$ of these tensors under an admissible infinitesimal bulk diffeomorphism $\xi^\mu$ by the change in their value after the diffeomorphsim is applied.  Any allowed diffeomorphism can then by definition be written
\begin{align}
\mathcal{L}_\xi g_{\tau \tau} & = -\frac{2\delta_\xi \sigma}{\tau}-\frac{2\sigma \delta_\xi \sigma}{\tau^2} + o(\tau^{-2}) \label{lietautau}\,, \\
\mathcal{L}_\xi g_{\tau a} & = o(\tau^{-1}) \label{lietaua}\,, \\
\mathcal{L}_\xi g_{ab} & = \tau (\delta_\xi k_{ab} - 2\delta_\xi\sigma h_{ab}) + \log\tau \delta_\xi i_{ab} + \delta_\xi j_{ab} + o(\tau^0)\,. \label{lieab} 
\end{align}
For consistency of notation, we can also write
\begin{align}
    \delta_\xi h_{ab}=0\,,
\end{align}
which is the condition that the hyperboloid metric be exactly preserved by an allowed diffeomorphism.  Eq.~\eqref{lietaua} can be solved in general as
\begin{equation}
 \xi^b =\chi^b(\phi^a)-\int g^{ab}\partial_a \xi^\tau g_{\tau\tau}\,, \label{xib}
\end{equation}
where the integration constant $\chi^a$ is any vector on $i^+$. To determine the allowed form of $\xi^\mu$, note that the $\tau$ component must take the form 
\begin{align}\label{xitau}
    \xi^\tau = H \log\tau + \omega + A\tau^{-1}\log\tau + B \tau^{-1} + o(\tau^{-1})\,,
\end{align}
where $H, \omega, A, B$ are all functions of $\phi^a$.  Plugging in the expansion \eqref{xitau},  we have
\begin{align}
    \xi^a &= \chi^a - D^aH\frac{\log\tau}{\tau} - D^a(\omega+H) \tau^{-1} + \frac{1}{4\tau^{2}}\bigg( -D^a(A+2B) + k^{ab} D_b(H + 2\omega) \nonumber \\
    &- 4\sigma D^a (H + 2\omega) -\log\tau \big( 2D^a A + 8\sigma D^a H - 2k^{ab}D_b H \big) \biggr) +o(\tau^{-2})\,.\label{xia}
\end{align}
Using \eqref{Beig-Schmidt}, we can now compute $\mathcal{L}_\xi g_{\tau\tau}$ as
\begin{align}
    \mathcal{L}_\xi g_{\tau\tau} &= -\frac{2}{\tau}(H + \chi^aD_a\sigma) +\frac{2}{\tau^2} \big(-A + B +(\omega- 2H)\sigma + D_a\sigma (D^aH + D^a \omega \nonumber\\
    &- \sigma \chi^a) \big) + \frac{2\log\tau}{\tau^2} (A + \sigma H + D_a\sigma D^aH) + o(\tau^{-2})\,,
\end{align}
and compare with \eqref{lietautau} to infer
\begin{align}
    \delta_\xi \sigma &= H + \chi^a D_a\sigma \label{deltaxi sigma}\,,\\
    A &= -\sigma H - D_a\sigma D^a H\,,\\
    B &= -\sigma \omega - D_a\sigma D^a (\omega + 2H)\,.
\end{align}
Finally, using these expressions in \eqref{xitau}, from \eqref{Beig-Schmidt} we can compute 
\begin{align}
    \mathcal{L}_\xi g_{ab} &= \tau^2\mathcal{L}_\chi h_{ab}+\tau\biggr(\mathcal{L}_\chi (k_{ab}- 2\sigma h_{ab})-2 \big((D_aD_b - h_{ab})\omega + D_aD_b H \big)\biggr)\nonumber\\ & - 2\tau \log\tau (D_aD_b - h_{ab})H + o(\tau)\,,\label{Liexigab}
\end{align}
and compare with \eqref{lieab} to determine
\begin{subequations}
\begin{align}
    \mathcal{L}_\chi h_{ab} & = 0\,, \label{KillingH}\\
    (D_aD_b - h_{ab})H & = 0\,,\label{H constraint}\\
    \delta_\xi k_{ab} & =  \mathcal{L}_\chi k_{ab} -2 (D_aD_b - h_{ab})\omega\,.\label{deltaxik}
\end{align}
\end{subequations}
This computation could be extended to $o(\tau^0)$ to obtain $\delta_\xi i_{ab}$ and $\delta_\xi j_{ab}$, but such variations are not required here.  We will present these quantities in Eqs. \eqref{deltaxiifinal}-\eqref{deltaxijfinal} below after adopting some additional assumptions.

The general solution to Eq.~\eqref{KillingH} is spanned by the 6 Killing vectors on the hyperboloid, written out explicitly below in Eq.~\eqref{chiY}.  These vectors satisfy $D_a \chi^a=0$ and $(D^2-2)\chi^a=0$. Further properties can be found in Appendix B of \cite{compere_asymptotic_2011}.  We can determine the general solution to \eqref{H constraint} for $H$ as follows. The trace of \eqref{H constraint} gives $(D^2-3)H=0$, whose solutions were analyzed in \ref{sec:equation}.  Of these solutions, only the $\ell=0,1$  modes of the origin-regular branch, $\psi^{\mathcal{O}}_0$ and $\psi^{\mathcal{O}}_1$, satisfy the original equation \eqref{H constraint}.  Noting the simple forms \eqref{psiO0} and \eqref{psiO1} of these modes,  the general solution for $H$ is the four-parameter family
\begin{align}\label{H}
    H = h^0 \cosh \rho - h^i n_i \sinh \rho\,,
\end{align}
where we name the constants $h^\mu=(h^0,h^i)$ and introduce 
\begin{align}
n_i=(\sin \theta \cos \phi, \sin \theta \sin \phi, \cos \theta)\,.
\end{align}
These are the Cartesian components of the unit normal to the unit two-sphere in three-dimensional Euclidean space, but here they function as a set of three scalars that span the $\ell=1$ spherical harmonics.  We will raise and lower such Cartesian indices $i,j,\dots$ with the Euclidean metric $\delta_{ij}$.

\subsection{Additional assumptions}

At present, the coordinate freedom in the formalism is ten parameters (six hyperboloid Killing fields $\chi^a$ and four numbers $h^\mu$ determining $H(\phi^a)$) and a free function $\omega(\phi^a)$ on $i^+$.  By contrast, the BMS group that we wish recover is described by six parameters (three rotations and three boosts) and and a free function on the sphere (translations and supertranslations).  We now find further assumptions cutting the freedom down to the BMS group.  Happily (but not coincidentally), we will see that these assumptions also enable a convenient match to null infinity.

In finding appropriate assumptions, we are motivated by the idea that at late times the bodies should no longer be interacting or evolving.  In particular, the $O(1/\tau)$ Beig-Schmidt tensors $\sigma$ and $k_{ab}$ should be determined by a sum of the $O(1/r)$ terms of stationary spacetimes that have boosted to the appropriate asymptotic velocity.  These terms share the same form as a boosted Schwarzschild spacetime, which we consider in \ref{sec:schwarzschild}.  We find that $\sigma$ has a pole at the point of $i^+$ corresponding to the boosted velocity and in fact satisfies an equation \eqref{eqndelta} with an effective point source.  We therefore promote the homogeneous equation \eqref{First Einstein} for $\sigma$ by adding an appropriate sum over point sources,
\begin{align}\label{sigma-point}
    (D^2-3)\sigma = \sum_{n=1}^N 4\pi M_n \frac{\delta^{(3)}(\phi-\phi_n)}{\sqrt{h}}\,.
\end{align}
We regard $M_n$ as the masses of $N$ outgoing bodies with asymptotic velocities corresponding to the points $\phi^a_n=(\rho_n,\theta_n,\phi_n)$ on $i^+$, an interpretation which will be confirmed when we define charges on $i^+$.  We emphasize that we are not modeling the massive bodies as point particles; the point sources in \textit{velocity} space $i^+$ reflect the presence of isolated, finite-size isolated bodies which approach definite velocities at late times.  It would be interesting to begin with a more fundamental assumption reflecting the presence of $N$ isolated bodies and prove Eq.~\eqref{sigma-point} directly as a consequence of the limiting process.  For now, it serves as a convenient assumption for the formulating the notion of a spacetime containing asymptotically isolated bodies.

We also find that, in Schwarzschild spacetime, it is possible to find a gauge where $\sigma$ vanishes at large $\rho$.  We will impose this condition in general,
\begin{align}\label{sigmavanish}
    \lim_{\rho \to \infty} \sigma = 0\,.
\end{align}
This assumption fixes a unique solution to Eq.~\eqref{sigma-point} (see \ref{sec:equation}), and helps us match to null infinity (see Sec.~\ref{sec:i+scri+}).  It also removes the logarithmic translation degree of freedom $H$, since $\sigma \to \sigma + H$ under a logarithmic translation \eqref{deltaxi sigma}, but $H$ blows up at infinity as is evident from \eqref{H}.

We also find that, in Schwarzschild spacetime, it is possible to find a gauge where $k_{ab}=0$.  However, we can see by two (related) lines of reasoning that imposing this condition in general would be too physically restrictive.  First, setting $k_{ab}=0$ would reduce our coordinate freedom to the Poincar\'e group instead of the BMS group (see discussion below \eqref{BMSexplicit}).  Second, we will see that $k_{ab}$ matches to the Bondi shear $C_{AB}$ at null infinity [Eq.~\eqref{kAB match} below].  This occurs both for $i^+$ matching with $\scri^+$ and for $i^-$ matching with $\scri^-$.  In a spacetime with gravitational memory, the Bondi shear will be non-zero either in the future or the past; thus imposing $k_{ab}=0$ everywhere would eliminate spacetimes with gravitational memory.  In our formalism, the change in $k_{ab}$ between $i^-$ and $i^+$ will encode the total gravitational memory [Sec.~\ref{sec:gravitational-memory}].

Rather than imposing that $k_{ab}$ must vanish, we will instead simply impose that it is pure gauge.  This is a physical condition on our spacetimes, motivated by the fact that $k_{ab}$ is pure gauge in Schwarzschild spacetime.  From Eq.~\eqref{deltaxik}, $k_{ab}$ must then take the form 
\begin{align}
\label{kfromPhi}
    k_{ab} = -2(D_aD_b - h_{ab})\Phi\,,
\end{align}
for some scalar $\Phi$.  We also assume that $\Phi$ (and hence $k_{ab}$) is smooth.

The curl of $k_{ab}$, $B^1_{ab}\equiv  \frac{1}{2}\epsilon_{acd}D^c k^{d}_{\;\;b}$, is proportional to the leading part of the magnetic part of the Weyl tensor along the $\partial_\tau$ direction in the expansion for large $\tau$, namely $\epsilon_{a\tau pq}C^{pq}_{\;\;\;\;b\tau}$ \cite{Ashtekar:1978zz}.  The field $B_{ab}^{1}$ identically vanishes under the assumption \eqref{kfromPhi}. The inclusion of spacetimes that admit a non-vanishing $B_{ab}^{1}$ (see e.g. \cite{Satishchandran:2019pyc}) would require relaxing our assumptions.

At present, this scalar $\Phi$ is an arbitrary function on $i^+$, since the gauge freedom in $k_{ab}$ is an arbitrary scalar $\omega$ \eqref{deltaxik} on $i^+$.  However, the BMS group should feature an arbitrary scalar \textit{on the sphere}, not this more general functional dependence.  It turns out that we can reduce down to the BMS group by imposing that the trace $k$ of $k_{ab}$ vanishes,
\begin{align}\label{tracek}
 k \equiv   k^a{}_a = 0 \qquad \Leftrightarrow \qquad (D^2-3)\Phi=0\,.
\end{align}
To see that this condition can always be imposed, suppose that $k_{ab}$ has some non-zero trace $k$.  Then we can set it to zero by solving the equation $(D^2-3)\omega=k/2$ for $\omega$ and making the corresponding coordinate transformation (see Eq.~\eqref{deltaxik}).  The analysis of \ref{sec:equation} shows that a solution always exists.  With \eqref{tracek} imposed, the freedom in the choice of $\omega$ is restricted to
\begin{align}\label{omegaeqn}
    (D^2 - 3)\omega = 0\,.
\end{align}
Since $\omega$ cannot have poles, it follows from the mode analysis (\ref{sec:equation}) that the solution is determined by a single function on the sphere, as expected for the BMS group.  We shall see shortly that the remaining coordinate freedom is indeed precisely the BMS group.

\subsection{Final coordinate freedom}

Let us now revisit the infinitesimal diffeomorphisms taking into account the vanishing of $H$ imposed by \eqref{sigmavanish}.  (The other new condition \eqref{omegaeqn} does not affect the form of the generators.)  The generator \eqref{xitau} and \eqref{xia} now takes the form 
\begin{align}
    \xi^\tau & = \omega - \tau^{-1} (\sigma \omega + D_a \sigma D^a \omega) + o(\tau^{-1})\,, \label{xitaufinal} \\
    \xi^a & = \chi^a - \tau^{-1} D^a \omega + \frac{1}{2\tau^{2}}\bigg( -D^a(\sigma \omega + D_b \sigma D^b \omega) +  k^{ab} D_b\omega - 4\sigma D^b \omega \biggr) +o(\tau^{-2})\,,\label{xiafinal}
\end{align}
where $\chi^a$ is a Killing field on the hyperboloid and $\omega$ is a scalar satisfying \eqref{omegaeqn}. We can now evaluate Eq.~\eqref{Liexigab} to order $o(\tau^0)$ to obtain the transformation laws for the second-order tensors $i_{ab}$ and $j_{ab}$.  Displaying these along-side the already computed laws \eqref{deltaxi sigma} and \eqref{deltaxik} for the first order quantities, we have 
\begin{subequations}\label{deltaxiall}
\begin{align}
    \delta_\xi \sigma & = \mathcal{L}_\chi \sigma \,,\label{deltaxisigmafinal} \\
    \delta_\xi k_{ab} & = \mathcal{L}_\chi k_{ab}-2 (D_a D_b - h_{ab}) \omega\,, \label{deltaxiomegafinal} \\
\delta_\xi i_{ab} &= \mathcal{L}_\chi i_{ab} \,,\label{deltaxiifinal} \\
    \delta_\xi j_{ab} &=\mathcal{L}_\chi j_{ab} -\omega^cD_ck_{ab} +k_{ab} \omega + \omega_c D_{(a}k_{b)}{}^c - k_{(a}{}^c\omega_{b)c}\nonumber\\&- 4\sigma\omega h_{ab} - 4\sigma_{(a}\omega_{b)} +D_aD_b (\sigma \omega + \sigma^c\omega_c)\,,\label{deltaxijfinal}
\end{align}
\end{subequations}
where $\omega_a = D_a\omega$, $\omega_{ab} = D_a\omega_b$, and similarly for $\sigma$. The latter variation \eqref{deltaxijfinal} matches with Eq. (B.23) of  \cite{chakraborty_supertranslations_2022} after the change of convention $\omega\rightarrow -\omega$, and after dropping the $O(\omega^2)$ terms since we only consider an infinitesimal variation.  Eqs. \eqref{kfromPhi}-\eqref{deltaxiomegafinal} imply 
\begin{equation}\label{deltaPhi}
\delta_\xi \Phi = \mathcal L_\chi \Phi +\omega\,.
\end{equation}

Eqs.~\eqref{xitaufinal}-\eqref{xiafinal} are the full infinitesimal coordinate freedom of the formalism, and Eqs.~\eqref{deltaxisigmafinal}--\eqref{deltaxijfinal} gives its action on the fields on $i^+$.  Given two generators $\xi_1$ and $\xi_2$, it is easy to check that the difference between applying $\xi_2$ then $\xi_1$ and doing the opposite order ($\xi_1$ then $\xi_2$) is equivalent to applying the generator given by the  ``adjusted'' Lie bracket
\cite{Barnich:2010eb} 
\begin{equation}
[\xi_1,\xi_2]_\star \equiv     [\xi_1,\xi_2] - \delta_{\xi_1} \xi_2 + \delta_{\xi_2} \xi_1\,, \label{star bracket}
\end{equation}
where $[,]$ is the ordinary commutator of vector fields.  The action of $\delta_{\xi_1}$ on another generator $\xi_2$ is defined by the chain rule: $\xi_2$ depends on the $i^+$ fields $X = \{\sigma,k_{ab},i_{ab},j_{ab} \}$ via Eqs.~\eqref{xitaufinal}--\eqref{xiafinal}, and their variation is then given by Eqs.~\eqref{deltaxisigmafinal}--\eqref{deltaxijfinal}. The Lie derivative acting on the metric reduces to a variation of the fields $X$: $\mathcal L_\xi g_{\mu\nu}=\delta_{\xi} g_{\mu\nu}(X)\equiv \frac{d}{d\epsilon}g_{\mu\nu}(X+\epsilon \delta_\xi X)$ where the variations $\delta_\xi X$ are listed in Eqs. \eqref{deltaxiall}. The Lie derivative of the adjusted Lie bracket then has the property $\mathcal L_{[\xi_1,\xi_2]_*} g_{\mu\nu} = \delta_{[\xi_1,\xi_2]} g_{\mu\nu}(X)$ with $\delta_{[\xi_1,\xi_2]}X=\delta_{\xi_1}\delta_{\xi_2}X - \delta_{\xi_2}\delta_{\xi_1}X$ where the variation $\delta_\xi$ only acts on $X$, not on the parameters $(\chi,\omega)$ of $\xi$.

Below we will define a charge associated with each allowed generator $\xi$.  Given such a definition, one can define the asymptotic symmetry group as the allowed transformations modulo the trivial transformations (with vanishing associated charge).  In the remainder of this section we will show that in our framework, to the order in $\tau^{-1}$ in which we work, all allowed transformations are non-trivial.  In this sense our framework is gauge-fixed, so the Lie algebra under the adjusted Lie bracket \eqref{star bracket} is the asymptotic symmetry algebra, to which we now turn our attention.

\subsection{Sphere parameterization and BMS algebra}

The general form of an allowed infinitesimal diffeomorphism is given in Eqs.~\eqref{xitaufinal} and \eqref{xiafinal}, parameterized by two tensors on $i^+$: a scalar $\omega$ satisfying \eqref{omegaeqn}, and a vector $\chi^a$ satisfying $\mathcal{L}_\chi h_{ab}=0$ (i.e., a Killing field of $i^+$).  To prepare for matching with null infinity and also to relate more directly to associated standard parameterizations of symmetries and conserved quantities, it is convenient to express these in terms of tensors on the celestial sphere.

First consider $\omega$.  Given some smooth function $T(\theta,\phi)$ on the sphere,
 \begin{align}\label{T}
     T = \sum_{\ell,m} T_{\ell m} Y_{\ell m}(\theta,\phi)\,,
 \end{align}
one can uniquely associate a (regular) solution to Eq.~\eqref{omegaeqn}
\begin{align}\label{omegaT}
    \omega_T =  \sum_{\ell,m} T_{\ell m} \psi^\mathcal{O}_\ell(\rho) Y_{\ell m}(\theta,\phi)\,,
\end{align}
where the modes $\psi_\ell^{\mathcal O}$ are given in Eq.~\eqref{psiellO}.
Conversely, all regular solutions to \eqref{omegaeqn} take this form: there is a one-to-one correspondence between functions on the sphere and valid choices of $\omega$.  We will usually represent $\omega$ by the corresponding choice of $T(\theta,\phi)$. 

Sometimes it is convenient to separately discuss individual modes.  Time translations are represented by the $\ell=0$ mode, while space translations are represented by appropriate real combinations of the $\ell=1$ modes,
\begin{align}\label{omega0}
    \omega_0 & = \cosh \rho \qquad (\textrm{time translation}) \,,\\
    \omega_i & = -n_i \sinh \rho \qquad (\textrm{space translations})\,.\label{omegai}
\end{align}
Here we introduced the Cartesian components of the unit normal to the sphere,
\begin{align}\label{ni}
     n_i=(\sin \theta \cos \phi, \sin \theta \sin \phi, \cos \theta)\,,
\end{align}
which are regarded as three scalar functions on the sphere.  These scalars span the space of smooth $\ell=1$ functions on the sphere.

To justify the names in \eqref{omega0} and \eqref{omegai}, one can check directly that in flat spacetime for these choices of $\omega_{\nu}$ with $\chi^a=0$, the corresponding asymptotic symmetry $\xi^\mu$ is time translation when $\nu=0$ and the $i^{\rm th}$ space translation when $\nu=i$.  The higher modes are by analogy called (pure) supertranslations,
\begin{align}
    \omega_{\ell m} = T_{\ell m} \psi^{\mathcal{O}}_\ell(\rho) Y_{\ell m}(\theta,\phi) \qquad (\ell\geq 2: \textrm{pure supertranslations})\,.
\end{align}
We will collectively denote translations and pure supertranslations as supertranslations.

Now consider $\chi^a$.  Given a conformal Killing field $Y^A$ on the sphere,
\begin{align}\label{Y}
Y^A=\nabla^A (b^i n_i)-\epsilon^{AB}\nabla_B(\kappa^i n_i)\,,
\end{align}
we may uniquely associate a Killing field $\chi_Y^a$ of the hyperboloid by
\begin{align}\label{chiY}
    \chi_Y^\rho = b^i n_i\,, \qquad
    \chi_Y^A =  \coth \rho \nabla^A (b^i n_i) - \epsilon^{AB} \nabla_B (\kappa^i n_i)\,.
\end{align}
Conversely, all hyperboloid Killing fields $\chi^a$ determine a sphere conformal Killing field $Y^A$.  We will usually represent $\chi^a$ by the corresponding choice of $Y^A$.  In these equations, the covariant derivative $\nabla_A$ could be replaced by a coordinate derivative $\partial_A$.

Sometimes it is convenient to separately discuss individual Killing fields.  We have chosen the parameterization so that $b_i$ represents boosts and $\kappa_i$ represents rotations.  For example, working in flat spacetime and picking $b_i=(0,1,0)$, $\kappa_i=(0,0,1)$, and $\omega=0$ makes Eq.~\eqref{xitaufinal}--\eqref{xiafinal} become a linear combination of a boost in the $y$ direction and a rotation about the $z$ axis.  Most explicitly, we can write in general using $\epsilon^{\theta\phi}=+\csc\theta$
\begin{align}
    \chi^a_{b_i} & = (n_i,\coth \rho \pd_\theta n_i, \coth \rho \csc^2 \! \theta \pd_\phi n_i) \qquad \textrm{(boosts)}\, , \\
    \chi^a_{\kappa_i} & = (0,- \csc \theta \pd_\phi n_i,  \csc \theta \pd_\theta n_i) \qquad \qquad \quad \!  \textrm{(rotations)}\,.
\end{align}

Finally, we can introduce the unified notation that $\xi_T$ represents the vector constructed from \eqref{xitaufinal}--\eqref{xiafinal} using $\omega=\omega_T$ \eqref{omegaT} with $\chi^a=0$, while $\xi_Y$ represents the vector constructed using $\chi^a=\chi^a_Y$ \eqref{chiY} with $\omega=0$.  The adjusted Lie brackets \eqref{star bracket} of these vector fields are then found to take the following form to the order in $\tau$ we consider,\footnote{$\,\,$ If we had gauge-fixed to all orders in $\tau$, these expressions would hold exactly.}
\begin{align}
 [\xi_{T_1},\xi_{T_2}]_\star &= 0\,, \label{beforemixedC}\\
 [\xi_{T_1},\xi_{Y_2}]_\star &= \xi_{Y_2(T_1)}\,, \label{mixedC}
 \\
  [\xi_{Y_1},\xi_{Y_2}]_\star &= \xi_{[Y_1,Y_2]}\,,\label{aftermixedC}
\end{align}
where $[Y_1,Y_2]$ is the commutator on the sphere of the conformal Killing fields, and we defined the scalar function over the sphere
\begin{align}\label{YT}
    Y(T) \equiv Y^A \partial_A T - \frac{1}{2} \nabla_A Y^A T\,.
\end{align}
This is the BMS algebra as defined in sphere-covariant treatments such as \cite{Campiglia:2020qvc,Compere:2020lrt}. One can recover the classical presentation of the BMS algebra in modes as follows. First, one expands $\xi_Y$ as a linear combination of rotations $\xi_{\kappa_i}$ and boosts $\xi_{b_i}$. Second, one expands 
the (super-)translations as a sum of symmetric trace-free tensors $T=\sum_{\ell \geq 0}T_L n_L(\theta,\phi)$ where we use  the multi-index notation $L$ to designate any symmetric set of $\ell$ indices, $L=A_1\dots A_\ell$, $n_L$ is defined as $n_{A_1}\dots n_{A_\ell}$, and $T_L$ is symmetric and trace-free.  A translation is a combination of the 4 lower harmonics and a pure supertranslation is a combination of the $\ell \geq 2$ harmonics. Expanding the adjusted Lie bracket \eqref{mixedC} one obtains 
\begin{subequations}\label{BMSexplicit}
\begin{align}
 [\xi_{T},\xi_{\kappa_i}]_\star &= \xi_{T_i'}\,, \qquad T_i'=\epsilon_{kij}n_k T_j+\sum_{\ell \geq 2}\ell  \epsilon_{kij}n_{k}n_{L-1} T_{jL-1}\,,\\
  [\xi_{T},\xi_{b_i}]_\star &= \xi_{T_i''}\,, \qquad T_i''=T_i + n_i T + \sum_{\ell \geq 2} (\ell T_{i L-1}n_{L-1}-(\ell-1)T_L n_i n_L)\,,\label{BMSexplicit2}
\end{align}
\end{subequations}
where we used $n_k\epsilon_{kij}=\epsilon^{AB}\partial_A n_i \partial_B n_j$. 

From these relations, one obtains that the Lorentz algebra is a subalgebra of the BMS algebra but it is not an ideal of the BMS algebra because pure supertranslations appear in the right-hand side of \eqref{BMSexplicit}. The pure supertranslations also do not form an ideal subalgebra because of the term  $2T_{ij}n_j$ in the right-hand side of Eq. \eqref{BMSexplicit2}: ordinary translations appear in the commutator of a supertranslation and a boost. One cannot therefore quotient out the supertranslations to define the angular momentum using only the BMS algebra generators.

Had we wanted to obtain the Poincar\'e group instead of the BMS group, we could have demanded the stronger condition $k_{ab}=0$ instead of $k^a_a=0$, following the original work at $i^0$ \cite{beig_einsteins_1982}.  Eqs.~\eqref{deltaPhi} and \eqref{kfromPhi} show that the supertranslations are then eliminated from the formalism, and the BMS commutators reduce to the Poincar\'e algebra.

\subsection{Einstein equations}\label{sec:EOM}

 Finally, for completeness we present the field equations through second order in the Beig-Schmidt expansion.  Away from any sources, we may use the vacuum Einstein equations. Taking into account the trace-free condition $k^a{}_a=0$ adopted in \eqref{tracek}, the expansion \eqref{Beig-Schmidt} implies that away from sources we have
\begin{align}
    & \ && \ && (D^2 - 3) \sigma = 0 \,,\label{sigmavac} \\
    &k_a{}^a = 0\,, && D^a k_{ab} = 0\,, && (D^2+3)k_{ab} = 0\,,\\
    & i_a{}^a = 0\,, && D^a i_{ab} = 0\,, && (D^2 + 2)i_{ab} = 0\,,
\end{align}
and
\begin{align}
    j_a{}^a& = 12\sigma^2 - D_c\sigma D^c \sigma + \frac{1}{4}k_{cd}k^{cd}- k_{cd}D^cD^d\sigma\,,\\
     D^aj_{ab}& = \frac{1}{2}D^bk_{ac}k_b{}^c + D_a\big(8\sigma^2 - D^c\sigma D_c\sigma - \frac{1}{8} k_{cd}k^{cd} - k_{cd}D^cD^d\sigma\big)\,,\\
     (D^2 + 2)j_{ab} &= -2 i_{ab} + \text{NL}^{(\sigma,\sigma)}_{ab} + \text{NL}^{(\sigma,k)}_{ab}+\text{NL}^{(k,k)}_{ab}\,,
\end{align}
where $\text{NL}_{ab}$ are quadratic terms in $\sigma$ and $k$,
\begin{align}
    \text{NL}^{(\sigma,\sigma)}_{ab} &= D_aD_b(5\sigma^2 - D_c\sigma D^c \sigma) + h_{ab} (18\sigma^2 + 4D_c\sigma D^c \sigma) + 4\sigma D_a D_b \sigma\,,\\
    \text{NL}^{(\sigma, k)}_{ab} & = - D_aD_b(k_{cd}D^cD^d\sigma) - 2 h_{ab}k_{cd}D^cD^d\sigma- 4\sigma k_{ab} + 4D^c\sigma\big(D_{(a}k_{b)c} - D_ck_{ab}\big) \nonumber\\&+ 4(D_{c}D_{(a}\sigma)k^c{}_{b)}\,,\\
    \text{NL}^{(k,k)}_{ab} &= - k_{ac}k^c{}_b + k^{cd} \big(- D_dD_{(a}k_{b)c} + D_cD_dk_{ab}\big) - \frac{1}{2}D_bk^{cd} D_ak_{cd} + D^dk_{c(a} D_{b)}k_d{}^c\nonumber\\&+ D_c k_{ad} D^ck_b{}^d - D_ck_{ad} D^dk^c{}_b\,. \label{nlkkvac}
\end{align}
These expressions may be derived from the analytic continuation \eqref{continuation} of the equations in appendix C of Ref.~\cite{compere_relaxing_2011}.  The results match with the result in appendix C in \cite{chakraborty_supertranslations_2022} after setting $i_{ab}=0$.

 These equations hold away from any poles in the fields.  As already discussed above  \eqref{sigma-point}, there are poles in $\sigma$ at the velocities corresponding to the outgoing bodes.  These poles are inherited by $j_{ab}$ via the appearance of $\sigma$ in its source terms, as demonstrated in the example of Schwarzschild  spacetime in \ref{sec:schwarzschild}.  We suspect that $i_{ab}$ will also have such poles.  By contrast, we have insisted that $k_{ab}$ be smooth everywhere (and in fact pure gauge \eqref{kfromPhi}).  We leave a more detailed analysis of $j_{ab}$ and $i_{ab}$ to future work.

\subsection{Conserved charges}

The charges associated with these symmetries may be derived from analytic continuation of results in \cite{compere_relaxing_2011}, as follows. We write $a \mapsto b$ where $a$ uses the notation of \cite{compere_relaxing_2011}, and $b$ uses the notation of this paper at $i^+$.  With this convention, the symmetry generators are mapped under the analytic continuation \eqref{continuation} as $\omega \mapsto -i\omega_T$, $\xi_{(0)}^a \mapsto -\chi^a_Y$, while $n^a\partial_a=\partial_\tau \mapsto r^a\partial_a=\partial_\rho$.  Using these expressions, the charges at spatial infinity, (4.88) and (4.95) in \cite{compere_relaxing_2011}, give the charges at $i^+$ as\footnote{Our Lorentz charge $Q_Y$ is the analytic continuation of $\mathcal{Q}_{+\xi_{(0)}}$ in Eq.~(4.95) in Ref.~\cite{compere_relaxing_2011}.}
\begin{align}
    Q^{i^+}_{T} & = \frac{1}{4\pi} \int_{C}\sqrt{q}d^2x r^a( \omega_T D_a \sigma - \sigma D_a \omega_T) \,,\label{QTi+} \\
    Q^{i^+}_{Y} & = \frac{1}{8\pi}\int_C \sqrt{q}d^2x r^a\chi_Y^b\biggr(-j_{ab} +\frac{1}{2}i_{ab} + \frac{1}{2}k_{ac}k_{b}{}^c \nonumber \\ &+ h_{ab}\big(-\frac{1}{8}k_{cd}k^{cd} + 8\sigma^2 -k_{cd}D^{c}D^d\sigma -D_c\sigma D^c\sigma\big)\biggr),\label{QYi+}
\end{align}
where the integral is taken over a closed two-surface $C$ in $i^+$ with induced metric $q$ and outward pointing normal vector $r^a$.  The first formula provides the translation and supertranslation charges, where $\omega_T$ is given by \eqref{omegaT}. The expression \eqref{QTi+} for the supertranslation charges is identical to those provided in  \cite{1982RSPSA.381..323P,1998JMP....39.6573G,chakraborty_supertranslations_2022}. The Lorentz charges \eqref{QYi+} match with \cite{chakraborty_supertranslations_2022} after setting $i_{ab}=0$ and with \cite{1982RSPSA.381..323P,1998JMP....39.6573G} after setting $i_{ab}=k_{ab}=0$.  The normalization is such that the energy, momentum, angular momentum, and mass moment\footnote{$\,\,$ For smooth stress-energy $T_{\mu \nu}$ in Cartesian coordinates in flat spacetime, the mass moment becomes $N^i = \int T_{00} x^i - P^i t$.  It can be interpreted as the energy multiplied by the center of mass at time $t=0$.} are given by 
\begin{align}
    E & = Q^{i^+}_{T}|_{\omega_T = \cosh \rho}\,, \label{chargeE} \\
    P^i & = Q^{i^+}_{T}|_{\omega_T = n_i \sinh \rho}\,, \label{chargep} \\
    L^i & = Q^{i^+}_{Y}|_{\chi_Y^\rho = 0, \ \chi_Y^A = - \epsilon^{AB} \pd_B n_i} \,,\label{chargeL} \\
    N^i & = Q^{i^+}_{Y}|_{\chi_Y^\rho =  n_i, \ \chi_Y^A = \coth \rho \pd^An_i }\,, \label{chargeN}
\end{align}
where $n_i$ was defined in \eqref{ni}.  In our conventions, the energy is conjugate to a future-directed time-translation and the mass moment is conjugate to a future-directed boost.  This requires the momentum to be conjugate to a space translation in the opposite direction, and the angular momentum to be conjugate to left-handed rotations.  For example, the $x$ component of momentum involves the choice of $\omega$ corresponding to the Killing field $-\pd_x$ of flat spacetime, and the $z$ component of angular momentum similarly involves $-\pd_\phi$. In \ref{sec:schwarzschild} we check that $E=M$ for a Schwarzschild black hole of mass $M$ (provided the integration surface encloses the black hole).  Sometimes it is useful to discuss individual supermomentum charges.    We define
\begin{align}
    P_{\ell m} = Q^{i^+}_T|_{\omega_T = f^\mathcal{O}_\ell(\rho) Y_{\ell m}} \qquad (\ell \geq 2)\,. \label{chargePlm}
\end{align}

The charge integrals \eqref{QTi+} and \eqref{QYi+} are analogous to Gauss' law in the presence of point charges. Indeed, they can be written as $\int_C \sqrt{q}d^2x J_a r^a$ where $D_a J^a =0$ outside of sources as a consequence of the vacuum Einstein equations \eqref{sigmavac}--\eqref{nlkkvac}. They are therefore independent of the surface $C$ when the deformation of the surface lies in a region where the fields are smooth (no poles), as independently shown in \cite{chakraborty_supertranslations_2022}.  We may define the charge associated with an individual body by choosing a surface enclosing only that body.  If the integration surface is pushed to large $\rho$, then it represents the total charge on $i^+$, which is just the linear sum of the charges of each body.  We may interpret this linearity as the fact that the bodies do not interact at late  times.\footnote{$\,\,$ In comparison, this linearity breaks down on $t={\rm const}$ slices in the electromagnetic case, where field interaction energy makes a contribution to the mass moment, even at late times \cite{Gralla2022b}.}

If one defines final Lorentz charges using the coordinate positions of outgoing point particles in the special-relativistic formulas for angular momentum and mass moment, the individual particle contributions are sometimes logarithmically divergent at late times (e.g., Eq. 3.4 of \cite{Sahoo:2018lxl}).   The final charges we consider here are manifestly finite, but it is interesting to point out that their derivation involves the subtraction of a logarithmic divergence of exactly this kind.  In particular, the large-$\rho$ cutoff used in \cite{compere_relaxing_2011} to define the action at $i^0$ (see Eq. (4.34) therein) maps to a large-$\tau$ cutoff at $i^+$ as $\Lambda \mapsto i \Lambda$, given the continuation $\rho \mapsto i\tau$ relating the two asymptotic regions.  Since $\tau$ can be interpreted as the proper time of particles reaching $i^+$, $\Lambda$ is now interpreted as a late-proper-time cutoff.  The procedure defined in \cite{compere_relaxing_2011} then implies that the Einstein-Hilbert action is regularized in the infrared by the addition of the counterterm
\begin{equation}\label{Sct}
S_{\rm ct}=-\frac{\log \Lambda}{8\pi }\int_{i^+} d^3 x \sqrt{h} (D_a \sigma D^a \sigma + 3 \sigma^2),
\end{equation}
which is the analytic continuation of Eq.~(4.34) of Ref.~\cite{compere_relaxing_2011}.\footnote{$\,\,\,$ The analytic continuation of the second counter-term $S^{(k)}$ in (4.36) of \cite{compere_relaxing_2011} does not contribute to the charges.  To see this, recall that that the leading magnetic part of the Weyl tensor $B^1_{ab}$ vanishes in our framework (see discussion below Eq.~\eqref{kfromPhi}).  The analytic continuation of Eq. (4.74) of \cite{compere_relaxing_2011} thus vanishes, implying that the associated counter-term contribution to the charges also vanishes.}  (Here the integration over $i^+$ is understood to exclude the locations of the bodies.) This counterterm contributes to the symplectic structure and therefore to the canonical charges, giving the contribution
\begin{equation}
Q^{i^+}_{Y,{\rm ct}} \equiv \frac{\log \Lambda}{8\pi }\int_C d^2 x\, \sqrt{q}  r^a \chi_Y^b  i_{ab},  \label{Qct}
\end{equation}
which is the analytic continuation of Eq.~(4.91) of \cite{compere_relaxing_2011}.  This contribution is taken into account in the derivation of the finite Lorentz charge Eq.~\eqref{QYi+}.  We shall see, however, the counter-term does not contribute to the total charge, when $C=i^+_\partial$ (see discussion after Eq.~\eqref{i+ Y match}).

We now consider the transformation laws for the charges under allowed diffeomorphisms.  According to general results \cite{Barnich:2007bf,Compere:2019qed}, the change $\delta_\xi Q_\chi$ in a conserved and integrable charge $Q_\chi$ under an allowed infinitesimal diffeomorphism $\xi$ is antisymmetric under $\xi \leftrightarrow \chi$ and in fact defines a Lie bracket $\{ Q_{\xi}, Q_{\chi}\}\equiv -\delta_\xi Q_\chi$\footnote{$\,\,$ The sign convention of the bracket is aligned with our convention $\delta_\xi g_{\mu\nu}=+\mathcal L_\xi g_{\mu\nu}$.}.  The charges obey the same BMS algebra of the generators,
\begin{align}
 \{ Q_{T_1},Q_{T_2}\} &= 0\,, \label{BMSTT}\\
 \{Q_{T_1},Q_{Y_2}\} &= Q_{Y_2(T_1)}\,, \label{BMSTY}\\
  \{Q_{Y_1},Q_{Y_2}\} &= Q_{[Y_1,Y_2]}\,. \label{BMSYY}
\end{align}
This is another expression of the BMS symmetry of the formalism.

\subsection{Solutions for first order fields}

In this section we find the explicit general solution for the first-order fields $\sigma$ and $k_{ab}$.  We also present their large-$\rho$ limits, which will help elucidate the matching to null infinity.  The analogous calculations for the second-order fields would be far more involved, and we do not perform them in this paper.

The unique solution to \eqref{sigma-point} that vanishes at infinity is (\ref{sec:schwarzschild}) 
\begin{align}\label{sigma-nbodies}
    \sigma = \sum_{n=1}^N M_n \left(2\chi_n - \frac{2\chi_n^2 - 1}{\sqrt{\chi_n^2 - 1}}  \right), \qquad
    \chi_n = \gamma_n(\cosh \rho - v_n^i n_i \sinh\rho)\,,
\end{align}
where $v_n^i$ are the Cartesian components of the outgoing velocity of each the $n^{\rm th}$ particle, while the sphere unit normal $n_i$ was given in Eq.~\eqref{ni}.  Eq.~\eqref{sigma-nbodies} has poles at the hyperboloid points corresponding to the outgoing velocities.  To see this, note that $\chi_n=1$ occurs in particular at the location $(\rho_n,\theta_n,\phi_n)$ reached by a particle hitting $i^+$ with velocity $v^i_n = v_n n_i(\theta_n,\phi_n)$ and rapidity $\sinh \rho_n = \gamma v_n$ (implying $\cosh \rho_n = \gamma_n$).  The large-$\rho$ behavior is
\begin{align}
    \sigma \sim \sum_{n=1}^N \frac{-2 M_n}{\gamma_n^3(1-v_n^in_i )^3}e^{-3\rho}, \quad \rho \to \infty\,.\label{sigmalargerho}
\end{align}
Readers familiar with null infinity will notice the appearance of the formula for the late-time Bondi mass aspect $m$.  The detailed matching is discussed in Sec.~\ref{sec:i+scri+}. 

The charges $Q_n$ of each body are defined by taking the integration surface to surround only that body.  The supertranslation charge $Q_T$ \eqref{QTi+} depends linearly on $\sigma$, meaning that each body's charge $Q_{T,n}$ may be computed only from only its own field, i.e., the $n^{\rm th}$ term in the sum \eqref{sigma-nbodies}.  Each such term is separately a homogeneous solution to \eqref{sigma-point} away from its pole, and hence the integral is independent of surface.  Pushing the surface to infinity, we may use the $n^{\rm th}$ term in \eqref{sigmalargerho} to compute $Q_{T,n}$. The integral \eqref{QTi+} then becomes
\begin{align}\label{QTn}
    Q_{T,n} = \frac{1}{4\pi}\int_{S^2} \frac{ M_n}{\gamma_n^{3}(1-v_n^i n_i(x^A) )^3}T(x^A) d\Omega\,.
\end{align}
In particular, the energy and momentum are
\begin{align}\label{EnPn}
    E_n = \gamma_n M_n\,, \qquad P^i_n = \gamma_n M_n v^i_n = E_n v^i_n\,,
\end{align}
as expected.

We now turn to $k_{ab}$, which satisfies $k_{ab}=-2(D_a D_b-h_{ab})\Phi$ \eqref{kfromPhi} with $(D^3-3)\Phi=0$ \eqref{tracek}. The general regular solution for $\Phi$ may be expressed a multipole expansion, analogously to \eqref{omegaT}.  Since the operator $D_a D_b - h_{ab}$ annihilates $\ell=0$ and $\ell=1$ modes in this expansion, without loss of generality we make take the sum to begin at $\ell=2$.  We thus have 
\begin{align}\label{Phisolution}
    \Phi(\rho,\theta,\phi) =\sum_{\ell \geq 2,m}C^{(0)}_{\ell m} \psi^{\mathcal{O}}_\ell(\rho) Y_{\ell m}(\theta,\phi)\,,
\end{align}
where $\psi^{\mathcal{O}}_\ell$ are given in Eq.~\eqref{psiellO} below.   At large $\rho$, we have $\psi^{\mathcal{O}}_\ell \sim \frac{1}{2}e^{\rho}$ \eqref{fObig}, such that 
\begin{align}\label{supertranslation}
    \Phi(\rho,\theta,\phi) \sim \frac{1}{2}e^{\rho} C^{(0)}(\theta,\phi)\,, \qquad \rho \to \infty\,,
\end{align}
where
\begin{align}\label{C0sphericaldecompi+}
    C^{(0)}(\theta,\phi) = \sum_{\ell\geq2,m} C^{(0)}_{\ell m} Y_{\ell m}(\theta,\phi)
\end{align}
matches to a corresponding quantity at $\scri^+$.  It follows from \eqref{kfromPhi} that
\begin{align}
    k_{\rho\rho} & = o(e^{-2\rho}) \,,\label{krhorho large rho}\\
    k_{\rho A} & = -2\nabla^BC^{(0)}_{AB}e^{-\rho} + o(e^{-\rho}) \label{krhoA large rho}\,,\\
    k_{AB} & = \frac{1}{2}e^{\rho}C^{(0)}_{AB} + o(e^0)\,, \label{kAB large rho}
\end{align}
where we introduce
\begin{align}
    C^{(0)}_{AB} = (-2\nabla_A\nabla_B + \gamma_{AB}\nabla^2)C^{(0)}.\label{CABtimelike}
\end{align}

Finally, we expand the generator functions $\omega$ and $\chi$ at large $\rho$.  From Eq.~\eqref{omegaT} and \eqref{chiY}, we find
\begin{align}
    \omega_T & = \frac{1}{2}e^{\rho}T + O(e^{-\rho})\,, \label{omegalargerho} \\
    \chi^A & = Y^A - \nabla^A(\nabla_BY^B)e^{-2\rho} + O(e^{-4\rho}) \,,\label{xiAlargerho} \\
    \chi^\rho & = -\frac{1}{2}\nabla_A Y^A + O(e^{-2\rho}). \label{xirholargerho}
\end{align}

\section{Future null infinity $\scri^+$}\label{sec:scri+}

The structure of null infinity has been extensively discussed in the literature \cite{bms1,bms2,1962PhRv..128.2851S,Newman:1966ub,65d3a01556474011975af11cd037b472,Barnich:2010eb,Barnich:2011mi,Campiglia:2015yka,Madler:2016xju}, and a standard treatment can be found in Ref.~\cite{flanagan_conserved_2017}.  We now review these results.  We include for now the possibility of non-zero stress-energy at null infinity; however, this stress-energy will be assumed to vanish in Section \ref{sec:i+scri+} for the purpose of matching to timelike infinity, where we assumed a vacuum spacetime outside of isolated massive bodies.  

A spacetime is said to be asymptotically flat at future null infinity if there exist coordinates $(u,r,x^A)$ such that 
\begin{align}
    ds^2& = \left(-1+\frac{2m}{r} + O(r^{-2})\right)du^2 + 2\left(-1 + \frac{1}{r^2}\left(\frac{1}{16}C_{AB}C^{AB} + 2\pi T_{rr}^{(4)}\right) + O(r^{-3})\right)dudr\nonumber\\
    &+ 2r\left(\frac{1}{2r}\nabla^BC_{AB} + \frac{2}{3r^2}\left(N_A + u\partial_Am -\frac{3}{32}\partial_A(C_{BC}C^{BC})\right) + O(r^{-3})\right)dudx^A\nonumber\\
    &+r^2\left(\gamma_{AB} + r^{-1} C_{AB} + O(r^{-2})\right)dx^Adx^B.\label{bondi}
\end{align}
The coordinates $(u,r,x^A)$ are named the Bondi coordinates which are characterized by the gauge-fixing conditions $g_{rr}=g_{rA}=0$ and $\partial_r \text{det}(g_{AB}/r^4)=0$.  Here $x^A$ are coordinates on the unit two-sphere with metric $\gamma_{AB}$, and $\nabla_A$ is the metric-compatible derivative operator.  The span of the coordinates $(u,x^A)$ defines $\scri^+$, which is the manifold $\mathbb R \times S^2$.  The fields $m, C_{AB}, N_A$ depend on $(u,x^A)$ and are regarded as tensors on $\scri^+$. These are named the Bondi mass aspect, shear tensor, and angular momentum aspect, respectively.\footnote{$\,\,$ Our convention for $N_A$ matches with \cite{hawking_superrotation_2017}. $N_A$ as defined in \cite{flanagan_conserved_2017} is equal to $N_A+u \partial_A m$ here.  As discussed later, our $N_A$ only diverges as $\log u$ for large $u$, meaning that the $N_A$ of \cite{flanagan_conserved_2017} diverges linearly.}  Notice that we keep to relative order $O(r^{-3})$ everywhere except for the $g_{uu}$ term.  The $O(r^{-2})$ piece of $g_{uu}$ and the $O(r^0)$ piece of $g_{AB}$ are non-zero but do not contribute to the leading equations for the trio $m, C_{AB}, N_A$, and do not appear in any calculations of this paper. 

In writing Eq.~\eqref{bondi}, we have assumed following 
 \cite{flanagan_conserved_2017} that the stress-energy takes the form 
\begin{align}
    T_{uu}& = T_{uu}^{(2)}r^{-2} + O(r^{-3})\,,\\
    T_{ur} & = O(r^{-4})\,,\\
    T_{rr}& = T_{rr}^{(4)}r^{-4} + O(r^{-5})\,,\\
    T_{uA}& = T_{uA}^{(2)}r^{-2} + O(r^{-3})\,,\\
    T_{rA} & = T_{rA}^{(3)}r^{-3} + O(r^{-4})\,,\\
    T_{AB} & = T^{(1)}\gamma_{AB} r^{-1} + O(r^{-2})\,.
\end{align}

The shear tensor $C_{AB}$ is symmetric and trace-free and hence has two degrees of freedom.  It is conventional to encode these in scalars $C$ and $\Psi$ that provide the electric (parity-even) and magnetic (parity-odd) parts, respectively, 
\begin{align}\label{Cdecomp}
    C_{AB} = (-2\nabla_A\nabla_B + \gamma_{AB}\nabla^2)C + \epsilon_{C(A}\nabla_{B)}\nabla^C \Psi\,.
\end{align}
Notice that the differential operators acting on $C$ and $\Psi$ annihilate their $\ell=0$ and $\ell=1$ parts, meaning that $C_{AB}$ contains only $\ell \geq 2$ components.  Without loss of generality, we will take $C$ and $\Psi$ to have no $\ell=0,1$ components.

The asymptotic symmetry group is the BMS group.  As we did at $i^+$, we parameterize the generators via a scalar function $T$ \eqref{T} on the sphere and a conformal Killing field $Y^A$ \eqref{Y} of the sphere.  The (super-)translations take the form
\begin{align}
    \xi_T = T\partial_u +\left(-\frac{1}{r}\nabla^AT + O(r^{-2})\right) \partial_A +\left(\frac{1}{2}\nabla^2T+O(r^{-1})\right) \partial_r\,,\label{xiTscri+}
\end{align}
while the rotations and boosts are written as
\begin{align}
    \xi_Y = \frac{1}{2}u\nabla_BY^B \partial_u + (Y^A+O(r^{-1}))\partial_A + (-\frac{1}{2}r\nabla_AY^A+O(r^0))\partial_r\,.
\end{align}
The associated charges are given by  \cite{1984CQGra...1...15D,Barnich:2011mi,flanagan_conserved_2017}\footnote{$\,\,$ The coefficient $-1/4$ in \eqref{QYscri+} is determined by the property of cross-section continuity of the angular momentum \cite{Chen:2022fbu}.}
\begin{align}
    Q^{\scri^+}_T(u) & = \frac{1}{4\pi}\int_{S^2}m T d\Omega \, , \label{QTscri+} \\
    Q^{\scri^+}_Y(u) & = \frac{1}{8\pi}\int_{S^2}Y^A \left( N_A-\frac{1}{4}C_{AB}D_C C^{BC} - \frac{1}{16}\partial_A (C_{BC}C^{BC}) \right) d\Omega\,. \label{QYscri+}
\end{align}
According to \eqref{QTscri+}, the translation and supertranslation charges are the spherical harmonic components of the mass aspect $m$.  In particular, the energy $T=1$ is the $\ell=0$ component, the momenta $T=n_i(\theta,\phi)$ are the $\ell=1$ components [see \eqref{ni} for the definition of $n_i$], and the supermomenta are the higher-$\ell$ components.  According to \eqref{QYscri+}, the Lorentz charges are $\ell=1$ components of the angular momentum aspect $N_A$ with quadratic corrections in the shear.  In particular, the angular momenta are the odd-parity $\ell=1$ components [controlled by $\kappa_i$ in \eqref{Y}], while the mass moments are the even-parity $\ell=1$ harmonics [controlled by $b_i$ in \eqref{Y}].  Note that the quadratic corrections do not contribute when the shear is purely electric parity,
\begin{align}\label{rel0}
\Psi = 0 \quad \rightarrow \quad \int_{S^2} d\Omega Y^A \left( C_{AB}\nabla_C C^{BC}+\frac{1}{4}\partial_A (C_{BC}C^{BC})\right) = 0\,.
\end{align}
This fact was proven in Appendix A.3. of \cite{compere_vacua_2016}.

In light of the first term in Eq.~\eqref{xiTscri+}, it is sometimes said that supertranslations are angle-dependent time translations on $\scri^+$.  However, all three terms in \eqref{xiTscri+} are at the same order in the Bondi expansion, and all three are necessary for the generator to be an asymptotic symmetry.  The $\ell=0$ supertranslation is timelike (it is a pure time translation in flat spacetime), the $\ell = 1$ supertranslations are spacelike (they are pure spatial translations in flat spacetime) and all higher $\ell \geq 2$ supertranslations also appear to be spacelike:  the norm $\xi^2=D_A T D^A T - T (\nabla^2+1)T$ at $r \rightarrow \infty$ at fixed $u$ is numerically seen to be positive. Viewed on $\scri^+$ (the manifold spanned by time $u$ and angles $x^A$), supertranslations act as a special combination of time and angle translation determined by a function $T$ on the sphere. The $\ell=0$ case is the pure time translation $\pd_u$, while all $\ell \geq 0$ involve mixed angle-dependent time translation $\pd_u$ and angular translation $\pd_A$.  The $\ell=1$ case can be understood in flat spacetime as the dependence of retarded coordinates on the choice of spatial origin \cite{Boyle2016}.

The evolution of the Bondi mass aspect $m$ and angular momentum aspect $N_A$ can be derived from the $uu$ and $uA$ components of Einstein's Equation---see Eqs. (2.11a)-(2.11b) in \cite{flanagan_conserved_2017} or Eqs. (2.4a)-(2.4b) in Ref.~\cite{compere_poincare_2020}.  The evolution equations are 
\begin{align}
    \dot{m}& = -\frac{1}{8}\dot{C}_{AB}\dot{C}^{AB} -\frac{1}{4}\nabla^2(\nabla^2+2)\dot{C} - 4\pi T_{uu}^{(2)} \label{mdot} \,,\\
    \dot{N}_A&= - u\pd_A \dot{m} + I_{A} +\frac{1}{4}\epsilon_{AB}\nabla^B\nabla^2(\nabla^2 + 2)\Psi - 8\pi T_{uA}^{(2)} +\pi\pd_A \dot{T}_{rr}^{(4)}\label{Ndot}\,,
\end{align}
where we define\footnote{$\,\,$ The analogous quantity obtained from Eqs. (2.3a)-(2.3b) in Ref.~\cite{capone_charge_2022} is  $I_{A} = \frac{1}{4}\partial_A(C_{BC}\dot{C}^{BC}) - \frac{1}{4}\nabla_B(C^{BC}\dot{C}_{CA}) + \frac{1}{2}C_{AB}\nabla_C\dot{C}^{BC}$ which is equivalent to Eq. \eqref{defI} after using the property $X_{AC}\nabla_B Y^{BC}=X^{BC}D_A Y_{BC}-X^{BC}\nabla_B Y_{AC}$ valid for any pair of symmetric tracefree tensors $X_{AB}$, $Y_{AB}$.}
\begin{align}\label{defI}
    I_{A} = \frac{1}{4}\nabla_B\left( \dot C^{BC}C_{CA}\right)+\frac{1}{2}C_{AB}\nabla_C \dot C^{BC}. 
\end{align}
In writing these evolution equations we used the following relationships, 
\begin{align}\label{useful eqn for shear}
     \nabla_A\nabla_BC^{AB}& = -\nabla^2(\nabla^2 + 2)C\,,\\
    \nabla^B(\nabla_B\nabla^CC_{AC}-\nabla_A\nabla^CC_{BC})&= -\epsilon_{AB}\nabla^B\nabla^2(\nabla^2 + 2)\Psi\,.
\end{align}

We will use the symbol $\Delta$ to denote the total change in a quantity between $u \mapsto + \infty$ and $u \mapsto - \infty$.  A non-zero change in Bondi shear, $\Delta C_{AB} \neq 0$, indicates a permanent displacement in gravitational-wave detectors at $\scri^+$ and is known as gravitational-wave memory.\footnote{$\,\,$ Inertial test masses have fixed Bondi coordinate $x^A$ at leading and subleading order, i.e., $\Delta x^A = O(1/r^2)$ \cite{Strominger:2014pwa}.  When $\Delta C_{AB} \neq 0$ there is an $O(1/r)$ change in proper distance between two such observers, the gravitational memory.}  Integrating Eq.~\eqref{mdot} gives a differential equation for $\Delta C$
\begin{align}\label{memory scri+}
    \frac{1}{4}\nabla^2(\nabla^2+2)\Delta C = -\Delta m - \int_{-\infty}^{\infty} \left( \frac{1}{8}\dot{C}_{AB}\dot{C}^{AB} +4\pi T_{uu}^{(2)} \right) du \,.
\end{align}
The first term on the right is called the ordinary memory \cite{1974SvA....18...17Z}, the second term is the non-linear or gravitational wave memory \cite{B90,Christodoulou:1991cr,Blanchet:1992br} whereas the third term is the null memory \cite{1978ApJ...223.1037E,1978Natur.274..565T}. This equation can be solved explicitly using the spherical harmonic decomposition or using the Green function of the operator $\frac{1}{4}\nabla^2(\nabla^2+2)$ \cite{Strominger:2014pwa}. The result is 
\begin{equation}
\Delta C = -2 \int_{S^2} d\Omega' (1-n_i n_i')\left(\log(1-n_j n'_j)-1\right)\int^{\infty}_{-\infty}\mathcal F(u,x^{\prime A}) du\, ,\label{changeC} 
\end{equation}
where $\mathcal F(u,x^A) =  \frac{1}{8}\dot{C}_{AB}\dot{C}^{AB} +4\pi T_{uu}^{(2)} +\partial_u m$.

From Eqs.~\eqref{QTscri+}, \eqref{QYscri+}, \eqref{mdot}, and \eqref{Ndot}, the evolution equations for the charges are
\begin{align}
        \dot{Q}^{\scri^+}_T & = -\frac{1}{16\pi} \frac{\pd}{\pd u}\int_{S^2} C \nabla^2(\nabla^2+2)T d\Omega - \mathcal{E}^{\scri^+}_T \,,\\
        \dot{Q}^{\scri^+}_Y & = -\frac{1}{32\pi}\frac{\pd}{\pd u}\int_{S^2} \left(C_{AB}D_C C^{BC} + \frac{1}{4}\partial_A (C_{BC}C^{BC}) \right)Y^A d\Omega - \mathcal{E}^{\scri^+}_Y\, , \label{QYdot}
\end{align}
where 
\begin{align}
\mathcal{E}^{\scri^+}_T & = \frac{1}{32\pi}\int_{S^2}\left( \dot{C}_{AB}\dot{C}^{AB}  + 32 \pi T_{uu}^{(2)} \right) T d\Omega \, , \\
\mathcal{E}^{\scri^+}_Y & = \frac{1}{8\pi} \int_{S^2} \left( u\partial_A \dot{m} - I_{A}  + 8\pi {T}_{uA}^{(2)} - \pi \pd_A \dot T_{rr}^{(4)}\right)Y^Ad\Omega \,.
\end{align}
The $\Psi$ term in \eqref{Ndot} has disappeared from this formula since it contains only $\ell \geq 2$ harmonics, whereas $Y^A$ is purely $\ell=1$.

The total change in the charges is then given by 
\begin{align}
    \Delta Q^{\scri^+}_T & = -\frac{1}{16\pi} \int_{S^2} d \Omega\Delta C \nabla^2(\nabla^2+2)T - \int_{-\infty}^{\infty} \mathcal{E}_T^{\scri^+} du \label{deltaQT}\,, \\
    \Delta Q^{\scri^+}_Y & = -\frac{1}{32\pi} \int_{S^2} \Delta \left(C_{AB}D_C C^{BC} + \frac{1}{4}\partial_A (C_{BC}C^{BC}) \right)Y^A d\Omega - \int_{-\infty}^{\infty} \mathcal{E}_Y^{\scri^+} du\,. \label{deltaQY}
\end{align}

The change in supertranslation charge \eqref{deltaQT} naturally decomposes into a memory term (the first term) that is non-zero only for pure supertranslations (\emph{i.e.} $\ell \geq 2$) and a flux term (the second term) contributing for all $\ell$.  These are also called the soft term and the hard term, respectively.  The change in Lorentz charges does not contain soft/memory terms.  Note that the first integral in \eqref{deltaQY} vanishes if the magnetic parity scalar $\Psi$ vanishes at the boundaries $\scri^+_+$ and $\scri^+_-$ [see Eq.~\eqref{rel0}].  We will see that this indeed happens in our framework, and the final formulas will involve only the second integral in \eqref{deltaQY}.

\section{Matching between $i^+$ and $\mathcal{I}^+$}\label{sec:i+scri+}
Above we have considered $i^+$ and $\scri^+$ as two distinct limits, with two distinct frameworks.  We wish to consider spacetimes that are asymptotically flat at both infinities, where they are sufficiently regular to permit an identification of the symmetry groups and associated charges.  To enforce this regularity we will need additional assumptions about the behavior of the Bondi quantities at large $u$ and the Beig-Schmidt quantities at large $\rho$.

To see what is required, consider expressing the Bondi metric \eqref{bondi} in Beig-Schmidt coordinates \eqref{Beig-Schmidt} at large $u$.  In flat spacetime, the canonical transformation to Bondi coordinates is $u=t-r$, whereas the canonical transformation to Beig-Schmidt coordinates is $t = \tau \cosh \rho$ and $r=\tau \sinh \rho$ [Eq.~\eqref{transi+}].  Composing these together gives a notion of transforming from Bondi coordinates $(u,r,x^A)$ to Beig-Schmidt coordinates $(u,r,y^A)$ with ``no additional Poincar\'e transformations'':
\begin{align}\label{composed}
u = \tau e^{-\rho}\,, \qquad r = \tau \sinh \rho\,, \qquad x^A = y^A\, .
\end{align}
In this section we use $y^A$ for the angular coordinates at $i^+$, to distinguish these from the the angular coordinates $x^A$ at $\scri^+$.  While the angular coordinates agree here in flat spacetime, they will disagree at higher order in the asymptotic expansions in curved spacetime.

Notice that according to \eqref{composed}, a function $f(u/r)$ is independent of $\tau$.  This means that orders at $\scri^+$ are infinitely mixed up at $i^+$.  For example, the terms [$1/r$, $u/r^2$, $u^2/r^3$, ...] from infinitely many orders in $1/r$ in the Bondi expansion all appear at same order $1/\tau$ in the Beig-Schmidt expansion.  This is a general pattern: terms higher order in $1/r$ in the Bondi expansion are permitted to blow up faster with $u$ at late times.  The details of the tensor transformation law determine which terms are permitted.  As a simple example, consider the $1/r$ portion of the $uu$ component of the Bondi metric (i.e., the mass aspect $m$).  A term with dependence $u^n$ for some integer $n$ becomes\footnote{$\,\,$ The $d\rho$ terms are not permitted by the Beig-Schmidt form; we will see that these can be eliminated by correcting the flat spacetime transformation \eqref{composed} at higher order in $1/\tau$.  Here we are just making illustrative points for the purposes of motivating assumptions.}
\begin{align}
\frac{u^n}{r} du^2 = \tau^{n-1} \frac{e^{(-n-2) \rho}}{\sinh \rho} d\tau^2 + \textrm{($d\rho$ terms)}\,.\label{justforillustration}
\end{align}
Comparing Eq.~\eqref{justforillustration} with the general form \eqref{Beig-Schmidt}, we see that $n>0$ is not allowed, and an $n=0$ term would match up with the $e^{-3\rho}$ behavior \eqref{sigmalargerho} of $\sigma$ at large $\rho$.  Thus we suspect that the Bondi mass aspect $m$ ought to approach a $u$-independent function on the sphere at late times $u$, which determines the large-$\rho$ behavior of the Beig-Schmidt field $\sigma$.  Other terms like $(u/r^2) du^2$, $(u^2/r^3) du^2$, etc., would then match to $\sigma$ at subleading orders in $e^{-\rho}$, and corresponding terms with lower powers of $u$ would match at subleading orders in the Beig-Schmidt expansion in $1/\tau$.

Based on this kind of reasoning, we assume the following large-$u$ behavior for the fundamental quantities of the Bondi metric,
\begin{align}
    m & = m^{(0)} + m^{(1)} u^{-1} + o(u^{-1}) \, ,\label{ansatzm} \\
    C_{AB} & = C_{AB}^{(0)} + C_{AB}^{(1)}u^{-1} + o(u^{-1}) \label{ansatzC} \, ,\\
    N_A & = N_A^{(-1)} u + N_A^{(\log)}\log u+N_A^{(0)} + o(1)\, . \label{ansatzN}
\end{align}
These equations are consistent with previous calculations of gravitational scattering. The leading shear $C_{AB}^{(0)}$ is implied by the displacement memory \cite{B90,Christodoulou:1991cr,Blanchet:1992br} or, equivalently \cite{Strominger:2014pwa}, by the Fourier transform of the leading soft graviton theorem \cite{PhysRev.140.B516}. The subleading $u^{-1}$ term $C_{AB}^{(1)}$ is implied by the Fourier transform of the logarithmic corrections to the subleading soft graviton theorem \cite{PhysRevD.1.1559,Ciafaloni:2018uwe,Addazi:2019mjh,Laddha:2018myi,Saha:2019tub,sahoo_classical_2022}.  
Our falloff assumption is also consistent with the bound $C_{AB} = C_{AB}^{(0)} + O(1+\vert u\vert)^{-1/2}$ derived in \cite{ChristodoulouKlainerman+1994}. 

The leading behavior of \eqref{ansatzm}--\eqref{ansatzN} is motivated by analysis similar to Eq.~\eqref{justforillustration}.  For the corrections, we adopt a pragmatic strategy of including only polynomials until forced to include logs.  We included a logarithm in $N_A$ in order to be compatible with known scattering results \cite{sahoo_classical_2022}.  To see this, note that the evolution equations~\eqref{mdot} and \eqref{Ndot} imply the relationships
\begin{align}
    m^{(1)} & = -\frac{1}{4}\nabla^2(\nabla^2+2) C^{(1)}\,,\label{m1}\\
    N_{A}^{\rm (log)} & = \partial_A m^{(1)} +\frac{1}{4}\epsilon_{AB}\nabla^B\nabla^2(\nabla^2+2) \Psi^{(1)} \,,\label{Nlog}\\
    N_A^{{(-1)}} & = \frac{1}{4} \epsilon_{AB} \nabla^B \nabla^2(\nabla^2+2) \Psi^{(0)}\,,\label{N-1}
\end{align}
where we now assume that the stress-energy near null infinity vanishes, $T^{(2)}_{uu}=T^{(2)}_{uA}=0$.  Here $C^{(1)}$ and $\Psi^{(1)}$ are the electric and magnetic potentials of $C^{(1)}_{AB}$, defined analogously to Eq.~\eqref{Cdecomp}. From \eqref{Nlog}, we see that the log term $N_A^{\rm (log)}$ is required unless both $m^{(1)}$ and $\Psi^{(1)}$ are vanishing.  Refs.~\cite{Saha:2019tub,sahoo_classical_2022} have shown that these terms do not vanish in general, and in fact found their precise form for point particle scattering (see Eq.~(1.5) of \cite{sahoo_classical_2022}).  We therefore conclude that the log term in $N_A$ is necessary to describe gravitational scattering.  As we shall see shortly in Eq.~\eqref{irhoA match}, $N_{A}^{\rm(log)}$ matches with the log term $i_{ab}$ in the Beig-Schmidt metric \eqref{Beig-Schmidt} at $i^+$, requiring the presence of that term as well.

Eq.~\eqref{N-1} shows that the linear divergence $N_{A}^{(-1)}$ in the angular momentum aspect is related to the magnetic parity shear $\Psi^{(0)}$.  We will see shortly that matching to our assumed form at future timelike infinity will force $\Psi^{(0)}$ to vanish, such that $N_A^{(-1)}$ vanishes as well. The late-time divergence in $N_A$ is therefore logarithmic rather than linear.  This divergence implies that the currently defined super-Lorentz charges \cite{Barnich:2011mi,Campiglia:2015yka} do not admit a limit at timelike and spatial infinity.  By contrast, the Lorentz charges are finite because $N_A^{(\rm log)}$ has no $\ell=1$ part.

\subsection{Asymptotic coordinate transformation}\label{sec:asymptotic coordinate transformation}

We now augment the flat spacetime coordinate transformation \eqref{composed} with perturbative corrections at large $\tau$ in order to express a general Bondi metric in Beig-Schmidt form.  As before, our strategy is to include only polynomials until forced to include logs.  We find that a single log is necessary, and assume the general form
\begin{align}
    &u = \tau e^{-\rho} + \alpha(\rho, y^A) + \tau^{-1}A(\rho, y^A) + o(\tau^{-1}) \label{ansatzu}\,, \\
    & r = \tau \sinh\rho + \beta(\rho, y^A)+ \tau^{-1}B(\rho, y^A) + o(\tau^{-1})\label{ansatzr}\,, \\
    & x^A = y^A + \tau^{-1}p^A(\rho, y^A) + \tau^{-2}q^A(\rho, y^A)+\tau^{-2}\log\tau q^A_{\log} (\rho, y^A) + o(\tau^{-2})\, .\label{ansatzA}
\end{align}
The Bondi metric \eqref{bondi} contains remainder terms denoted $O(r^{-n})$ for integers $n$.  These refer to functions which scale to zero at least as fast as $r^{-n}$ as $r \to \infty$ at fixed $u,x^A$.  However, without a corresponding assumption about the $u$-dependence of these functions, the remainder terms at arbitrarily high order in $1/r$ can contribute at any order in the Beig-Schmidt expansion.  In keeping with the discussion surrounding Eq.\eqref{justforillustration}, it would be natural to allow one additional power of $u$ for each order in $1/r$.  However, in practice there will be logarithmic terms or even more complicated dependence, the form of which is unknown.  Rather than specifying some rather complex set of assumptions about the specific form of the unknown remainder terms at all orders in $1/u$ and $1/r$, we will simply assume a scaling in the Beig-Schmidt expansion (in $1/\tau$ and $e^{-\rho}$) which guarantees that they do not contribute at the order of our calculations.  

In particular, let $\epsilon_{uu}=O(r^{-2})$ denote the specific remainder terms in the Bondi metric component $g_{uu}$ [exact $g_{uu}$ minus displayed terms in Eq.~\eqref{bondi}], and let $\epsilon_{ur}=O(r^{-3})$ and $\epsilon_{uA}=O(r^{-2})$ and $\epsilon_{AB}=O(r^{0})$ denote analogous quantities for other components of Eq.~\eqref{bondi}.  Then we assume 
\begin{align}
\epsilon_{uu} & = \frac{o(e^{-2\rho})}{\tau} + o(e^{-\rho})\frac{\log\tau}{\tau^2}+ \frac{o(e^{-\rho})}{\tau^2} + o(1/\tau^2) \label{epsilonuu} \,, \\
    \frac{\epsilon_{AB}}{r^2} & =  \frac{o(e^{-4\rho})}{\tau} + o(e^{-3\rho})\frac{\log\tau}{\tau^2} + \frac{o(e^{-\rho})}{\tau^2} + o(1/\tau^2)\,, \label{epsilonAB} \\
\epsilon_{ur}, \frac{\epsilon_{uA}}{r} & =  \frac{o(e^{-4\rho})}{\tau} + o(e^{-3\rho})\frac{\log\tau}{\tau^2} + \frac{o(e^{-3\rho})}{\tau^2} + o(1/\tau^2)\, . \label{epsilonother}
\end{align}
The $uu$ remainder term has a lower order than the remaining simply because we have kept to lower order in that term in the Bondi metric \eqref{bondi}.

We now plug Eqs.~\eqref{ansatzu}--\eqref{epsilonother} into the Bondi metric \eqref{bondi} in order to find a metric in $(\tau,\rho,y^A)$ coordinates. 
We will refer to terms in \eqref{bondi} as leading, subleading, and subsubleading as follows.  The leading terms are $O(\tau^0)$ in $g_{\tau \tau}$, $O(\tau)$ in $g_{\tau A}$, and $O(\tau^2)$ in $g_{AB}$; the subleading terms are $O(\tau^{-1})$ in $g_{\tau \tau}$, $O(\tau^0)$ in $g_{\tau A}$, and $O(\tau)$ in $g_{AB}$; and the subsubleading terms are $O(\tau^{-2})$ in $g_{\tau \tau}$, $O(\tau^{-1})$ in $g_{\tau A}$, and $O(\tau^0)$ or $O(\log \tau)$ in $g_{AB}$.  Terms called leading, subleading, and subsubleading all have the same order in $\tau^{-1}$ (possibly multiplied with a $\log \tau$) when expressed in an orthonormal frame using the hyperboloid metric $h_{ab}$.

At leading order, where the metric is flat, we find that the Beig-Schmidt form is automatically satisfied, as it must be since we assumed the canonical transformation \eqref{composed} as our leading coordinate transformation in \eqref{ansatzu}--\eqref{ansatzA}.  At subleading order, we find that $g_{\tau \tau}$ takes the proper form with
\begin{align}\label{sigma match}
\sigma = - 2 m^{(0)} e^{-3\rho} + o(e^{-4\rho})\, .
\end{align}
However, we find that in general there are $\tau A$ and $\tau \rho$ cross-terms, which are not permitted by the Beig-Schmidt form.  In order to keep the metric free of such terms at subleading order, we find that we must have
\begin{align}
0 & = e^{-\rho}\frac{1}{2}\nabla^B C_{AB}^{(0)} - \sinh^2\!\rho \ \! p^C \gamma_{AC} -   \cosh\rho \ \! \partial_A \alpha - e^{-\rho} \partial_A\beta+ o(e^{-2\rho}) \label{subleading tau yA constraint}\,,\\
0 & = -4m^{(0)} e^{-3\rho} - e^{-\rho}\partial_\rho \beta - \cosh\rho \ \! \partial_\rho \alpha+o(e^{-4\rho})\, . \label{subleading tau rho constraint}
\end{align}
From Eq.~\eqref{subleading tau rho constraint} we see that it is consistent\footnote{$\,\,$ We expect that a different non-zero choice of $\alpha$ and $\beta$ at this order will correspond to introducing a supertranslation between $\mathcal{I}^+$ and $i^+$, which could cause the charges to fail to match.} to take $\alpha= o(e^{-3\rho})$ and $\beta= o(e^{-\rho})$, in which case the terms involving $\alpha$ and $\beta$ may be neglected from Eq.~\eqref{subleading tau yA constraint}.  We thus choose
\begin{align}
    \alpha & = o(e^{-3\rho}) \,,\label{alpha} \\
    \beta & = o(e^{-\rho})\,, \label{beta} \\
    p^A & = 2 \nabla_B C^{(0)AB} e^{-3\rho} + o(e^{-4\rho})\, . \label{p}
\end{align}
With these choices, we find that the subleading $g_{ab}$ components take the required form $k_{ab}-2 \sigma h_{ab}$ with $k_{ab}$ properly trace-free, with 
\begin{align}
    k_{\rho\rho}& =  o(e^{-2\rho}) \,,\label{krhorho match}\\
    k_{\rho A} & = -2e^{-\rho}\nabla^B C^{(0)}_{AB} + o(e^{-\rho})\,, \label{krhoA match}\\
    k_{AB} &=\frac{1}{2}e^{\rho}C^{(0)}_{AB} +o(e^{0})\, .\label{kAB match}
\end{align}

Eqs.~\eqref{sigma match}--\eqref{kAB match} relate the late-time behavior of the subleading Bondi quantities $m$ and $C_{AB}$ to the large-radius behavior of the subleading Beig-Schmidt quantities $\sigma$ and $k_{ab}$.  Comparing Eqs.~\eqref{kAB match} and Eq.~\eqref{kAB large rho}, we confirm that the symbol $C_{AB}^{(0)}$ introduced previously is indeed the same quantity as used in this section.  
From Eq.~\eqref{CABtimelike} we see that $C^{(0)}_{AB}$ has only a electric parity component, and hence we learn that the magnetic parity component vanishes,
\begin{align}\label{nopsi0}
\Psi^{(0)} = 0\, .
\end{align}
In other words, matching to our assumed form at timelike infinity ensures that the Bondi shear has no magnetic parity portion at late times on null infinity.  By eq.~\eqref{N-1} we thus see that the angular momentum aspect has no linear divergence,
\begin{align}\label{noN-1}
N^{(-1)} = 0\,,
\end{align}
which means that its expansion \eqref{ansatzN} becomes
\begin{align}
N_A =  N_A^{(\log)}\log u+N_A^{(0)} + o(1)\, . \label{ansatzN2}
\end{align}

We now turn to subsubleading order in $1/\tau$.  Performing a similar analysis, we find that by choosing 
\begin{align}
        A &= 2m^{(1)}e^{-3\rho} + o(e^{-3\rho}) \,,\label{A} \\
        B &= (-\frac{1}{8}C^{(0)}_{AB}C^{(0)AB} - 3m^{(1)})e^{-\rho} + o(e^{-\rho}) \label{B} \,,\\
         q^A &= \nabla_B C^{(1)AB}e^{-2\rho} + \big(2\nabla_B C^{(1)AB}+\frac{1}{3}(8N^{(0)A} -6C^{(0)AB}\nabla_C C^{(0)BC} \nonumber \\
        &+20\nabla^A m^{(1)})-\frac{4}{3}(2\rho-1) N^A_{\log}\big)e^{-4\rho} + o(e^{-4\rho}) \label{q} \,,\\
        q^{A}_{\log}& = \frac{8}{3}N^{(\rm{log}) A}e^{-4\rho}+o(e^{-4\rho})\,, \label{qlog}
\end{align}
then the metric conforms to Beig-Schmidt form with the identifications
\begin{align}
    i_{\rho\rho}& = o(e^{-2\rho}) \,,\label{irhorho match}\\
    i_{\rho A}& = -4e^{-2\rho}N_A^{(\log)} + o(e^{-3\rho})\,,\label{irhoA match}\\
    j_{\rho\rho}& = 16m^{(1)}e^{-2\rho} + o(e^{-2\rho})\,, \label{jrhorho match} \\
    j_{\rho A}& = -\nabla^BC^{(1)}_{AB}+\big(-4N_A^{(0)} - C^{(0)}_{AC}\nabla_BC^{(0)BC}-12\partial_A m^{(1)}\nonumber\\& + (4\rho-2)N_A^{(\log)}-\nabla^BC^{(1)}_{AB}\big) e^{-2\rho} +o(e^{-3\rho})\,. \label{jrhoA match}
\end{align}
We have only displayed the components and orders thereof that contribute to the charges \eqref{QTi+} and \eqref{QYi+}.  In particular, the forms of $j_{AB}$ and $i_{AB}$ are not needed for our purposes.

\subsection{Matching of charges}

In the previous section we found a coordinate transformation that brings a general Bondi metric of the form \eqref{bondi} into a Beig-Schmidt metric of the form \eqref{Beig-Schmidt}, working in the $u\to \infty, \rho \to \infty$ region of overlap between the two expansions.  Since the BMS charges are defined at each infinity via integrals involving definite coordinate expressions for generators, the coordinate transformation provides a one-to-one identification of the charges in the region of overlap.  For example, the $x$ component of angular momentum at $i^+$ is given by choosing $Y^A=(-\cos y^2,\cot y^1 \sin y^2)$ in Eq.~\eqref{QYi+} for $Q^{i^+}_Y$, where $y^A=(y^1,y^2)$ are the spherical coordinates in the Beig-Schmidt frame.  Likewise, the $x$-component of angular momentum at $\scri^+$ is given by choosing $Y^A=(-\cos x^2,\cot x^1 \sin x^2)$ in Eq.~\eqref{QYscri+} for $Q_Y^{\scri^+}$, where $x^A=(x^1,x^2)$ are the spherical coordinates in the Bondi frame.  These two expressions are considered the ``same'' charge $L_x$ under the identification provided by our coordinate system.  This identification makes physical sense only if the expressions also have the same numerical value in the region of overlap, which can be interpreted as saying that our coordinate transformation is trivial, simply relating the two infinities without introducing any BMS transformations along the way.  We now show that this is indeed the case.  One may also reverse the logic, taking the matching of charges to be the definition of the ``same'' BMS frame, and regarding our calculation as an existence proof that a shared BMS frame can be constructed.

Using Eqs.~\eqref{ansatzm} and \eqref{ansatzN2} together with Eqs.~\eqref{m1}, \eqref{Nlog}, in Eqs.~\eqref{QTscri+} and \eqref{QYscri+}, the Bondi charges at late times are given by 
\begin{align}
    Q_T^{\scri^+_+} & \equiv \lim_{u \to \infty} Q^{\scri^+}_T = \frac{1}{4\pi}\int_{S^2}Tm^{(0)} d\Omega \label{QTfinal}\,, \\
    Q_Y^{\scri^+_+} & \equiv \lim_{u \to \infty} Q^{\scri^+}_Y = \frac{1}{8\pi}\int_{S^2}Y^AN_A^{(0)} d\Omega\,. \label{QYfinal}
\end{align}
The logarithmic term $N_A^{\rm (log)}$ does not contribute to \eqref{QYfinal} on account of the $\nabla^2(\nabla^2+2)$ operator in Eqs.~\eqref{m1} and \eqref{Nlog}, which annihilates the $\ell=1$ factor $Y^A$ in \eqref{QYscri+} after integration by parts.  The shear terms of \eqref{QYscri+} disappear after using \eqref{rel0} evaluated at late times where the magnetic shear vanishes \eqref{nopsi0}.\footnote{$\,\,$ We note in passing that the extension of the Lorentz algebra to the $\text{Diff}(S^2)$ algebra labeled by smooth functions $Y^A$ over the sphere \cite{Campiglia:2015yka} leads to logarithmic divergent charges using the definition \eqref{QYscri+}.}

We may compare this to the total Beig-Schmidt charges, which are computed by taking the integration surface in \eqref{QTi+} to surround all poles.  We will choose the surface to be a sphere of constant constant $\rho$ evaluated at arbitrarily large $\rho$.  For the supertranslation charges we have
\begin{align}\label{Future null future timelike momentum Match}
    Q_T^{i^+_\partial} & = \lim_{\rho \to \infty}\frac{1}{4\pi}\sinh^2\! \rho \int_{S^2}(\omega_T\partial_\rho \sigma - \sigma \partial_\rho \omega_T)d\Omega = \frac{1}{4\pi}\int_{S^2}m^{(0)} T d\Omega\,,
\end{align}
where in the last step we use Eqs.~\eqref{omegalargerho} and Eq.~\eqref{sigma match}.  Comparing with \eqref{QTfinal}, we see that the charges indeed match,
\begin{align}\label{i+ T match}
Q_T^{i^+_\partial} = Q_T^{\scri_+^+}\, .
\end{align}
For the Lorentz charges, a similar calculation gives
\begin{align}\label{Future null future timelike Lorentz Match}
    Q_Y^{i^+_\partial}  =\lim_{\rho \to \infty} \Bigg[ & \frac{1}{8\pi}  \int_{S^2}d\Omega N_A^{(0)} Y^A + \frac{1}{8\pi}\int_{S^2} d\Omega \left(-\frac{1}{4}C^{(0)}_{AC}\nabla_BC^{(0)BC}Y^A + \frac{1}{16}\nabla_AY^A C^{(0)}_{BC}C_{(0)}^{BC}\right)\nonumber\\
    &+ \frac{\sinh^2\!\rho (1+e^{-2\rho})}{8\pi}\int_{S^2}\left(Y^A - e^{-2\rho}\nabla^A\nabla_CY^C\right)\nabla^BC^{(1)}_{AB}\nonumber\\
    & + \frac{1}{8\pi}\int_{S^2} d\Omega Y^A\left(3\partial_Am^{(1)} - \rho N_A^{(\log)}\right) + \frac{1}{8\pi}\int_{S^2}d\Omega 2m^{(1)}\nabla_AY^A \Bigg]\, ,
\end{align}
where Eq. \eqref{xiAlargerho} was used. 
Given the lack of $\ell = 1$ harmonics in $m^{(1)},C_{AB}^{(1)}$ and $N_{A}^{(\log)}$ (see \eqref{m1}, \eqref{Cdecomp}, and \eqref{Nlog}, respectively), the last three terms will identically vanish. The second term also vanishes,
as shown in appendix A of Ref.~\cite{compere_vacua_2016} and in Eq. \eqref{rel0}.  The lone surviving first term is exactly what appears in \eqref{QYfinal}, showing that the Lorentz charges indeed properly match,
\begin{align}\label{i+ Y match}
    Q_Y^{i^+_\partial} = Q_Y^{\scri_+^+}.
\end{align}
We have shown that our coordinate transformation has the desired property of relating $\scri^+$ and $i^+$ such that identified BMS charges agree.

We can now justify the statement made previously after Eq. \eqref{Qct} that $Q_Y^{i^+_\partial}$ is finite without any infrared regulation.  We have found from the evolution equations \eqref{Nlog} that $N_A^{\rm (log)}$ has no $\ell=1$ part.  From Eq. \eqref{irhoA match}, it follows that $i_{\rho A}$ has no $\ell=1$ part at large $\rho$.  Together with the faster falloff of $i_{\rho \rho}$ shown in Eq.~\eqref{irhorho match}, it then follows that the integral in Eq. \eqref{Qct} vanishes at large $\rho$, as claimed.

\section{Past infinities: $\scri^-$, $i^-$ and matching}\label{sec:past-infinities}

As explained in Sec.~\ref{sec:flat}, the mathematical descriptions of $i^-$ and $\scri^-$ can be obtained from those of those of $i^+$ and $\scri^+$, respectively, by applying time-reversal and performing a coordinate transformation.  We will fix our conventions at $i^-$ and $\scri^-$ by doing this procedure and keeping the names of the symbols the same, with one important exception: we will modify the definitions of the charges $Q_T$ and $Q_Y$ so that the same choice of $T$ or $Y^A$ gives rise to the same charge at all infinities, i.e., the charges match across boundaries with a single choice of $T$ or $Y$.  Without these redefinitions, it would be necessary to use different (antipodally related) choices of $T$ or $Y^A$ to describe the same physical charge in the future and in the past.

\subsection{Past null infinity $\mathcal{I}^-$}\label{sec:scri-}
As explained in Sec.~\eqref{sec:flati-}, sending $u \to -v$ converts equations at $i^+$ to corresponding equations at $i^-$.  The Bondi expansion \eqref{bondi} becomes
\begin{align}
    ds^2& = \left(-1+\frac{2m}{r} + O(r^{-2})\right)dv^2 + 2\left(1- \frac{1}{r^2}\left(\frac{1}{16}C_{AB}C^{AB} + 2\pi T_{rr}^{(4)}\right) + O(r^{-3})\right)dvdr\nonumber\\
    &+ 2r\left(-\frac{1}{2r}\nabla^BC_{AB} - \frac{2}{3r^2}\left(N_A -v\partial_Am -\frac{3}{32}\partial_A(C_{BC}C^{BC})\right) + O(r^{-3})\right)dvdx^A\nonumber\\
    &+r^2\left(\gamma_{AB} + r^{-1} C_{AB} + O(r^{-2})\right)dx^Adx^B.\label{bondii-}
\end{align}
As described at the start of this section, for the $\scri^-$ version of the charges \eqref{QTscri+} and \eqref{QYscri+} associated with a given $T$ or $Y^A$, we modify the formulas to ensure a convenient match across $i^0$.  The needed modifications are a parity transformation $\Upsilon$ on $T$ and $Y^A$ together with a minus sign for $Q_Y$,
\begin{align}
    Q^{\scri^-}_T(v) & = \frac{1}{4\pi}\int_{S^2}m \Upsilon^*T d\Omega \label{QTscri-} \,,\\
    Q^{\scri^-}_Y(v) & = -\frac{1}{8\pi}\int_{S^2}\left( N_A-\frac{1}{4}C_{AB}D_C C^{BC} - \frac{1}{16}\partial_A (C_{BC}C^{BC}) \right)\Upsilon^*Y^A  d\Omega\, . \label{QYscri-}
\end{align}
The evolution equations \eqref{mdot} and \eqref{Ndot} become
\begin{align}
    \dot{m}& = \frac{1}{8}\dot{C}_{AB}\dot{C}^{AB} -\frac{1}{4}\nabla^2(\nabla^2+2)\dot{C} + 4\pi T_{vv}^{(2)}\,,\\
    \dot{N_A}&=  v\partial_A \dot{m} + I_{A} -\frac{1}{4}\epsilon_{AB}\nabla^B\nabla^2(\nabla^2 + 2)\Psi - 8\pi T_{vA}^{(2)}  +\pi \pd_A \dot{T}_{rr}^{(4)}\,,
\end{align}
where $I_{A}$ takes the same form as at $\scri^+$,
\begin{align}
    I_{A} =  \frac{1}{4}\nabla_B(\dot{C}^{BC}C_{CA}) + \frac{1}{2}C_{AB}\nabla^C\dot{C}^{BC}.
\end{align}
The gravitational-wave memory is
\begin{align}\label{memory scri-}
    \frac{1}{4}\nabla^2(\nabla^2+2)\Delta C = -\Delta m + \int_{-\infty}^{\infty} \left( \frac{1}{8}\dot{C}_{AB}\dot{C}^{AB} +4\pi T_{vv}^{(2)} \right) dv \,.
\end{align}
The charges evolve as
\begin{align}
    \Delta Q^{\scri^-}_T & = -\frac{1}{16\pi} \int_{S^2} d \Omega\Delta C \nabla^2(\nabla^2+2)\Upsilon^*T + \int_{-\infty}^{\infty} \mathcal{E}^{\scri^-}_T dv \label{deltaQTscri-}\,, \\
    \Delta Q^{\scri^-}_Y &=  \frac{1}{8\pi}\int_{S^2}\Delta \left( \frac{1}{4}C_{AB}D_C C^{BC} +\frac{1}{16}\partial_A (C_{BC}C^{BC}) \right)\Upsilon^*Y^A  d\Omega + \int_{-\infty}^{\infty}\mathcal{E}^{\scri^-}_Y dv \,,  \label{deltaQYscri-}
\end{align}
where
\begin{align}
\mathcal{E}^{\scri^-}_T &= \frac{1}{32\pi}\int_{S^2}\left( \dot{C}_{AB}\dot{C}^{AB} + 32 \pi T_{vv}^{(2)} \right) \Upsilon^*T d\Omega\,, \\
\mathcal{E}^{\scri^-}_Y & = -\frac{1}{8\pi} \int_{S^2} \left( v\partial_A \dot{m} + I_{A}  - 8\pi T_{vA}^{(2)} + \pi\partial_u \pd_A T_{rr}^{(4)}\right)\Upsilon^*Y^Ad\Omega\,.
\end{align}

\subsection{Past timelike infinity $i^-$  }
As described in section \ref{sec:flati-}, sending $\tau \to -\tau$ converts equations at $i^+$ to corresponding equations at $\scri^+$.\footnote{$\,\,$ In Sec.~\ref{sec:flati-} we used bars to distinguish quantities at $i^-$ from corresponding quantities at $i^+$.  We will not use that notation here, simply allowing context to determine whether $i^+$ or $i^-$ is being considered.}  The Beig-Schmidt expansion \eqref{Beig-Schmidt} becomes
\begin{align}
    ds^2 &= -(1 - \frac{2\sigma}{\tau}+\frac{\sigma^2}{\tau^2}+ o(\tau^{-2}))d\tau^2 + o(\frac{1}{\tau})d\tau d\phi^a \\
    &+\tau^2 \bigg( h_{ab} -\tau^{-1}(k_{ab}-2\sigma h_{ab}) + \frac{\log (-\tau)}{\tau^2}i_{ab} + \tau^{-2}j_{ab} + o(\frac{1}{\tau^2})\bigg) d\phi^a d\phi^b\,,
\end{align}
now as an expansion for $\tau \to -\infty$.  

We also change the names for the sources in the equation \eqref{sigma-point} for $\sigma$ to reflect the physical fact that the momentum direction of a massive particle entering from $i^-$ is antipodally related to its angular coordinates $(\theta,\phi)$.  This is accomplished by inserting an antipodal map on the pole location,
\begin{align}\label{sigma-pointi-}
    (D^2-3)\sigma = \sum_{n=1}^N 4\pi M_n \frac{\delta^{(3)}(\phi-\Upsilon\phi_n)}{\sqrt{h}}\,.
\end{align}
Here the parity transformation $\Upsilon$ refers only to the angular coordinates $(\theta,\phi)$ of the hyperboloid coordinates $(\rho,\theta,\phi)$.  The pole at the $i^-$ hyperboloid is at the antipodal point of the momentum direction of the $n^{\rm th}$ body.

Making the same changes to the formulas for the charges as done for $\scri^-$ in \eqref{QTscri-} and \eqref{QYscri-}, the charges at $i^-$ are
\begin{align}
    Q^{i^-}_{T} & = \frac{1}{4\pi} \int_{C}\sqrt{q}d^2x r^a(D_a\sigma \omega_{\Upsilon^*  T} - \sigma D_a\omega_{\Upsilon^* T}) \,,\label{QTi-} \\
    Q^{i^-}_{Y} & = -\frac{1}{8\pi}\int_C \sqrt{q}d^2x r^a\chi_{\Upsilon^* Y}^b\biggr(-j_{ab} +\frac{1}{2}i_{ab} + \frac{1}{2}k_{ac}k_{b}{}^c \nonumber \\&+ h_{ab}\big(-\frac{1}{8}k_{cd}k^{cd} + 8\sigma^2 -k_{cd}D^{c}D^d\sigma -D_c\sigma D^c\sigma\big)\biggr)\, ,\label{QYi-}
\end{align}
where $\omega_T$ and $Y_T$ are defined in equation \eqref{omegaT} and \eqref{chiY} respectively.

\subsection{Matching between $i^-$ and $\scri^-$}
The matching of $i^-$ and $\scri^-$ follows from the matching of $i^+$ and $\scri^+$ by sending $u \to -v$ and $\tau \to -\tau$.  
In particular, we find that the magnetic parity shear vanishes at early times $v\to - \infty$ 
\begin{align}\label{noPsi-}
    \Psi = O(v^{-1})\,,
\end{align}
and the Bondi fields are given by
\begin{align}
    m & = m^{(0)} - m^{(1)}v^{-1} + o(v^{-1})\,,\\
    C_{AB} & = C_{AB}^{(0)} -C_{AB}^{(1)}v^{-1} + o(v^{-1})\,,\\
    N^A & = \log(- v) N_A^\text{log}+ N_A^{(0)}+ o(v^0)\,.
\end{align}
The results of matching to Beig-Schmidt fields at large $\rho$ on $i^-$ are the same exact equations \eqref{m1}-\eqref{N-1},\eqref{sigma match},\eqref{krhorho match}-\eqref{kAB match}, and \eqref{irhorho match}-\eqref{jrhoA match} presented at $i^+$ and $\scri^+$.

The matching of charges then follows analogously.  In particular, 
The $v\to -\infty$ behavior of the charges is [analogously to \eqref{QTfinal} and \eqref{QYfinal}]
\begin{align}
    Q_T^{\scri^-_-} & \equiv \lim_{v \to -\infty} Q^{\scri^-}_T = \frac{1}{4\pi}\int_{S^2}(\Upsilon^*T)m^{(0)} d\Omega \label{QTinitial} \,,\\
    Q_Y^{\scri^-_-} & \equiv \lim_{v \to -\infty} Q^{\scri^-}_Y = \frac{1}{8\pi}\int_{S^2}(\Upsilon^*Y^A)N_A^{(0)} d\Omega\,, \label{QYinitial}
\end{align}
and the charges match as
\begin{align}
 Q_T^{i^-_\partial} &= Q_T^{\scri^-_-}\,,\label{i- T match}\,\\ 
Q_Y^{i^-_\partial} &= Q_Y^{\scri^-_-}\label{i- Y match}\,.
\end{align}

\section{Spatial infinity $i^0$}\label{sec:i0}

We next turn to spatial infinity.  We will follow the treatment of Refs.~\cite{compere_relaxing_2011}, with minor adjustments that make the charge algebra be the BMS algebra.  An important difference from our treatment of $i^+$ is that we retain the logarithmic translations as trivial coordinate transformations, which help us match properly to both $\scri^+$ and $\scri^-$.

From Eq.~(3.14) of Ref.~\cite{compere_relaxing_2011}, the Beig-Schmidt ansatz for spatial infinity is
\begin{align}
    ds^2 &= \left(1+\frac{2\sigma}{\rho} +\frac{\sigma^2}{\rho^2} + o(\rho^{-2}) \right)d\rho^2 +  o(\rho^{-2})\rho d\rho d\phi^a\nonumber\\
    & + \rho^2\biggr(h_{ab} + \rho^{-1}\big(k_{ab}-2\sigma h_{ab}\big) +\rho^{-2}\log\rho \, i_{ab}+ \rho^{-2}j_{ab} +o(\rho^{-2}) \biggr)d\phi^a d\phi^b\,,\label{Beig-Schmidt-spatial}
\end{align}
where $h_{ab}$ is the Lorentzian hyperboloid metric, (denoted $h^0_{ab}$ in \eqref{h0})
\begin{align}
    h_{ab}d\phi^a d\phi^b& = -d\tau^2 + \cosh^2\!\tau \gamma_{AB}dx^Adx^B\, .
\end{align}
It was shown in \cite{compere_relaxing_2011} that one may always use the coordinate freedom to set the trace of $k$ to zero, and we shall make that choice here,
\begin{align}\label{tracek0}
    k_{ab}h^{ab} = 0\,,
\end{align}
as we did at at $i^+$ in Eq.~\eqref{tracek} above. 

It is convenient to organize tensors on $i^0$ according to their behavior under the discrete parity symmetry $\Upsilon_\mathcal{H}$ of $i^0$ defined as 
\begin{align}\label{parity}
    \Upsilon_\mathcal{H} (\tau,\theta,\phi) = (-\tau, \pi - \theta, \phi + \pi)\,.
\end{align}

We include the subscript $\mathcal{H}$ (for ``hyperboloid'') to distinguish from parity on the celestial sphere,
\begin{align}\label{parity-sphere}
    \Upsilon(\theta,\phi) = (\pi - \theta, \pi + \phi)\,.
\end{align}
We say that a tensor $\mathcal{T}$ on $i^0$ has definite (even or odd) parity when $\Upsilon_\mathcal{H}^*\mathcal{T} = \pm \mathcal{T}$, with the upper/lower sign corresponding to even/odd parity, and similarly for tensors on the sphere.  In particular we have
\begin{align}
    \Upsilon^* Y_{\ell m} = (-1)^\ell Y_{\ell m}\, ,\qquad \Upsilon^* n_i = -n_i\, .
\end{align}

\subsection{Coordinate freedom}

The infinitesimal diffeomorphisms that preserve the Beig-Schmidt form \eqref{Beig-Schmidt-spatial} take the form 
\begin{align}
    \xi^\rho & = H \log \rho + \omega + o(\rho^0)\,, \\
    \xi^a &= \chi^a + \frac{\log \rho}{\rho} D^a H + \frac{1}{\rho}D^a (H+\omega) + o(\rho^{-1})\, . 
\end{align}
These equations are analogous to Eqs.~\eqref{xitau} and \eqref{xia}, except we write only the leading and subleading components.  

We will discuss each of $\chi^a$, $H$ and $\omega$ in turn.  The vector $\chi^a$ is a Killing field of $i^0$, which we relate to a conformal Killing field $Y^A$ on the sphere by
\begin{align}
    \chi_Y^\tau = b^in_i, \qquad
    \chi_Y^A  = -\epsilon^{AB}\partial_B (\kappa^in_i) + \tanh\tau \partial^A (b^in_i)\, ,\label{chiY0}
\end{align}
using Eq.~\eqref{Y}.  This equations is analogous to \eqref{chiY}, whose surrounding discussion explains the meaning of $b^i$ and $\kappa^i$ as boosts and rotations.  The parity of the hyperboloid Killing field $\chi^a_Y$ is even,
\begin{align}
    \Upsilon_\mathcal{H}^*\chi_Y^a = \chi_{Y}^a\, ,\label{paritychiYa}
\end{align}
while the sphere-parity of the associated conformal Killing field is even for rotations ($\Upsilon^* Y^A = Y^A$) and odd for boosts ($\Upsilon^* Y^A = -Y^A$). The asymptotic behavior of the generator $\chi_Y$ is given by 
\begin{align}
    \chi^\tau_Y &\sim -\frac{1}{2}\nabla_A Y^A, && \tau \to +\infty \,,\label{chitau+} \\
    \chi^\tau_Y &\sim \frac{1}{2}\nabla_A \Upsilon^* Y^A, && \tau \to -\infty \label{chitau-}\,,\\
    \chi^A_Y &\sim  Y^A, && \tau \to +\infty \label{chiA+}\,,\\
    \chi^A_Y &\sim  \Upsilon^*Y^A, && \tau \to -\infty\, . \label{chiA-}
\end{align}

The scalar $H$ satisfies
\begin{align}\label{H0eqn}
    (D_a D_b + h_{ab})H=0\, ,
\end{align}
analogously to \eqref{H constraint}.  Making arguments similar to those leading to \eqref{H}, the solutions are a four-parameter family
\begin{align}\label{H0}
H = h^0 \sinh \tau + h^i n_i \cosh \tau\, ,
\end{align}
written in terms of $h^\mu=(h^0,h^i)$, analogously to \eqref{H}.  Finally, the scalar $\omega$ satisfies
\begin{align}\label{i0supertranslation}
    (D^2+3)\omega = 0\, ,
\end{align}
analogously to Eq.~\eqref{omegaeqn}.  Later in Sec.~\ref{sec:einstein0} we will adopt an additional assumption [see \eqref{Phiodd}] that restricts $\omega$ to odd-parity solutions,
\begin{align}\label{omegaodd0}
    \Upsilon^*_{\mathcal H} \omega = - \omega\, .
\end{align}
We associate a solution $\omega_T$ to each scalar $T(x^A)=\sum T_{\ell m} Y_{\ell m}$ \eqref{T} using the odd-parity mode functions of \ref{sec:equation},
\begin{align}\label{omegaT0}
    \omega_T(\tau,x^A) = \sum_{\ell,m} T_{\ell m} \psi_\ell^O(\tau)Y_{\ell m}(x^A)\, ,
\end{align}
where we drop the hats in Eq.~\eqref{psiellO}.
From the asymptotic behavior of $\psi_\ell^O(\tau)$, we obtain
\begin{align}
    \omega_T(\tau, x^A)& \sim -\frac{1}{2}e^{\tau} T(x^A), \qquad \tau \to +\infty\,,\\
    \omega_T(\tau, x^A) & \sim +\frac{1}{2}e^{-\tau} T(\Upsilon x^A), \qquad \tau \to -\infty\,,\label{omegatauminus}
\end{align}
where $\Upsilon$ is the antipodal mapping of the sphere \eqref{parity-sphere}. 
The transformation law of the metric elements are given in Eqs. (4.104)-(4.111) in \cite{compere_relaxing_2011} 
\begin{align}
    \delta_\xi \sigma &= \mathcal{L}_{\chi}\sigma + H\, , \label{deltaxisigma0}\\
    \delta_\xi k_{ab} &= \mathcal{L}_{\chi}k_{ab} + 2 (D_aD_b + h_{ab})\omega\, ,\\
    \delta_\xi i_{ab} & = \mathcal{L}_{\chi} i_{ab} + D_c\big((D^cH)(D_aD_b\sigma + h_{ab}\sigma)\big)\, ,\\
    \delta_\xi j_{ab} &= \mathcal{L}_{\chi} j_{ab} + k_{c(a}D_{b}D^c\omega + k_{ab} \omega + D^c\omega(D_ck_{ab} - D_{(a}k_{b)c}) \nonumber \\ &+ \big(-4 \sigma \omega h_{ab} + 4 D_{(a}\sigma D_{b)}\omega + D_aD_b(- \sigma \omega+ D_c\sigma D^c \omega_T)\big)\nonumber\\ &+  10 \sigma H h_{ab} - 4 H_{(a} \sigma_{b)} + 2H D_aD_b\sigma + 2 h_{ab} D_c\sigma D^c H\, \nonumber\\
    &+ \frac{1}{2}D_c(k_{ab}D^c H)+\frac{3}{2}D_c ((D_a D_b\sigma + h_{ab}\sigma) D^c H)\, .
\end{align}

\subsection{Einstein equations}\label{sec:einstein0}

The vacuum Einstein equations imply that $\sigma$ satisfies
\begin{align}\label{sigmaeqn0}
    (D^2 + 3)\sigma = 0\,,
\end{align}
analogously to Eq.~\eqref{First Einstein}.  At $i^+$ we  adopted the condition that $\sigma$ that vanish at large $\rho$, which eliminated the logarithmic translation degrees of freedom.  At $i^0$ there are two disconnected asymptotic regions $\tau \to \pm \infty$ and it is not possible to make $\sigma$ vanish simultaneously at both.  Instead of working with a single unique $\sigma$, we will introduce two functions $\sigma^\pm$ that are related by a logarithmic translation \eqref{H0}.  We will see in Sec.~\ref{charges and representation at i0} that this logarithmic translation is trivial in that it does not affect the BMS charges; hence, $\sigma^\pm$ are both valid representatives of the same physics.  

Let us see how to construct these functions $\sigma^\pm$.  The general solution to Eq. \eqref{sigmaeqn0} is a linear combination of even and odd parity harmonics defined in \ref{sec:equation},
\begin{align}\label{sigmagen}
    \sigma = \sum_{\ell,m} \left( a^E_{\ell m} \psi_{\ell}^E(\tau) + a^O_{\ell m}\psi_{\ell}^O(\tau) \right)Y_{\ell m}(x^A)\,.
\end{align}
The odd-parity harmonics blow up at both infinities for all $\ell$.  The even-parity harmonics also blow up at both infinities for $\ell=0,1$, but vanish at both infinities for $\ell \geq 2$.  The general solution vanishing at $\tau \to + \infty$ is constructed by using the $\ell=0,1$ odd-parity modes to eliminate the divergence in the even-parity modes,
\begin{align}
    \sigma^+ = \sum_{\ell,m} a^E_{\ell m} \psi_{\ell}^E(\tau) Y_{\ell m}(x^A) + 2 a^E_{00}\psi_{0}^O(\tau)Y_{00} + 2 \sum_{m=1}^3 a^E_{1m}\psi_{m}^O(\tau) Y_{1m}(x^A)\,.
\end{align}
The solution $\sigma^-$ vanishing at $\tau \to -\infty$ is the same with the second and third terms subtracted instead of added.  Dropping the superscript $E$ on the coefficients $a_{\ell m}$, we can write both together as 
\begin{align}\label{sigmapm}
    \sigma^\pm = \sum_{\ell,m}a_{\ell m} \psi_{\ell}^E(\tau) Y_{\ell m}(x^A) \pm  H_0\,,
\end{align}
where 
\begin{align}
    H_0 & = -2 a_{00} Y_{00} \sinh \tau -2 \sum_{m=1}^3 a_{1 m}Y_{1 m} \cosh \tau \\
     & = -2 E \sinh \tau + 2 P^i n_i \cosh \tau\, . \label{cute}
\end{align}
In the second line we note that the low-$\ell$ harmonics are directly related to the energy $E$ and momentum $P^i$ according to the charges defined below in Sec.~\ref{charges and representation at i0}. 

We have chosen the name $H_0$ since the form of \eqref{cute} is precisely the allowed form of the logarithmic translation \eqref{H0}.  Indeed, according to \eqref{deltaxisigma0}, Eq.~\eqref{sigmapm} can be interpreted as the statement that $\sigma^\pm$ is even parity up to gauge.  The particular gauge transformations $H=\pm H_0$ are required to modify the purely even parity solution to be regular at $\tau \to  \pm\infty$, respectively.  

At $i^+$ we assumed that $k_{ab}$ derives from a potential $\Phi$ via $k_{ab}=-2(D_a D_b - h_{ab})\Phi$ \eqref{kfromPhi}, motivated by the idea that $k_{ab}$ should be pure gauge, as it is in the Schwarzschild spacetime.  Although there is no analogous motivation at $i^0$, the success of this assumption in giving us the BMS group at $i^+$ suggests that we make the analogous assumption at $i^0$,
\begin{align}
\label{kfromPhi0}
    k_{ab} =-2 (D_aD_b + h_{ab})\Phi\, .
\end{align}
According to Sec.~4.2 of Ref.~\cite{compere_relaxing_2011} using Eq.~(A.56) of Ref.~\cite{compere_asymptotic_2011}, this assumption is equivalent to the vanishing of the magnetic portion of the Weyl tensor at first order in the Beig-Schmidt expansion. 
Eq.~\eqref{tracek0} implies that $\Phi$ satisfies
\begin{align}\label{Phieqn0}
(D^2+3)\Phi=0\, .
\end{align}
This is the same equation as satisfied by $\sigma$ and has the same general solution \eqref{sigmagen} in terms of even- and odd-parity solutions.  At $i^+$ we selected a single family of mode solutions via the requirement that $\Phi$ be smooth. However, this requirement is trivial at $i^0$, since both families are smooth. We will exclude the even-parity solutions, for the following reasons.  First, we will see in Sec. \ref{charges and representation at i0} that the $\ell=0,1$ even-parity solutions would give rise to non-zero charge associated with logarithmic translations.  Allowing these solutions would thus make the symmetry group larger than the BMS group we wish to recover; without a compelling physical reason to include the logarithmic translation charges, we choose to discard them. The $\ell \geq 2$ even parity solutions fall off as $e^{-3|\tau|}$ and can consistently be taken to vanish without affecting the matching of all fields on $i^0$ and $\scri^\pm$.  Note also that the analytic continuations of the smooth solutions at $i^+$ are exactly the odd-parity solutions we use at $i^0$ [see Eq.~\eqref{psii0i+}].  Thus we demand that $\Phi$ and hence $k_{ab}$ has purely odd parity,
\begin{align}\label{Phiodd}
    \Upsilon_\mathcal{H}^* \Phi = - \Phi \, ,\qquad \Upsilon_\mathcal{H}^* k_{ab} = - k_{ab}\, .
\end{align}
This implies our earlier analogous condition \eqref{omegaodd0} on $\omega$, which satisfies the same equation as $\Phi$.  The $\Phi$ mode solutions are the same as those of $\omega$ in \eqref{omegaT0}--\eqref{omegatauminus}, except that (as with $i^+$) we assume without loss of generality that $\Phi$ has no $\ell=0,1$ modes,    
\begin{align}\label{Phisolution0}
    \Phi(\tau, x^A) =\sum_{\ell\geq 2,m}C^{(0)}_{\ell m} \psi^{O}_{\ell m}(\tau) Y_{\ell m}(x^A)\, .
\end{align}
The asymptotic behavior is 
\begin{align}
    \Phi(\tau,\theta,\phi) &\sim -\frac{1}{2}C^{(0)} e^{\tau}, \qquad \tau \to +\infty \label{Phi0+}\,, \\
    \Phi(\tau,\theta,\phi) &\sim +\frac{1}{2}\Upsilon^*C^{(0)} e^{|\tau|}, \qquad \tau \to -\infty\label{Phi0-} \,,
\end{align}
where 
\begin{align}\label{C0eq}
    C^{(0)}(x^A) = \sum_{\ell\geq 2,m}C^{(0)}_{\ell  m} Y_{\ell m}(x^A)\,.
\end{align}
Note that the odd parity solutions have the correct $e^{|\tau|}$ blowup as $\tau \to \pm \infty$ required for $k_{ab}$ to match to $C_{AB}$ at $\scri^\pm$ [Eqs.~\eqref{ki0+} and \eqref{ki0-}].

The subleading metric components $i_{ab}$ and $j_{ab}$ obey the following differential equations
\begin{align}
    (D^2-2)i_{ab} = 0 \,,&& D^ai_{ab} = 0\,, && h^{ab}i_{ab} = 0\, .
\end{align}
with 
\begin{align}\label{Evolution}
    &(D^2-2)j_{ab} = 2i_{ab} + S_{ab}\,,\\
    &D^aj_{ab} = \frac{1}{2}k^{bc} D_bk_{ac}+D_a((D_c\sigma)D^c\sigma+8\sigma^2-\frac{1}{8}k_{cd}k^{cd}+k_{cd}D^cD^d\sigma)\,,\\
    &h^{ab}j_{ab} = 12\sigma^2 + \sigma_c\sigma^c+\frac{1}{4}k_{cd}k^{cd} + k_{cd}D^cD^d\sigma\,.
\end{align}
These equations are from Ref.~\cite{compere_relaxing_2011}, equation (C.168) to (C.173).
The form of $S_{ab}$ is quadratic in $k_{ab}$ and $\sigma$. We refer to equation (C.174) in Ref.~\cite{compere_relaxing_2011}.\\

\subsection{Charges and BMS representation}\label{charges and representation at i0}

At $i^+$, the charges must surround sources to be non-zero.  At $i^0$ we do not have sources, but the topology allows surfaces that cannot be continuously shrunk to a point and hence may have non-zero ``topological'' charges.  These surfaces arise as limits of large surfaces enclosing the bulk spacetime and give the charges of physical interest.

The charges and their representation on $i^0$ are described in detail in Sec.~4.6 of Ref.~\cite{compere_relaxing_2011}. Here we summarize the results that are useful for our purposes.  The charges are given by 
\begin{align}
     Q_{T}^{i^0}& = \frac{1}{4\pi}\int_{S^2}d^2x \sqrt{-h}n_a(\omega_T D^a\sigma - \sigma D^a \omega_T) \,,\label{QTi0} \\
    Q_{Y}^{i^0} & = -\frac{1}{8\pi} \int_{S^2}d^2x \sqrt{-h}\chi_Y^b n^a \bigg(-j_{ab} + \frac{1}{2}i_{ab}+\frac{1}{2}k_{a}{}^ck_{cb} \nonumber\\ &+ h_{ab} \big(8\sigma^2 + \sigma_c\sigma^c -\frac{1}{8}k_{cd}k^{cd} + k_{cd}D^cD^d\sigma\big)\bigg)\label{QYi0}\,,\\
    Q_{H} & = \frac{1}{16\pi}\int_{S^2} d^2x\sqrt{-h}n^a k_{ab}D^b H \label{QH0}\,,
\end{align}
where $h = \text{det}(h_{ab})$ and $n_a$ is the future-directed unit normal vector to two-sphere $S^2$.  Since these integrals are independent of the surface, we choose the two-sphere for simplicity.  Using $\sigma=M \cosh2\tau \,\text{sech}\tau$ and $T=-\sinh \tau$ we can check that $Q^{i^0}_T=+M$ (at any $\tau$).  These are analogous to Eqs.~\eqref{QTi+} and \eqref{QYi+} at $i^+$, except that now we have a third charge, $Q_H$, since we have not excluded the logarithmic transformation at $i^0$.  However, it turns out that this charge vanishes under our assumptions:
\begin{align}
    Q_H& = \frac{1}{16\pi}\int_{S^2} d^2x\sqrt{-h}n^a (D_aD_b + h_{ab})\Phi D^b H\nonumber\\
    & = \frac{1}{16\pi}\int_{S^2} d^2x\sqrt{-h}n^a\big(\Phi D_aH + (D_bH)(D_a D^b \Phi) \big)\nonumber\\
 & = \frac{1}{8\pi}\int_{S^2} d^2x\sqrt{-h}n^a\big(-\Phi D_aH+ H D_a \Phi - \!\!\underbrace{D^b(D_{[a}H D_{b]} \Phi)}_\text{total divergence, vanishes} \big)\nonumber\\
    & = 0\, .\label{0}
\end{align}
In the first line we used Eq.~\eqref{kfromPhi0}, and in the third line  we used Eqs.~\eqref{H0eqn} and \eqref{Phieqn0}.  For the last equality, we use the orthogonality of the spherical harmonics together with the fact that $H$ \eqref{H0} only contains $\ell \leq 1$  harmonics while $\Phi$ \eqref{Phisolution0} only contains $\ell \geq 2$ harmonics. If we had included the even-parity harmonics of $\Phi$, the $\ell=0,1$ terms would contribute and the logarithmic charge $Q_H$ would be non-zero in general.

Finally we present the charge algebra.  As with $i^+$, the change in a charge defines a Poisson bracket \cite{Barnich:2007bf} 
\begin{align}\label{poisson0}
    \bigl\{ Q_{\xi} ,Q_{\xi'} \bigr\}= -\delta_\xi Q_{\xi'} = \delta_{\xi'} Q_{\xi}\, .
\end{align}
The supertranslation and Lorentz charges form the BMS algebra,
\begin{align}
    \bigl\{Q_{T_1},Q_{T_2} \bigr\} &= 0 \,,\label{qq1} \\
    \bigl\{ Q_{T_1},Q_{Y_2}\bigr\} & = Q_{Y_2(T_1)}\,, \label{qq2}\\
    \bigl\{ Q_{Y_1},Q_{Y_2}\bigr\} & = Q_{[Y_1,Y_2]} \,,\label{qq3}
\end{align}
where $Y(T)$ was given in \eqref{YT}. The logarithmic translation charges commute with all BMS charges,
\begin{align}
    \bigl\{Q_{H_1}, Q_{Y_2}\bigr\} =
    \bigl\{Q_{H_1},Q_{H_2} \bigr\} = 
    \bigl\{Q_{T_1},Q_{H_2} \bigr\} 
    =0 \label{qq6}\,,
\end{align}
which follow from the vanishing of $Q_H$ \eqref{0} together with \eqref{poisson0}. Therefore our asymptotic symmetry group is the BMS group. One can also obtain the BMS group at $i^0$ using Hamiltonian methods \cite{Henneaux:2018cst}.
 
It is useful to contrast with our approach at $i^\pm$.  At $i^\pm$ we removed the logarithmic degree of freedom at the outset by requiring that $\sigma$ vanish at large $\rho$, as needed to match to $\scri^\pm$.  Here at $i^0$ we included the logarithmic degree of freedom as a trivial gauge transformation, since this transformation is needed to switch between $\sigma^+$ and $\sigma^-$, which vanish at $\tau \to \infty$ and $\tau \to -\infty$ (respectively) and are needed to match to $\scri^+$ and $\scri^-$ (respectively). In both cases we obtain the BMS group.

\section{Matching between $i^0$ and $\mathcal{I}^{\pm}$}\label{sec:i0scripm}

It remains to match $i^0$ together with $\scri^\pm$.  The assumptions and calculations are precisely analogous to those of Sec.~\ref{sec:i+scri+}, and we present just a few key equations and results. We use a vertical bar with the symbols introduced in Sec.~\ref{sec:overlaps} to distinguish different quantities with the same name.
For example, $m|_{\scri^+_-}$ refers to the mass aspect of $\scri^+$ considered at $u \to -\infty$, while $m|_{\scri^-_+}$ refers to the mass aspect of $\scri^-$, considered at $v \to \infty$.  

As before, we find that the shear $C_{AB}$ matches to $k_{ab}$ and hence must be purely electric at both $\scri^-_+$ and $\scri^+_-$ [the analog of Eq.~\eqref{nopsi0}], 
\begin{align}\label{nopsipm}
    \Psi^{(0)}|_{\scri^+_-} = \Psi^{(0)}|_{\scri^-_+} = 0\, .
\end{align}
This implies the lack of a linear divergence in the angular momentum aspect [the analog of Eq.~\eqref{noN-1}].  The Bondi quantities take the form
\begin{align}
    m|_{\scri^+_-} & = m^{(0)}\big|_{\mathcal{I}_-^+} + m^{(1)}\big|_{\mathcal{I}_-^+}u^{-1} + o(u^{-1})\,,\\
    C_{AB}|_{\scri^+_-} & = C_{AB}^{(0)}\big|_{\mathcal{I}_-^+} +C_{AB}^{(1)}\big|_{\mathcal{I}_-^+}u^{-1} + o(u^{-1})\,,\\
    N^A|_{\scri^+_-} & = \log(-u) N_A^\text{(log)}\big|_{\mathcal{I}_-^+} + N_A^{(0)}\big|_{\mathcal{I}_-^+} + o(u^0)\, \label{NAansatzsnegativeu}.
\end{align}
and
\begin{align}
    m|_{\scri^-_+} & = m^{(0)}\big|_{\mathcal{I}^-_+} - m^{(1)}\big|_{\mathcal{I}^-_+}v^{-1} + o(v^{-1})\,,\\
    C_{AB}|_{\scri^-_+} & = C_{AB}^{(0)}\big|_{\mathcal{I}^-_+} -C_{AB}^{(1)}\big|_{\mathcal{I}^-_+}v^{-1} + o(v^{-1})\,,\\
    N^A|_{\scri^-_+} & = \log v N_A^\text{log}\big|_{\mathcal{I}^-_+} + N_A^{(0)}\big|_{\mathcal{I}^-_+} + o(v^0)\, .
\end{align}
The evolution equations for these quantities are identical, but we write them twice for consistency:
\begin{align}
   m^{(1)}\big|_{\mathcal{I}^+_-} & = -\frac{1}{4}\nabla^2(\nabla^2+2)C^{(1)}\big|_{\mathcal{I}^+_-}\,,\\
    N_{A}^\text{(log)}\big|_{\mathcal{I}^+_-} & = -\partial_A m^{(1)}\big|_{\mathcal{I}^+_-} -\frac{1}{4}\epsilon_{AB}\nabla^B\mathcal{D}\Psi^{(1)}\big|_{\mathcal{I}^+_-}\,,
\end{align}
and
\begin{align}
   m^{(1)}\big|_{\mathcal{I}^-_+} & = -\frac{1}{4}\nabla^2(\nabla^2+2)C^{(1)}\big|_{\mathcal{I}^-_+}\,,\\
    N_{A}^\text{(log)}\big|_{\mathcal{I}^-_+} & = -\partial_A m^{(1)}\big|_{\mathcal{I}^-_+} -\frac{1}{4}\epsilon_{AB}\nabla^B\mathcal{D}\Psi^{(1)}\big|_{\mathcal{I}^-_+}\, .
\end{align}
Performing a coordinate transformation analogous to \eqref{ansatzu}--\eqref{ansatzA}, now relating $\scri^+_-$ to $i^0_+$, we find that the metric can be put into Beig-Schmidt form \eqref{Beig-Schmidt-spatial} with
\begin{align}
    &\sigma\big|_{i^0_+} = 2m^{(0)}\big|_{\mathcal{I}_-^+}e^{-3\tau} + o(e^{-3\tau}) \label{sigmai0+} \,,\\
    &k_{AB}\big|_{i^0_+} = \frac{1}{2}e^{\tau}C_{AB}^{(0)}\big|_{\mathcal{I}_-^+} + o(e^{+\tau}) \label{ki0+}\,,\\
    &i_{\tau\tau}\big|_{i^0_+} =o(e^{-2\tau})\,,\\
    &i_{\tau A}\big|_{i^0_+} = 4e^{-2\tau}N_A^\text{(log)}\big|_{\mathcal{I}_-^+}+o(e^{-2\tau})\,,\\
    &j_{\tau\tau}\big|_{i^0_+} = -16m^{(1)}\big|_{\mathcal{I}_-^+}e^{-2\tau} + o(e^{-2\tau})\,,\\
    &j_{\tau A}\big|_{i^0_+} = -\nabla^BC^{(1)}_{AB}\big|_{\mathcal{I}_-^+} + \big(4N_A^{(0)}\big|_{\mathcal{I}_-^+} +C_{AB}^{(0)}\big|_{\mathcal{I}_-^+}\nabla_CC^{(0)BC}\big|_{\mathcal{I}_-^+}+ (-4\tau + 2)N_A^\text{(log)}\big|_{\mathcal{I}_-^+}\nonumber\\
    &+\nabla^BC^{(1)}_{AB}\big|_{\mathcal{I}_-^+}+12\partial_Am^{(1)}\big|_{\mathcal{I}_-^+}\big)e^{-2\tau}+ o(e^{-2\tau})\,,
\end{align}
holding  for $\tau \to \infty$.  The analogous matching of $\scri^-_+$ with $i^0_-$ gives the same equations with $\tau \to -\tau$,
\begin{align}
    &\sigma\big|_{i^0_-} = 2m^{(0)}\big|_{\mathcal{I}^-_+}e^{-3|\tau|} + o(e^{-3|\tau|}) \label{sigmai0-}\,,\\
    &k_{AB}\big|_{i^0_-} = \frac{1}{2}e^{|\tau|}C_{AB}^{(0)}\big|_{\mathcal{I}^-_+} + o(e^{+|\tau|}) \label{ki0-}\,,\\
    &i_{\tau\tau}\big|_{i^0_-}=o(e^{-2|\tau|})\,,\\
    &i_{\tau A}\big|_{i^0_-} = -4e^{-2|\tau|}N_A^\text{(log)}\big|_{\mathcal{I}^-_+}+o(e^{-2|\tau|})\,,\\
    &j_{\tau\tau}\big|_{i^0_-} = -16m^{(1)}\big|_{\mathcal{I}^-_+}e^{-2|\tau|} + o(e^{-3|\tau|})\,,\\
    &j_{\tau A}\big|_{i^0_-} = \nabla^BC^{(1)}_{AB}\big|_{\mathcal{I}^-_+}+\big(-4N_A^{(0)}\big|_{\mathcal{I}^-_+}-C_{AB}^{(0)}\big|_{\mathcal{I}^-_+}\nabla_CC^{(0)BC}\big|_{\mathcal{I}^-_+}-(-4|\tau|+2)N_A^\text{(log)}\big|_{\mathcal{I}^-_+}\nonumber \\
    &-\nabla^BC^{(1)}_{AB}\big|_{\mathcal{I}^-_+}-12\partial_Am^{(1)}\big|_{\mathcal{I}^-_+}\big)e^{-2|\tau|}+ o(e^{-2|\tau|})\,,\label{end}
\end{align}
now holding for $\tau \to -\infty$.  

By same reasoning leading to Eqs.~\eqref{QTfinal} and \eqref{QYfinal}, we have
\begin{align}
    Q_T^{\scri^+_-} & \equiv \lim_{u \to -\infty} Q^{\scri^+}_T = \frac{1}{4\pi}\int_{S^2}Tm^{(0)}\big|_{\mathcal{I}_-^+} d\Omega \label{QTscri+-} \,,\\
    Q_Y^{\scri^+_-} & \equiv \lim_{u \to -\infty} Q^{\scri^+}_Y = \frac{1}{8\pi}\int_{S^2}Y^AN_A^{(0)}\big|_{\mathcal{I}_-^+} d\Omega \label{QYscri+-} \,,\\    
    Q_T^{\scri^-_+} & \equiv \lim_{v \to +\infty} Q^{\scri^-}_T = \frac{1}{4\pi}\int_{S^2}(\Upsilon^*T) m^{(0)}\big|_{\mathcal{I}_+^-} d\Omega \label{QTscri-+} \,,\\
    Q_Y^{\scri^-_+} & \equiv \lim_{v \to +\infty} Q^{\scri^-}_Y = -\frac{1}{8\pi}\int_{S^2}(\Upsilon^*Y^A)N_A^{(0)}\big|_{\mathcal{I}_+^-} d\Omega\,, \label{QYscri-+}
\end{align}
where the appearance of the antipodal map $\Upsilon$ arises from our definitions \eqref{QTscri-} and \eqref{QYscri-} of the charges at past null infinity.

Using the formulas for the $i^0$ charges \eqref{QTi0} and \eqref{QYi0} together with the asymptotic behaviors \eqref{chitau-} and \eqref{chiA+} as well as the matching equations \eqref{sigmai0+}--\eqref{end}, we find that the charges properly match,
\begin{align}\label{i0match}
    Q^{\scri^+_-} = Q^{i^0} = Q^{\scri^-_+}\,,
\end{align}
holding both for $Q_T$ and $Q_Y$.  The calculation is similar to that described surrounding Eqs.~\eqref{Future null future timelike momentum Match} \eqref{Future null future timelike Lorentz Match}.

We now discuss a subtlety.  According to \eqref{sigmai0+} and \eqref{sigmai0-}, it appears that $\sigma$ is required to vanish at both $\tau \to \infty$ and $\tau \to -\infty$, which would be inconsistent with Einstein's equation [see discussion below Eq.~\eqref{sigmaeqn0}] .  However, Eq.~\eqref{sigmai0+} and Eq.~\eqref{sigmai0-} are in service of the matching of gauge-invariant charges, and there is no requirement to use the same gauge for $\sigma$ in the two different matching locations.  In other words, the relevant consistency check is that there exists a gauge transformation switching between the behaviors \eqref{sigmai0+} at $\tau \to -\infty$ and \eqref{sigmai0-} at $\tau \to \infty$.  This gauge transformation was identified in Sec.~\ref{sec:einstein0} as the logarithmic translation degree of freedom, and the needed form of $\sigma$ at $i^0_\pm$ was denoted $\sigma_\pm$.  The logarithmic transformation does not affect $k_{ab}$, and the needed behavior \eqref{ki0+} and \eqref{ki0-} can be seen from the end behavior of $k_{ab}$ from \eqref{kfromPhi0}, \eqref{Phi0-}, and \eqref{Phi0+}.  Similar consistency checks could be performed on the other fields, with enough effort analyzing their detailed equations of motion.  We have confined ourselves to the minimal analysis required to demonstrate the matching of charges.

\subsection{Antipodal matching across $i^0$}\label{sec:antipodal}

There has been considerable discussion of potential antipodal relationships between fields at $\scri^+_-$ and $\scri^-_+$.  We can check whether this occurs in our formalism as follows.  According to the equations of motion at $i^0$, the field $\sigma^+$ used to match at $i^0_+$ is related to the field $\sigma^-$ used to match at $i^0_-$ by a hyperboloid parity transformation \eqref{parity},
\begin{align}
    \Upsilon^*_\mathcal{H}\sigma^+ = \sigma^-\,,
\end{align}
as can be seen from \eqref{sigmapm}.  Since these match to $m$ at $\scri^+_-$ and $\scri^-_+$ via Eqs.~\eqref{sigmai0+} and \eqref{sigmai0-}, we find
\begin{align}\label{mmatch}
    \Upsilon^*m^{(0)}\big|_{\mathcal{I}_-^+} = m^{(0)}\big|_{\mathcal{I}_+^-} \,,
\end{align}
noting that when restricted to the sphere, $\Upsilon_\mathcal{H}$ reduces to the sphere parity transformation $\Upsilon$ \eqref{parity-sphere}.  Thus the mass aspect indeed matches antipodally.

For the shear, we note that the field $k_{ab}$ has odd parity \eqref{Phiodd}, so Eqs.~\eqref{ki0+} and \eqref{ki0-} imply
\begin{align}\label{Cmatch}
    \Upsilon^*C_{AB}^{(0)}\big|_{\mathcal{I}_-^+} & = -C_{AB}^{(0)}\big|_{\mathcal{I}_+^-}\,.
\end{align}
Eqs.~\eqref{mmatch} and \eqref{Cmatch} derive the antipodal relations used by Strominger \cite{Strominger2014} as a consequence of Einstein's equation in our framework.

An antipodal relationship has also been conjectured for the angular momentum aspect $N_A$ \cite{hawking_superrotation_2017}.  We have seen that $N_A$ actually diverges logarithmically as a consequence of the logarithmic correction to the subleading soft graviton theorem \cite{Saha:2019tub}, and hence cannot match directly.  We could still investigate whether the coefficient of the divergence matches, but this would require analyzing Einstein's equation at second order in the Beig-Schmidt expansion (solving for $i_{ab}$ and $j_{ab}$), which is outside the scope of this paper.  However, we can at least analyze the portion of $N_A$ that contributes to the charges, which we have shown to match.  In particular, we know from \eqref{i0match} that the right-hand-sides of Eqs.~\eqref{QYscri+-} and \eqref{QYscri-+} must be equal for all conformal Killing fields $Y_A$.  Since these span the $\ell=1$ vector harmonics on the sphere, it follows that the $\ell=1$ components of $N^{(0)}_A$ must be antipodally related with a minus sign,
\begin{align}\label{antipodalNA}
    \Upsilon^* N_A^{\ell = 1}|_{\scri^+_-}=-N_A^{\ell = 1}|_{\scri^-_+}\,.
\end{align}
This differs by a minus sign from the conjectured relationship (2.13) in Ref.~\cite{hawking_superrotation_2017} and from the work of \cite{capone_charge_2022} offered in support of that conjecture.  
In the appendices we provide two examples confirming the sign in \eqref{antipodalNA}---the translated Schwarzschild spacetime (\ref{sec:schwarzschild}) and the Kerr spacetime (\ref{sec:Kerr})---and in Sec.~\ref{sec:comparison} we discuss errors in Ref.~\cite{capone_charge_2022}. 

\subsection{Comparison with previous work}\label{sec:comparison}

Significant previous work has been devoted to the matching of charges between $\scri^-$ and $\scri^+$.  Early work matched momenta and angular momenta, while imposing boundary conditions that discarded supertranslations \cite{Ashtekar:1978zz,PhysRevLett.43.649,1979JMP....20.1362A}.  Later, the method of conformal geodesics was used to match Newman-Penrose constants across $\scri^+$ and $\scri^-$ \cite{1998JGP....24...83F,Friedrich:1999ax,Friedrich:1999wk}.  More recently, Strominger conjectured an antipodal match of all BMS charges between $\scri^+$ and $\scri^-$ \cite{Strominger2014}. BMS was first matched across $i^0$ by Troessaert in linearized theory (which is sufficient to fully match the supermomenta in the non-linear theory)  \cite{troessaert_bms4_2018} using Refs.~\cite{compere_relaxing_2011,1998JGP....24...83F} (see also \cite{Prabhu:2019fsp,Mohamed:2021rfg}). Ref.~\cite{Prabhu:2021cgk} matched all BMS charges across $i^0$ after gauge-fixing supertranslations close to $i^0$.

Most recently, Ref.~\cite{capone_charge_2022} performed an analysis of matching across $i^0$ that is quite similar to ours.  However, they assumed that $\sigma$ at $i^0$ has no $\ell=0,1$ modes [see their (5.14)].  As mentioned below Eq.~\eqref{cute}, the $\ell=0,1$ modes of $\sigma$ in fact encode the total energy and momentum of the spacetime.  By the positive mass theorem \cite{schon_proof_1979,schoen_positivity_1979,schon_proof_1981,witten_new_1981}, the only spacetime satisfying the assumptions of \cite{capone_charge_2022} is flat spacetime.   Ref.~\cite{capone_charge_2022} obtained the matching equation \eqref{antipodalNA} without the minus sign, which is consistent with the correct relation \eqref{antipodalNA} only when  $N_A=0$. They also did not include the $\log u$ term in $N_A$ [See our \eqref{ansatzN2} as compared with their (2.4b)], which contradicts the logarithmic correction to the subleading soft theorem \cite{Saha:2019tub}.

\section{Flux balance laws and scattering formalism}\label{sec:flux-balance}

In the previous sections we established assumptions and definitions such that BMS group is the asymptotic symmetry group of all five infinities and demonstrated how the different BMS groups can be identified such that charges with the same representatives $T$ and $Y^A$ are numerically equal in the regions of overlap.  We can thus think of a single BMS group with globally conserved charges.  In particular, consider a spacetime with $N^-$ incoming particles and $N^+$ outgoing particles, in the sense of the sources present in \eqref{sigma-point} and \eqref{sigma-pointi-}, respectively.  Let $Q^{i^+}_n$ be the final Beig-Schmidt charge of the $n{}^{\rm th}$ particle [Eqs.~\eqref{QTi+} and \eqref{QYi+} on a sphere surrounding the particle], and let $Q^{i^-}_n$ be the initial Beig-Schmidt charge of the $n{}^{\rm th}$ initial particle [Eqs.~\eqref{QTi-} and \eqref{QYi-} on a sphere surrounding the particle].  As discussed below \eqref{chargePlm}, the sum of all the $Q_n^{i^+}$ is equal to the total charge $Q^{i^+_{\pd}}$ defined in \eqref{Future null future timelike momentum Match}, and similarly for $i^-$. Then from the matching equations~\eqref{i+ T match}, \eqref{i+ Y match}, \eqref{i- T match}, \eqref{i- Y match} and \eqref{i0match}, it follows that
\begin{align}\label{trivial}
    \sum_{n=1}^{N^+} Q^{i^+}_{n} + \Delta Q^{\scri^+} = Q^{i^0} =  \sum_{n=1}^{N^-} Q^{i^-}_n + \Delta Q^{\scri^-},
\end{align}
for any charge $T$ or $Y$.  The formulas for $\Delta Q^{\scri^\pm}$ are given in Eqs.~\eqref{deltaQT}, \eqref{deltaQY}, \eqref{deltaQTscri-} and \eqref{deltaQYscri-}.  Since the magnetic parity shear vanishes at $u \to \pm \infty$ on $\scri^+$ and $v \to \pm \infty$ on $\scri^-$ [Eqs.~\eqref{nopsi0}, \eqref{noPsi-}, \eqref{nopsipm}], the first terms in \eqref{deltaQY} and \eqref{deltaQYscri-} in fact vanish on account of \eqref{rel0}.

Eq.~\eqref{trivial} holds for any choice of scalar $T$ or conformal Killing field $Y^A$, where the appropriate formulas for the BMS charges are \eqref{QTi+} and \eqref{QYi+} for $i^+$, \eqref{QTscri+} and \eqref{QYscri+} for $\scri^+$, \eqref{QTi0} and \eqref{QYi0} for $i^0$,  \eqref{QTscri-} and \eqref{QYscri-} for $\scri^-$, and \eqref{QTi-} and \eqref{QYi-} for $i^-$.  The choices of $T$ and $Y$ corresponding to traditional names are 
\begin{table}[h!]
    \centering
    \begin{tabular}{|c|c|c|}
        \hline
        Charge & name & generator \\
        \hline
       $E$  &  Energy &$\!\!\!\!\!\!\!\!\!\!\!\!\!\!\!\!\!\!\!\!\!\!\!\!\!\! T = 1$\\
       $\,P^i$ & Momentum &$\!\!\!\!\!\!\!\!\!\!\!\!\!\;  \; T  = n^i(\theta,\phi)$\\
       $\;\,\;P_{\ell m}$ & Supermomentum & $\!\!\!\!\!\!\!\!\!\!\!\!\!\qquad \!\!\!\!\! T = Y_{\ell m}(\theta, \phi)$\\
       $L^i$ & Angular Momentum &$\!\!\!\!\!\!\!\!\!\!\!\!\qquad\;\;\;\; Y^{A} = - \epsilon^{AB} \partial_Bn^i(\theta,\phi) $\\
       $\,N^i$ & Mass Moment &$\qquad \!\!\!\!\!\!\!\!\!\!\!\!\!\!\!\!\!\!\!Y^{A} =  \partial^An^i(\theta,\phi) $\\
       \hline
    \end{tabular}
    \label{sign convention}
\end{table}

We use the standard orientations $\epsilon_{t r \theta \phi}=\epsilon_{r \theta \phi}=+r^4\sin^2\theta$ and  $\epsilon_{\theta\phi}=+\csc\theta$. The conventions are all standard except possibly for mass moment; here we use a definition that reduces to $\vec{N}=E \vec{x} - \vec{P} t$ for a point particle in flat spacetime.

\subsection{Gravitational memory}\label{sec:gravitational-memory}
As reviewed in Sec.~\ref{sec:scri+}, a change in Bondi shear physically leads to a finite permanent displacement of inertial observers at $\scri^\pm$ known as gravitational memory.  In our case the memory is purely electric parity, characterized by a change in the potential $C$ of \eqref{Cdecomp}, which is written in explicit form in Eq. \eqref{changeC}.  The changes across $\scri^\pm$ can be written 
\begin{align}
    \Delta C|_{\mathcal{I}^+} &= C|_{\mathcal{I}^+_+} -  C|_{\mathcal{I}^+_-} \label{DeltaC+} \,,\\
    \Delta C|_{\mathcal{I}^-} &= C|_{\mathcal{I}^-_+} -  C|_{\mathcal{I}^-_-}\,. \label{DeltaC-}
\end{align}
Formulas for $\Delta C|_{\mathcal{I}^+}$ and $\Delta C|_{\mathcal{I}^-}$ are given in Eqs.~\eqref{memory scri+} and \eqref{memory scri-}, respectively.  Since the shear matches antipodally across $i^0$ with a minus sign \eqref{Cmatch}, the natural notion of total displacement memory is
\begin{align}\label{Ctotal}
\Delta C^{\rm total} & = \Delta C|_{\scri^+} + (- \Upsilon^*\Delta C|_{\scri^-}) \\
& = C|_{\mathcal{I}^+_+} - (-\Upsilon^*C|_{\mathcal{I}^-_-})\,,
\end{align}
where in the second line we use Eqs.~\eqref{DeltaC+}, \eqref{DeltaC-}, and \eqref{Cmatch}.  From Eqs.~\eqref{kfromPhi}, \eqref{Phisolution}, and \eqref{supertranslation}, noting that $C^{(0)}$ in these equations is equal to $C|_{\mathcal{I}^+_+}$ defined here, we see that $C|_{\mathcal{I}^+_+}$ uniquely determines $k_{ab}|_{i^+}$.  Similarly, $C|_{\mathcal{I}^-_-}$ uniquely determines $k_{ab}|_{i^-}$.  We say the presence of non-zero memory forces a non-zero change in $k_{ab}$ from $i^-$ to $i^+$,
 \begin{align}
     k_{ab}^{i^+} - (-\Upsilon^*k^{i^-}_{ab}) \propto \Delta C^{\rm total}\,,
 \end{align}
 where the left-hand side is the relevant notion of ``change''. (The manifolds $i^\pm$ are identified by the matching across $\scri^\pm$ and $i^0$ performed in this paper.)  
 While $k_{ab}$ is pure gauge at both $i^+$ and $i^-$ individually, its change is gauge-invariant, encoding the physical effect of gravitational memory.  

\subsection{Spin and its supertranslation invariance}

In scattering problems, one usually discusses the mass and spin of individual bodies.  The BMS charges can be used to define these quantities in a BMS-covariant manner.  
The mass of a body is defined by
\begin{align}\label{mass}
M = \sqrt{E^2 - P^i P_i}\,,
\end{align}
which is fully BMS-invariant.  To define the spin, note that a point particle at position $x^i$ in special relativity has angular momentum $L^i = S^i +\epsilon^i{}_{jk}x^jP^k$ and mass moment $N^i =  E x^i - P^i t$, where $S^i$ is its spin.  The formula for spin in terms of charges is thus
\begin{align}\label{spin}
    S^i = L^i - \frac{1}{E}\epsilon^i{}_{jk}N^jP^k\,,
\end{align}
which defines the spin of a body in our framework.  The spin is invariant under translations and transforms in the expected way under rotations and boosts.  We now show that the spin is in fact invariant under supertranslations as well.

The infinitesimal change in the spin $S$ under a supertranslation $T$ is given by
\begin{align}
\delta_T S^i & = \delta_T L^i - \frac{P^k}{E}\epsilon^i{}_{jk}\delta_T N^j \\
& = -Q_{Y_{L^i}(T)} - \frac{P^k}{E}\epsilon^i{}_{jk}(- Q_{Y_{N^j}(T)})\,,\label{Hongji special}
\end{align}
where in the second line we use \eqref{BMSTT} and \eqref{BMSTY}, and from \eqref{YT} we have 
\begin{align}
    Y_{L^i}(T) & =  Y^A_{L^i} \pd_A T - \frac{1}{2} \nabla_A Y^A_{L^i} T = -\epsilon^{AB}\partial_B n_i \pd_A T\,,\\
    Y_{N^i}(T) & =  Y^A_{N^i} \pd_A T - \frac{1}{2} \nabla_A Y^A_{N^i} T = \gamma^{AB} \partial_B n_i \pd_A T +n_i T\,.
\end{align}
The important point is that only translation and pure supertranslation charges appear in \eqref{Hongji special}.  As noted above \eqref{QTn}, for the purpose of computing such a charge one may use the $\sigma$ due to the body alone.  From \eqref{sigmalargerho}, this field has the large-$\rho$ behavior
\begin{align}
    \sigma \sim -2m(x^A)e^{-3\rho}\,,
\end{align}
where we introduce the mass aspect of the body $m(x^A)$ as
\begin{align}\label{aspectonebody}
    m(x^A) = \frac{M}{\gamma^3(1-v_i n^i(x^A))^3}\,,
\end{align}
with $v^i=P^i/E$ and $\gamma = 1/\sqrt{1-v_iv^i}$.  Using these expressions, we may compute
\begin{align}
    \delta_{T}L_i &= \{L_i,Q_T\} = - Q_{Y_{L^i}(T)} =- \frac{1}{4\pi}\int_{S^2}d\Omega(\epsilon^{AB}\partial_A m \partial_B n_i) T\,, \\
    \delta_{T} N_i &= \{N_i,Q_T\} = - Q_{Y_{N^i}(T)}  = \frac{1}{4\pi}\int_{S^2}d\Omega (\partial_A m \partial^A n_i- 3 n_i m)T\,.
\end{align}
Plugging in to Eq. \eqref{Hongji special}, and using $P^i/E = v^i$, we have
\begin{align}
   &\delta_T S^i =\frac{1}{4\pi}\int_{S^2}d\Omega T \biggr(-\epsilon^{AB}\partial_Am\partial_B n_i - \epsilon^{ijk}(\partial_Am\partial^A n_j - 3n_j m)v_k\biggr) \nonumber\\
   &=\frac{1}{4\pi}\int_{S^2}d\Omega T\frac{3M}{\gamma^3(1-v_ln^l)^4}\biggr(-\epsilon^{AB}v_j\partial_An^j\partial_Bn^i +\epsilon^{ijk}v_j\big(v_p\partial_A n_p \partial^A n_k -n_k(1-v_pn^p)\big)\biggr)\,,
\end{align}
where in the last equality we used \eqref{aspectonebody}.  
It can be easily shown that $\partial_An_i\partial^A n_j = \delta_{ij} - n_i n_j$ and $\epsilon^{AB}\partial_An_i\partial_B n_j = \epsilon_{ijk}n^k$. Hence after a direct substitution we obtain 
\begin{align}
     &\delta_{T}S^i =\frac{1}{4\pi}\int_{S^2}d\Omega T\frac{3M}{\gamma^3(1-v_ln^l)^4}(\epsilon^i{}_{jk}v^jv^k) = 0,
\end{align}
showing that spin angular momentum is supertranslation invariant.

\subsection{Choice of BMS frame}\label{sec:scattering}

Even in covariant theories of physics, it is often convenient to perform calculations and/or present results in special frames.  In special relativistic scattering, one has ten Poincar\'e degrees of freedom available to help simplify the description.  One common choice is to make the total initial momentum vanish,
\begin{align}
P^i\vert_{i^-_\partial} = 0, \qquad \textrm{``initial center of momentum frame''}\,, \label{cm}
\end{align}
fixing the boost degrees of freedom.  One could further choose the total initial mass moment to vanish, 
\begin{align}
N^i\vert_{i^-_\partial} = P^i\vert_{i^-_\partial} = 0, \qquad \textrm{``initial center of energy frame''}\,, \label{ce}
\end{align}
which further fixes the spatial translations.  In some cases it is also convenient to make definite choices fixing the remaining rotational and time-translation degrees of freedom.

In general relativistic scattering there is one more kind of freedom to fix, the pure supertranslations.  This can be done by making a definite choice of the function $C(\theta,\phi)$ (defined with only $\ell \geq 2$ harmonics) at some boundary.  One natural choice is to fix it to vanish initially,
\begin{align}
C|_{\scri^-_-} & = 0 \qquad \textrm{``good cut''} \, ,\label{goodcut} 
\end{align}
where the name ``good cut'' derives from older work focusing on future null infinity \cite{Newman:1966ub}. In the absence of incoming radiation, the ``good cut'' gauge is equivalent to $C|_{\scri^+_-}=0 $ after using trivially the antipodal map \eqref{Cmatch}.

Another gauge that has been used in Post-Minkowski (PM) calculations, \emph{e.g.} Ref.~\cite{Damour:2020tta}, as noted in \cite{Javadinezhad:2022ldc}, corresponds instead to the condition 
\begin{align}
\left.\left(\frac{1}{4} \nabla^2(\nabla^2+2)C + m_{\ell \geq 2}\right)\right|_{\scri^+_-}& = 0 \qquad \textrm{``nice cut''}\, , \label{moreschi}
\end{align}
where $m_{\ell \geq 2}$ denotes the $\ell \geq 2$ part of the mass aspect $m$. 
This condition (or its analog at $\scri^+_+$) has also been used to compare numerical waveforms obtained from different methods \cite{Mitman:2021xkq,Mitman:2022kwt}.  The  spherical harmonic components of the LHS of \eqref{moreschi} are the ``Moreschi supermomenta'' $\frac{1}{4\pi}\int_{S^2}d\Omega \, T (m-D_AD_B C^{AB})$ \cite{OMMoreschi_1988} evaluated at $\scri^+_-$ for $T=4\pi Y^*_{\ell m}$, and the name ``nice'' cut derives from that reference.  Note, however, that the Moreschi supermomenta  do not obey an antipodal relationship across $i^0$: the shear potential $C$ matches antipodally with a minus sign \eqref{Cmatch}, while the mass aspect matches antipodally with a plus sign \eqref{mmatch}.  Therefore, the relationship \eqref{moreschi} will not be true at $\scri^-_+$ if it is assumed at $\scri^+_-$.

Any conditions imposed at $i^-$ will not in general be true at $i^+$.  For example, suppose that we impose vanishing initial total momentum \eqref{cm}, mass moment \eqref{ce} and shear \eqref{goodcut}.  These quantities will in general have non-zero final values.  When there is change in momentum we say there was ``recoil'' equal to $P_i^{i^+_\pd}$, since the body system is now moving relative to the initial center-of-momentum frame.  When there is a change in mass moment we say that there was a ``scoot'' equal to $N_i^{i^+_\pd}$, since the body system is now displaced from the original.  When there is a change in shear, we say that there was ``gravitational memory'' equal to  $C|_{i^+_\pd}(\theta,\phi)=\Delta C^{\text{total}}(\theta,\phi)$ [see Eq.~\eqref{Ctotal}], as already discussed in Sec.~\ref{sec:gravitational-memory}. 
In conclusion, once one fixes a ``canonical'' gauge fixing condition at one asymptotic boundary such as $i^+$, $i^-$ or $i^0$, then the same ``canonical'' gauge condition will \emph{not} in general hold at the two other boundaries because of the recoil, scoot and displacement memory caused by radiation reaching $\scri^+$ or originating from $\scri^-$.  In our framework, choosing a BMS frame at one infinity automatically fixes BMS frames at all others; these do not in general obey the ``same'' conditions as imposed at the original infinity. BMS covariance is necessary to relate ``local canonical gauges'' at individual boundaries $i^+$, $i^-$ or $i^0$.

\subsection{Scattering angles and impact parameter}

In two-body scattering it is conventional to discuss scattering angles as a function of impact parameter.  These can be defined using the BMS charges.  The scattering angles $\chi^{(n)}$ can be defined using the directions of initial and final particle momenta,
\begin{equation}
    \cos\chi^{(n)}=\frac{\mathbf{P}^{(n)}|_{i^+} \cdot \mathbf{P}^{(n)}|_{i^-}}{P^{(n)}|_{i^+} P^{(n)}|_{i^-}}\,,
\end{equation}
where we use the notation $\mathbf{A}$ for a vector with $A=\sqrt{\mathbf{A}\cdot\mathbf{A}}$ its magnitude.  If there is no recoil, then we can use a center of momentum frame such that $\mathbf{P}^{(1)}=-\mathbf{P}^{(2)}$ at both $i^-$ and $i^+$, and there is a single scattering angle $\chi=\chi^{(1)}=\chi^{(2)}$.\footnote{$\,\,$ The distinction of two independent scattering angles is not necessary in the current PM scattering literature because the recoil is a very high-order effect, $P_i^{i^+_\pd}=O(G^6)$ \cite{Damour:2022ybd}.}

In the initial center of energy frame \eqref{ce}, 
the impact parameter $b$ can be defined from the charges at $i^-$ as
\begin{align}\label{impactparameter}
    b = \sqrt{s_is_jP^{ij}}\,,
\end{align}
where $P_{ij}$ projects orthogonally to the initial momentum direction $P^-_i=P^{(1)}_i|_{i^-}$,
\begin{align}
    P_{ij} &= \delta_{ij} - \frac{P^-_iP^-_j}{{P^-}^2}\,, \label{projectionmatrix}
\end{align}
and the relative distance $s_i$ is defined as
\begin{align}
     s_i &= \left. \left(\frac{N^{(1)}_i}{E^{(1)}}-\frac{N^{(2)}_i}{E^{(2)}}\right) \right|_{i^-} = \frac{N^-_i}{E^-}\,, \label{separationvector}
\end{align}
with $N^-_i=N^{(1)}_i|_{i^-}$ and $E^-$ the relativistic reduced energy,
\begin{align}
E^- \equiv \left.\frac{E^{(1)}E^{(2)}}{E^{(1)}+E^{(2)}}\right|_{i^-}.
\end{align} 
If we define the relative orbital angular momentum as
\begin{align}
    L^{\text{orb}-}_i = \frac{1}{E^{-}}\epsilon_{ijk} N^-_{j} P^-_k=\epsilon_{ijk} s_{j} P^-_k\, ,
\end{align}
then the impact parameter can be alternatively written as 
\begin{align}\label{impactparameter alternative}
   b= \frac{L^{\text{orb}-}}{P^{-}}\,,
\end{align}
where $L^{\text{orb}-}$ is the magnitude of the relative orbital angular momentum.  To see this, we compute [suppressing the index $-$] that 
\begin{align}
    (L^\text{orb})^2 &= \epsilon_{ijk}\epsilon_{ilm}s^js^lP^kP^m =  (\delta_{jl}\delta_{km} - \delta_{jm}\delta_{kl})s^js^lP^kP^m \\
    & = \bigg(s^2P^2 - (s_jP^j)^2\bigg) = P^2 s_i s_j \bigg(\delta^{ij} - \frac{P^iP^j}{P^2}\bigg) \\ 
    &= P^2 s_is_j P^{ij}\,,
\end{align}
where we used the projection \eqref{projectionmatrix}, which implies that $b = \sqrt{s_is_jP^{ij}} = L_\text{orb}/P$.

The impact parameter is invariant under translations but \textit{not} invariant under $\ell \geq 2$ supertranslations.  Comparisons between calculations done with different gauge-fixing conditions [e.g. \eqref{goodcut} and \eqref{moreschi}] will have to take this difference into account.

\section{Conclusion}\label{sec:conclusion}

We conclude with a summary of the novel aspects of this work.  The first is completeness: as far as we are our aware, this is the first framework for asymptotically flat gravity with a BMS representation at all five infinities such that the BMS charges match not by assumption, but rather as a consequence of Einstein's equations and boundary conditions that are consistent with all known scattering results.  We showed how $i^\pm$ encodes the asymptotic states of massive bodies, in which finite-sized bodies appear as point sources for fields on $i^\pm$.  For example, particles, stars and black holes are described with the same mathematics.  Each individual ingoing or outgoing body has a full set of BMS charges, which provide rigorous definitions for important quantities like initial impact parameter and final spin, in a general asymptotically flat context.  In particular, we showed that the spin of each body is supertranslation-invariant in an arbitrary BMS frame. We proved that the BMS charges are globally conserved in the sense of Eq.~\eqref{trivial}, which now holds as a theorem. 

We also introduced some new points of interpretation and visualization.  In particular, asymptotic massive-body states can be thought of as insertions on $i^\pm$ and therefore drawn as dots on hyperbolae at the start and end of spacetime.  The infinities $\scri^+$/$\scri^-$ can be regarded as networks of distant stationary observers/emitters and hence may be drawn as vertical lines where light rays end/start.  Spatial infinity can also be drawn as a hyperboloid representing superluminal asymptotic velocities, providing separation between ingoing and outgoing null rays on the diagram.  The result is a new ``puzzle piece'' diagram (Figs.~\ref{fig:cool}--\ref{fig:PenrosePuzzle}) that seems naturally suited to the structure of classical gravitational scattering and may have advantages for quantum and/or non-gravitational scattering as well.
 
Finally, our calculations have revealed a number of new details that may be notable to practitioners in the field.  Many of these are related to logarithmic terms that are required for compatibility with the classical limit of logarithmic corrections to the subleading soft theorem  \cite{Laddha:2018myi,Saha:2019tub,sahoo_classical_2022}.  First, the Bondi angular momentum aspect generically diverges logarithmically for large $\vert u\vert \rightarrow \infty$ [Eqs.~\eqref{ansatzN2} and \eqref{NAansatzsnegativeu}].  As a consequence, the super-Lorentz charges as defined in \cite{Barnich:2011mi,Campiglia:2015yka} and in subsequent work fail to have limits to $i^0$ and $i^+$ (see discussion just before Sec.~\ref{sec:asymptotic coordinate transformation}).  Similarly, the angular momentum aspect cannot be directly matched across $i^0$ as conjectured in \cite{hawking_superrotation_2017} (see Sec.~\ref{sec:antipodal}).  The finite $\ell=1$ piece that determines the Lorentz charges does match [Eq.~\eqref{antipodalNA}], but with the opposite sign of the conjectured relation.  A conceptually separate logarithm is the (pure gauge) logarithmic translation that we employed at $i^0$ to enable matching to both $\scri^+_-$ and $\scri^-_+$ (see Eq.~\eqref{sigmapm} and discussion below).  While we were able to gauge-fix away these translations at $i^\pm$, we were forced to retain them at $i^0$ in order to consistently match charges with our method.  Finally, the finiteness of Lorentz charges of individual bodies at $i^\pm$ can be understood as a consequence of an infrared renormalization of the action on $i^\pm$; however, no such renormalization is needed for the total charge (see Eq.~\eqref{Sct} and surrounding discussion). 

\appendix
\addtocontents{toc}{\fixappendix}

\section*{Acknowledgements}
G.C. would like to thank A. Fiorucci and R. Ruzziconi for related discussions. G.C. is Senior Research Associate of the F.R.S.-FNRS and acknowledges support from the FNRS research credit J.0036.20F and the IISN convention 4.4503.15.  This work was supported in part by NSF grant
PHY–1752809 to the University of Arizona.

\section{Hyperboloid scalar wave equations}\label{sec:equation}
In this appendix we analyze the mode solutions for $\sigma$ and $\omega$ at $i^\pm$ and $i^0$.  

\subsection{Timelike infinity $i^\pm$}

At $i^\pm$, the quantities $\sigma$ and $\omega$ satisfy
\begin{align}\label{psieqn}
    (D^2-3)\psi = 0\,,
\end{align}
where $D_a$ is compatible with the metric \eqref{h+}.  We first decompose in spherical harmonics and make the ansatz
\begin{align}
    \psi(\rho,x^A) = \psi_{\ell}(\rho) Y_{\ell m}(\theta,\phi)=\frac{1}{2\sinh\rho}f_{\ell}(\coth \rho) Y_{\ell m}(\theta,\phi)\,,
\end{align}
which gives rise to the mode equation
\begin{align}\label{feqn}
    (x^2-1)\frac{d^2f_\ell(x)}{dx^2} + 2x \frac{df_\ell(x)}{dx} - \big(\ell(\ell+1)+ \frac{4}{x^2-1}\big)f_\ell(x) = 0\, .
\end{align}
This is the associated Legendre equation with $m=2$. Since $\coth\rho \in [1,+\infty)$, the branch cut is set at $(-\infty,1]$.\footnote{$\,\,$  This corresponds to the associated Legendre equation of type 3 in the Mathematica implementation.}  We will write the solutions as two families $\psi^\infty_\ell$ and $\psi^{\mathcal{O}}_\ell$ regular at infinity and the origin, respectively.  For $\ell \geq 2$ these are the associated Legendre functions $P_\ell^2$ and $Q_\ell^2$, respectively.  For $\ell = 0,1$, the Legendre function $P_\ell^2$ is not defined, but one can find the appropriate solution by inspection.  The results are:
\begin{align}
    \psi_\ell^\infty(\rho) &= \begin{cases} 2\cosh\rho - 2\sinh\rho - \frac{1}{\sinh\rho}, & \ell = 0\,;\\  2 \sinh\rho-2 \cosh\rho + \frac{\coth\rho}{\sinh\rho} , & \ell = 1\,; \\ \frac{1}{2\sinh\rho}P_\ell^2(\coth\rho), & \ell \geq 2\,, \end{cases} \label{psiellinfty}\\
    \psi_\ell^\mathcal{O}(\rho)&=\frac{1}{2\sinh\rho}Q_\ell^2(\coth\rho)\,.\label{psiellO}
\end{align}
The low-$\ell$ origin-regular solutions take a simple form,
\begin{align}
    \psi^\mathcal{O}_0 = \cosh\rho\,, \label{psiO0}\\
    \psi^\mathcal{O}_1 = \sinh\rho\,. \label{psiO1}
\end{align}
The end behaviors are
\begin{align}
    \psi_\ell^\mathcal{O} & \sim 2^{-\ell - 2}\frac{\Gamma(\frac{1}{2})\Gamma(\ell+3)}{\Gamma(\ell + \frac{3}{2})}\sinh^\ell\rho , &&\rho \to 0 \,,\label{fOsmall} \\
     \psi_\ell^\mathcal{O} & \sim \frac{1}{2}e^\rho , && \rho \to \infty \,,\label{fObig} \\
     \psi_0^\mathcal{\infty} & \sim -\sinh^{-1}\rho, && \rho \to 0\,,\\
     \psi_1^\mathcal{\infty} & \sim \sinh^{-2}\rho, && \rho \to 0\,,\\
     \psi_0^\mathcal{\infty} & \sim -2e^{-3\rho}, && \rho \to\infty\,,\\
     \psi_1^\mathcal{\infty} & \sim 6e^{-3\rho}, && \rho \to\infty\,,\\
     \psi_{\ell\geq 2}^\infty & \sim  2^{\ell - 1}\frac{\Gamma(\ell + \frac{1}{2})}{\Gamma(\ell-1)\Gamma(\frac{1}{2})}\sinh^{-\ell-1}\rho, &&   \rho \to 0 \label{fIsmall} \,,\\
     \psi_{\ell\geq 2}^\infty & \sim \frac{1}{2}(\ell-1)\ell(\ell+1)(\ell + 2)e^{-3\rho}, && \rho \to \infty.\label{FfIbig}
\end{align}

Notice that there are no mode solutions which are simultaneously regular at the origin and vanishing at infinity.  By the completeness of the spherical harmonics, the same is true of solutions $\psi$ to the original equation \eqref{psieqn}.  If we instead consider the equation with some non-zero source on the right-hand-side, this shows the uniqueness of solutions vanishing at infinity.

\subsection{Spatial infinity $i^0$}

In this section, we follow the convention introduced in Sec.~\ref{sec:flati0} that consists in using hats for all functions and coordinates at spatial infinity $i^0$.  We are interested in the equation
\begin{align}
    (\hat{D}^2+3)\hat{\psi} = 0\, .
\end{align}
Its solutions may be determined by analytic continuation \eqref{magic} together with appropriate field redefinitions, but we present each independently. Our treatment overlaps but differs from earlier work \cite{troessaert_bms4_2018,capone_charge_2022}. Our ansatz is
\begin{align}
    \hat{\psi}(\hat{\tau},x^A) =\hat{\psi}_{\ell}(\hat{\tau}) Y_{\ell m}(\theta,\phi)= \frac{1}{2\cosh\hat{\tau}}f_\ell(\tanh\hat{\tau}) Y_{\ell m}(\theta,\phi)\, .
\end{align}
This ansatz gives the same differential equation \eqref{feqn} for $f(x)$. We will organize the solutions based on their even (E) or odd (O) parity, 
\begin{align}
    \hat{\psi}^E_\ell(\hat{\tau}) &= \begin{cases} 2\cosh\hat{\tau} - \frac{1}{\cosh\hat{\tau}}, & \ell = 0\, ; \\ 2 \sinh\hat{\tau} + \frac{\tanh\hat{\tau}}{\cosh\hat{\tau}}, & \ell = 1\, ; \\ \frac{1}{2\cosh\hat{\tau}}P_\ell^2(\tanh\hat{\tau}), & \ell \geq 2 \, ,\end{cases}\\
    \hat{\psi}_\ell^O(\hat\tau)&=-\frac{1}{2\cosh\hat{\tau}}Q_\ell^2(\tanh\hat{\tau})\, .
\end{align}
In particular,
\begin{align} \Upsilon_\mathcal{H}^*[\hat{\psi}_\ell^{E}(\hat{\tau})Y_{\ell m}(x^A)] & = \hat{\psi}_\ell^{E}(\hat{\tau})Y_\ell^m(x^A)\, ,\\
\Upsilon_\mathcal{H}^*[\hat{\psi}_\ell^O(\hat{\tau})Y_{\ell m}(x^A)] &= - \hat{\psi}_\ell^O(\hat{\tau})Y_{\ell m}(x^A)\, ,
\end{align}
where $\Upsilon_{\mathcal{H}}$ was given in \eqref{parity}. Here, the associated Legendre functions are the standard ones defined in the unit circle of the complex plane\footnote{$\,\,$ Their implementation in Mathematica is the type 1 associated Legendre functions.}. Note that $\hat\psi^O_0=-\sinh\hat\tau$, $\hat\psi^O_1=-\cosh\hat\tau$. 

The end behavior of the even and odd solutions is summarized as
\begin{align}
    \hat{\psi}^E_0(\hat{\tau})&\sim e^{|\hat{\tau}|} ,\qquad &\hat{\tau} \to \pm\infty\, ,\\
    \hat{\psi}^E_1(\hat{\tau})&\sim \pm e^{|\hat{\tau}|}, \qquad &\hat{\tau} \to \pm\infty\, , \\
    \hat{\psi}^E_{\ell\geq2}(\hat{\tau}) &\sim (\pm1)^\ell \frac{(\ell-1)\ell(\ell+1)(\ell+2)}{2} e^{-3|\hat{\tau}|},\qquad  &\hat{\tau} \to \pm\infty\, ,\\
    \hat{\psi}^O_\ell(\hat{\tau}) &\sim  -(\pm1)^{\ell+1} \frac{1}{2}e^{|\hat{\tau}|},\qquad &\hat{\tau} \to \pm\infty\, .
\end{align}

For $\ell \geq 2$, the even-parity solutions $\hat{\psi}^E_\ell$ vanish at $\hat\tau \to \pm \infty$, while the odd-parity solutions $\hat{\psi}^O_\ell$ diverge at $\tau \to \pm \infty$.  For $\ell=0$ and $\ell=1$, both solutions diverge as $\hat\tau \to \pm \infty$.  We instead define linear combinations that are regular at one infinity, 
\begin{align}
    \hat{\psi}^{\pm \infty}_{0,1} = \hat{\psi}^E_{0,1}\pm 2\hat{\psi}^O_{0,1}\, .
\end{align}
The $\hat{\psi}^{+ \infty}$ solution diverges at $\tau \to -\infty$ and vanishes at $\hat\tau \to +\infty$; instead, the $\hat{\psi}^{-\infty}$ solution vanishes at $\tau \to -\infty$ and diverges at $\hat\tau \to +\infty$.  Writing these explicitly, and including the regular solutions $\hat{\psi}_\ell^E$ for higher $\ell$, we have the solution families
\begin{align}
    \hat{\psi}_\ell^{\pm\infty}(\hat{\tau}) &= \begin{cases} 2\cosh\hat{\tau} - \frac{1}{\cosh\hat{\tau}} \mp 2\sinh\hat{\tau}, & \ell = 0\,;\\ 2 \sinh\hat{\tau} + \frac{\tanh\hat{\tau}}{\cosh\hat{\tau}} \mp 2 \cosh\hat{\tau}, & \ell = 1 \,;\\ \frac{1}{2\cosh\hat{\tau}}P_\ell^2(\tanh\hat{\tau}), & \ell \geq 2\,  .\end{cases}\label{psiellpminfty}
\end{align}
These families are not independent, since they share the same $\ell\geq 2$ mode functions. However, they each provide a basis for constructing solutions regular at their respective values of large $|\hat\tau|$.

Using the analytic continuation $\hat{\tau} = \rho - i \frac{\pi}{2}$, we have
\begin{align}
    \hat{\psi}_\ell^{+\infty}\left(\rho - i\frac{\pi}{2}\right) &= i\psi_\ell^\infty(\rho)\,,\\
    (-1)^{\ell+1}\hat{\psi}_\ell^{-\infty}\left(\rho -  i\frac{\pi}{2}\right) &=  i\psi_\ell^\infty(-\rho)\,,\\
  -\hat{\psi}_\ell^O\left(\rho - i\frac{\pi}{2}\right) &= i\psi_\ell^\mathcal{O}(\rho)\label{psii0i+}\,,\\
    \hat{\psi}_{\ell\geq 2}^{E}\left(\rho - i\frac{\pi}{2}\right) &= i\psi_{\ell\geq 2}^\infty(\rho)\, .
\end{align}
In this sense, the solution $\hat{\psi}_\ell^{+\infty}$ at $i^0$ is analogous to $\psi_\ell^\infty$ at $i^+$. Similarly the parity-odd solution $\hat{\psi}_\ell^O$ at $i^0$ is analogous to the smooth solution $\psi_\ell^\mathcal{O}$ at $i^+$.

\section{Schwarzschild metric at $i^+$}\label{sec:schwarzschild}

In this section we put the Schwarzschild metric in Beig-Schmidt form at future timelike infinity and show that $\sigma$ satisfies an equation with a point source.  We begin with the metric in Schwarzschild coordinates,
\begin{align}
    ds^2 = -\left(1-\frac{2M}{r}\right)dt^2 + \left(1-\frac{2M}{r}\right)^{-1}dr^2 + r^2\gamma_{AB}dx^Adx^B.
\end{align}
Appendix A of Ref.~\cite{chakraborty_supertranslations_2022} gives a recipe for putting a metric in Beig-Schmidt form.  Following these steps, we find that a suitable coordinate transformation is 
\begin{align}
    t & = \Bar{\tau} \cosh\Bar{\rho}, \qquad r = \Bar{\tau} \sinh\Bar{\rho} \,,
\end{align}
with
\begin{align}
\bar{\tau} & = \tau + M \bar{\tau}_0(\rho) + \frac{M^2}{\tau}\bar{\tau}_{1}(\rho) + O(\tau^{-2}) \,,
\\
\bar{\rho} & = \tau + \frac{M}{\tau} \bar{\rho}_1(\rho) + \frac{M^2}{\tau^2}\bar{\rho}_{2}(\rho) + O(\tau^{-3}) \,, 
\end{align}
and 
\begin{align}
    \bar{\tau}_0 & = \sinh \rho +2 \rho  \cosh \rho\,, \\
    \bar{\tau}_1 & = -2 \rho ^2 \sinh ^2\rho + \rho  \sinh (2 \rho )-\frac{1}{2}\cosh ^2\rho+\frac{1}{2}\cosh ^2(2 \rho ) \text{csch}^2\rho -2\,, \\
    \bar{\rho}_1 & = -2 \rho  \sinh \rho -3 \cosh \rho \,,\\
    \bar{\rho}_2 & = \frac{1}{16} \text{csch}^3\rho \Big(4 \rho ^2 \cosh (5 \rho )+\left(8 \rho ^2-6\right) \cosh \rho +\left(13-12 \rho ^2\right) \cosh
   (3 \rho )\\ \nonumber &\qquad +24 \rho  \sinh \rho-8 \rho  \sinh (3 \rho )+\cosh (5 \rho )\Big)\,.
\end{align}
After this transformation to $(\tau,\rho,\theta,\phi)$ coordinates, the metric takes the Beig-Schmidt form \eqref{Beig-Schmidt} with 
\begin{align}
    \sigma& = - M (2\sinh\rho + \frac{1}{\sinh\rho})\,,\\
    k_{ab}& = 0\,,\\
    i_{ab} &= 0\,,\\
    j_{\rho\rho}& = \frac{1}{2} M^2 (-24 \cosh (2 \rho )+5 \cosh (4 \rho )+13) \coth ^2\rho\text{csch}^2\rho\,,\\
    j_{\rho A} &= 0\,,\\
    j_{AB}& = \frac{1}{8} M^2 (25 \cosh (2 \rho )-2 \cosh (4 \rho )+3 (\cosh (6 \rho )-6)) \text{csch}^2\rho \gamma_{AB}\,.
\end{align}
However, this form does not satisfy our additional requirement \eqref{sigmavanish} that $\sigma$ falls off at large $\rho$.  We may fix this using the logarithmic transformation discussed in Sec.~\ref{sec:original-coordinate-freedom}.  The relevant choice of $H$ is $H = 2 M \cosh \rho$, which is $h^\mu=(2M,0,0,0)$ in the notation of \eqref{H}.  However, Sec.~\ref{sec:original-coordinate-freedom} only considered infinitesimal diffeomorphisms at first order, whereas we must preserve the Beig-Schmidt form at subsubleading order in $\tau^{-1}$, which corresponds to second order in $H$.  Augmented with the relevant nonlinear corrections, the last coordinate transformation we need is
\begin{align}
    \tau &\to \tau + H\log\tau +\tau^{-1}\big(A_2\log^2\tau+A_1\log\tau+A_0\big)+o(\tau^{-1})\,,\label{slog1} \\
    \rho &\to \rho + \frac{\log\tau+1}{\tau}\partial_\rho H + \tau^{-2}\big(B_2\log^2\tau+B_1\log\tau+B_0\big)+o(\tau^{-2})\,,\label{slog2}
\end{align}
with
\begin{align}
    H &= 2M\cosh\rho\,,\\
    A_2& = -2M^2\sinh^2\rho\,,\\
    A_1& = -4M^2(\sinh^2\rho-\sinh(2\rho))\,,\\
    A_0& = \frac{1}{2} M^2 \text{csch}^2\rho  (-8 \sinh (2 \rho )+\sinh (4 \rho )+4 \cosh (2 \rho )-3)\nonumber\\&-\frac{1}{2} (M \cosh (2 \rho ) \text{csch}\rho
   -2 M \cosh \rho )^2\,,\\
   B_2& = 2M^2\sinh(2\rho)\,,\\
   B_1& = B_2\,,\\
   B_0& = -\frac{1}{4} M^2 \text{csch}^2\rho  (6 \sinh (2 \rho )-3 \sinh (4 \rho )-8 \cosh (2 \rho )+4 \cosh (4 \rho )+12)\, .
\end{align}
These functions are fixed completely by the ansatz \eqref{slog1} and \eqref{slog2} and the vanishing of cross-terms in the Beig-Schmidt metric.  More geometrically, we have obtained the integral curve of the generator in Eqs.~\eqref{xitau} and \eqref{xia}.
Applying the transformation \eqref{slog1} and \eqref{slog2}, we find that the metric remains in Beig-Schmidt form, with the new values 
\begin{align}
    \sigma& = - M (\frac{1}{\sinh\rho}-2e^{-\rho}) \label{newsigma} \,,\\
    k_{ab}& = 0\,,\\
    i_{ab} &= 0\,,\\
    j_{\rho\rho} & = \frac{1}{2} M^2 e^{-4 \rho } \text{csch}^4\rho (-4 \sinh (2 \rho )+\cosh (2 \rho )-7)\,,\\
    j_{\rho A} &= 0\,,\\
    j_{AB} & = M^2 e^{-4 \rho } \left(\text{csch}^2\rho+3\right)\gamma_{AB}\,.
\end{align}
The fields $\sigma$ and $j_{ab}$ have poles at the origin $\rho = 0$, while the remainder are smooth.  It is easy to check that, for any surface surrounding the origin, the energy \eqref{chargeE} is equal to the mass $M$, while all other charges vanish.  

\subsection{Distributional description}

Treated as an ordinary function, the field $\sigma$ blows up at the origin $\rho=0$ and satisfies $(D^2-3)\sigma=0$ everywhere else.  This suggests that, treated as a distribution, it may satisfy an inhomogeneous equation with a point source.  We now show that this is indeed the case.

We denote the action of distribution $D$ on a test function $f$ by $\langle f|D \rangle$.  A linear partial differential operator $\mathcal{L}$ defines a distribution $\mathcal{L}D$ by $\langle f|\mathcal{L}D\rangle$ = $\langle \mathcal{L}^\dagger f|D \rangle$, where $\mathcal{L}^\dagger$ is the adjoint operator.  The adjoint operator is uniquely defined by the expression 
\begin{align}
    g \mathcal{L}^\dagger f - f \mathcal{L} g = D_a s^a[g,f]\,, \label{adjoint}
\end{align}
where $s^a$ is the non-unique symplectic current.  Our operator $D^2-3$ is self-adjoint,
\begin{align}
    \mathcal{L} = \mathcal{L}^\dagger = D^2-3, \qquad s^a[g,f] = g D^a f - f D^a g\,.
\end{align}

We use the natural volume element $\sqrt{h} d^3\phi = \sinh^2 \! \rho \ \! d\rho d\Omega$ to define our distributions.  The distribution $\mathcal{L} \sigma$ acts on test functions $f$ as 
\begin{align}
    \langle f| \mathcal{L}\sigma \rangle =  \langle \mathcal{L} f| \sigma \rangle & = \int \sigma \,   \mathcal{L} f\,  \sqrt{h} d^3\phi\, .
\end{align}
Notice that for $\rho>0$ we have $\sigma \mathcal{L} f = D_a s^a$, since the other term $f \mathcal{L} \sigma$ in \eqref{adjoint} vanishes by $\mathcal{L}\sigma=0$. We therefore write the integral 
as the $\epsilon \to 0$ limit of the integral over the region $\rho>\epsilon$ and use this identity, finding 
\begin{align}
    \langle \mathcal{L}\sigma \vert f \rangle 
    & = - \lim_{\epsilon \to 0} \int_{\rho=\epsilon}\left( \sigma D_a f - f D_a \sigma \right) \sinh^2 \! \epsilon \ \! r^a d\Omega \\
    & = 4\pi M f(0)\,,
\end{align}
where $r_a=\partial_a \rho$ is the \textit{outward} normal.  (The minus sign arises because the normal vector appearing in Stokes' theorem is actually the radially inward normal $-r_a$, since the original integration region is  $\rho>\epsilon$.)  The second line follows by direct computation using the form of $\sigma$ given in Eq.~\eqref{newsigma}.  We have thus shown that
\begin{align}\label{eqndelta}
    (D^2- 3)\sigma = 4\pi M\frac{\delta^{(3)}(\phi^a)}{\sqrt{h}}\,.
\end{align}
The combination $\delta^{(3)}/{\sqrt{h}}$ is the covariant delta function.

\subsection{Boosted Schwarzschild}\label{sec:boosted schwarzschild}

Beginning with the Schwarzschild metric in the ordinary Schwarzschild coordinates, we were naturally led to a Beig-Schmidt frame with a single body at rest.  However, the velocity of a body is a coordinate choice, and we could consider alternative Beig-Schmidt frames for the Schwarzschild metric where the body has some non-zero asymptotic velocity.  This exercise will help motivate our assumptions on the behavior of the mass aspect $\sigma$ in the context of multi-body scattering, and also provide useful formulas for its manipulation.

In flat spacetime, the boost that makes a stationary body move with velocity $v^i$ in the primed coordinates is 
\begin{align}
    t & = \gamma (t' - v_i x'^i) \,,\label{t1} \\
    x^i& = x'^i + (\gamma - 1)\frac{v^iv_j}{v^2}x'^j -\gamma v^i t\,,
\end{align}
where $\gamma=1/\sqrt{1-v^2}$.  The hyperbolic coordinates are related by 
\begin{align}
    t & = \tau \cosh \rho \,, \quad \ \! \qquad t' = \tau' \cosh \rho' \,,\\
    x^i & = \tau n^i \sinh \rho\,, \qquad x'^i = \tau' n'^i \sinh \rho'\,.
\end{align}
Recalling that $\tau=\tau'$ is Lorentz-invariant and comparing with \eqref{t1}, we see that
\begin{align}
    \tau = \tau'\,, \qquad 
    \cosh\rho = \gamma (\cosh\rho' - v^i n'_i \sinh\rho')\,. \label{coshy}
\end{align}
Based on the interpretation of $\rho$ as the rapidity of a particle moving in the $(\theta,\phi)$ direction, we expect the origin $\rho=0$ to be mapped to the unique point such that 
\begin{align}\label{findparticle}
    v^i = \tanh \rho' n'^i\,,
\end{align}
for which $\cosh \rho' = \gamma$.  This is easily confirmed by plugging in to \eqref{coshy}.

Although these formulas were derived in flat spacetime, we may use them unchanged on $i^+$ since the boost symmetry acts directly as a true symmetry on $i^+$ (Sec.~\ref{sec:i+}).  The mass aspect $\sigma$ \eqref{newsigma} is a tensor under Lorentz transformations as a consequence of \eqref{deltaxisigmafinal}. It thus becomes
\begin{align}
    \sigma &= - M (2\sinh\rho + \frac{1}{\sinh\rho}- 2\cosh\rho)\nonumber\\
    & = - M (\frac{2\cosh^2\rho - 1}{\sqrt{\cosh^2\rho-1}}- 2\cosh\rho)\nonumber\\
    &= -M(\frac{2\chi^2-1}{\sqrt{\chi^2 -1}}-2\chi)\,,\label{wegotit}
\end{align}
where
\begin{align}
\chi = \gamma (\cosh\rho' - v_i n'_i \sinh\rho')\,.
\end{align}
The equation \eqref{eqndelta} becomes
\begin{align}\label{eqndeltai-}
    (D'^2 - 3)\sigma = 4\pi M\frac{\delta^{(3)}(\phi'^a-\phi_0'^a)}{\sqrt{h}}\,,
\end{align}
where $\phi_0'^a$ encodes the velocity of the particle in the boosted frame as the point $(\rho'_0,\theta'_0,\phi'_0)$ determined by Eq.~\eqref{findparticle}.  

In the main body we promote the right-hand-side of \eqref{eqndelta} to a sum over sources with different velocities [Eq.~\eqref{sigma-point}].  (We also drop the prime in the notation, since the expression is covariant.)  By the arguments of~\ref{sec:equation}, the solution is unique if $\sigma$ vanishes at large $\rho$.  This unique solution is easily constructed by superposing terms of the form \eqref{wegotit}, each with the corresponding value for $v_i$, as given in Eq.~\eqref{sigma-nbodies} in the main body.

\section{Translated Schwarzschild at $\scri^+$ and $\scri^-$}\label{sec:translated-schwarzschild}

The translated Schwarzschild metric provides a useful example of the antipodal matching of the angular momentum aspect as well as a convenient check of the sign convention for the mass moment.  Future null infinity is described by the outgoing  Eddington-Finkelstein coordinates
\begin{align}
    & ds^2 = -(1-\frac{2M}{r}) du^2 - 2du dr + r^2 d\theta^2 + r^2\sin^2\theta d\phi^2\, .
\end{align}
We perform a translation $z \to z -\Delta z$, so that the ``position'' of the black hole in the new coordinate system is $z=\Delta z$. The translation leads to the following shift of the outgoing coordinates
\begin{align}
    & u \to u + \Delta z\cos\theta -\frac{\Delta z^2\sin^2\theta}{2r} + O(r^{-2})\,,\\
    & r \to r - \Delta z\cos\theta + \frac{\Delta z^2\sin^2\theta}{2r} + O(r^{-2})\,,\\
    & \theta \to \theta + \frac{\Delta z\sin\theta}{r} + \frac{\Delta z^2\sin\theta \cos\theta}{2r^2} + O(r^{-3})\,,\\
    & \phi \to \phi\,.
\end{align}
The resulting metric is
\begin{align}\label{ts+}
    ds^2 &= \left( -1 + \frac{2M}{r} + \frac{2M \Delta z\cos\theta}{r^2} + O(r^{-3})\right)du^2 + 2 (-1 + O(r^{-3}))du dr \nonumber\\ & 
    + 2 \left( -\frac{2M\Delta z\sin\theta}{r} + O(r^{-2})\right)dud\theta + \left(r^2 + O(r^{-1})\right)d\theta^2 \nonumber\\ &+  \left(r^2\sin^2\theta + O(r^{-1})\right)d\phi^2\,.
\end{align}
Past null infinity is described by the ingoing Eddington-Finkelstein coordinates
\begin{align}
    & ds^2 = -(1-\frac{2M}{r}) dv^2 + 2dv dr + r^2 d\theta^2 + r^2\sin^2\theta d\phi^2\,.
\end{align}
Performing the same translation $z\to z-\Delta z$,
\begin{align}
    & v \to v - \Delta z \cos\theta +\frac{\Delta z^2\sin^2\theta}{2r} + O(r^{-2})\,,\\
    & r \to r -\Delta z \cos\theta + \frac{\Delta z^2\sin^2\theta}{2r} + O(r^{-2})\,,\\
    & \theta \to \theta + \frac{\Delta z\sin\theta}{r} + \frac{\Delta z^2\sin\theta \cos\theta}{2r^2} + O(r^{-3})\,,\\
    & \phi \to \phi\,,
\end{align}
the resulting metric is
\begin{align}\label{ts-}
    ds^2 &= \left(-1 + \frac{2M}{r} + \frac{2M\Delta z \cos\theta}{r^2} + O(r^{-3})\right)dv^2 + 2 (1 + O(r^{-3}))dv dr \nonumber\\ & 
    + 2 \left( \frac{2M\Delta z\sin\theta}{r} + O(r^{-2})\right)dvd\theta + \left(r^2 + O(r^{-1})\right)d\theta^2 \nonumber\\ &+  \left(r^2\sin^2\theta + O(r^{-1})\right)d\phi^2\, .
\end{align}
Comparing Eqs.~\eqref{ts+} and \eqref{ts-} with \eqref{bondi} and \eqref{bondii-} respectively, the angular momentum aspects are seen to be
\begin{align}
    &N_A\big\vert_{\mathcal{I}^+} = - 3M \Delta z\sin\theta (d\theta)_A \,,&& N_A\big\vert_{\mathcal{I}^-} = - 3M \Delta z\sin\theta (d\theta)_A\,,
\end{align}
which obey
\begin{align}
    N_A\big\vert_{\mathcal{I}^+} = - \Upsilon^* N_A\big\vert_{\mathcal{I}^-}.
\end{align}
This is consistent with our Eq.~\eqref{antipodalNA} but not with Eq.~(2.13) of Ref.~\cite{hawking_superrotation_2017}.

We may use the Killing field $Y_z^A = \partial^A \cos\theta$ to compute the mass moment $N^i$.  From Eqs.~\eqref{QYscri+} and \eqref{QYscri-}, we have
\begin{align}
    & Q_{Y_z}\big\vert_{\mathcal{I}^+} = \frac{1}{8\pi}\int N_A\vert_{\mathcal{I}^+}Y^A_z d\Omega = M \Delta z\,,\\
    & Q_{Y_z}\big\vert_{\mathcal{I}^-} = -\frac{1}{8\pi} \int N_A\vert_{\mathcal{I}^-}\Upsilon^*Y^A_z d\Omega = M \Delta z\,,\label{Lorentz Charge past null schwarzschild}
\end{align}
showing explicitly that the charges match and have the expected value $M \Delta z$.

\section{Asymptotic fields of Kerr}
\label{sec:Kerr}

In this appendix, we will investigate the Kerr metric in Beig-Schmidt form at timelike infinity. (Earlier work on Beig-Schmidt form of the Kerr metric focused on spatial infinity \cite{Mann:2008ay,Virmani:2011gh} and in these works the tensor $j_{ab}$ was never fully computed.) We will start with Boyer-Lindquist coordinates \cite{boyer_maximal_1967}, 
\begin{align}
    ds^2 &= -\big(1-\frac{2Mr}{\Sigma}\big)dt^2-\frac{4Mar\sin^2\theta}{\Sigma}dtd\phi + \frac{\Sigma}{\Delta}dr^2 +\Sigma d\theta^2 \nonumber\\
    &+ \big(r^2 + a^2 + \frac{2Ma^2r\sin^2\theta}{\Sigma}\big)\sin^2\theta d\phi^2,\\
    \Sigma& = r^2 + a^2 \cos^2\theta\,,\qquad 
    \Delta = r^2 - 2Mr + a^2\,.
\end{align}
We make use of the coordinate transformation
\begin{align}
t \to \tau \cosh\rho\,,\qquad 
r \to \tau \sinh\rho\,, 
\qquad 
\theta \to \theta\,,\qquad \phi \to \phi\,,
\end{align}
with the additional adjustment
\begin{align}
    \tau &\to \tau - F_0 + H\log\tau + \frac{1}{\tau}\big(A_2\log^2\tau + A_1\log\tau + A_0\big) + o\big(\frac{1}{\tau}\big)\,,\\
    \rho &\to \rho +\frac{1}{\tau}\big(4M\cosh\rho + \partial_\rho F_0 - (1+\log\tau)\partial_\rho H\big) + \frac{1}{\tau^2}\big(B_2\log^2\tau \nonumber\\& + B_1\log\tau + B_0\big) + o\big(\frac{1}{\tau^2}\big)\,,\\
    \theta &\to \theta  -\tau^{-2}\frac{a^2\sin\theta\cos\theta}{2\sinh^2\rho}+o\big(\frac{1}{\tau^2}\big)\,,\\
    \phi & \to \phi -\tau^{-2}\frac{aM\coth\rho}{\sinh^2\rho} + o\big(\frac{1}{\tau^2}\big)\,,
\end{align}
where
\begin{align}
    F_0 &= - M (\sinh\rho + 2\rho \cosh\rho)\,,\\
    H &= 2M\cosh\rho\,,\\
    A_2 & = -2M^2\sinh^2\rho\,,\\
    A_1 & = -2 M^2 \sinh ^2\rho  (2 \rho -\coth \rho +2)\,,\\
    A_0 & = \frac{1}{4} \biggr(-a^2 \cos (2 \theta )+a^2+M^2 (4 \rho  (\rho +2)+11)+24 \coth\rho-4 \text{csch}^2\rho\nonumber\\
    &-M^2 \big(-4 (\rho -1) \sinh (2 \rho )+(4 \rho  (\rho +2)-3) \cosh (2
   \rho ) \big)\biggr)\nonumber\\
   &-\frac{1}{2} M^2 (\sinh\rho+\cosh\rho (\coth\rho-2))^2\\
   B_2 & = 2M^2\sinh(2\rho)\,,\\
   B_1 & = \frac{1}{2} M^2 ((8 \rho +4) \sinh (2 \rho )-4)\,,\\
   B_0 & = \frac{1}{2} \biggr(-\frac{1}{2} \coth\rho \big(-a^2 \cos (2 \theta )+a^2-16 M^2\big)+M^2 (\coth\rho-4)\nonumber\\
   &+M^2 (4 \rho  (\rho +1)+7) \sinh (2 \rho )-2 M^2 (2
   \rho +3 \cosh (2 \rho )+1) \text{csch}^2\rho\biggr)\, .
\end{align}
This puts the metric in Beig-Schmidt form \eqref{Beig-Schmidt} with 
\begin{align}
    \sigma &=  - M (\frac{1}{\sinh\rho}-2e^{-\rho})\,,\\
    k_{ab} & = i_{ab} = 0\,,\\
    j_{\rho\rho} & = \frac{1}{2} M^2 e^{-4 \rho } \text{csch}^4\rho (-4 \sinh (2 \rho )+\cosh (2 \rho )-7) \,,\label{jrhorhokerr} \\
    j_{\rho \theta}& = 0,\\
    j_{\rho \phi} &= 3 a M \sin ^2\theta \text{csch}^2\rho \label{jrhothetakerr}\,, \\
    j_{AB} & = M^2 e^{-4 \rho } \left(\text{csch}^2\rho+3\right)\gamma_{AB}\,.
\end{align}
With a surface surrounding the origin $\rho=0$ of $i^+$, the angular momentum charges \eqref{chargeL} are calculated to be
\begin{align}
    L_x = 0, \qquad L_y = 0, \qquad L_z = aM\, .
\end{align}

Eqs.~\eqref{jrhorho match} and \eqref{jrhoA match} relate $j_{ab}$ to the angular momentum aspect $N_A$ at $\scri^+$.  From Eqs.~\eqref{jrhorhokerr} and \eqref{jrhothetakerr} we conclude that 
\begin{align}
    N_\theta = 0,\qquad     N_\phi  = -3aM \sin^2\theta\, .
\end{align}
This matches with equation (5.8) in \cite{compere_poincare_2020}, which was found directly in Bondi gauge at $\scri^+$.  At past null infinity we can set the Kerr metric in the analogous Bondi gauge and obtain
\begin{align}
    N_\theta = 0,\qquad 
    N_\phi  = +3aM \sin^2\theta\, .
\end{align}
This is a non-trivial check of the matching conditions $N_A\vert_{\mathcal I^+}=-\Upsilon^* N_A\vert_{\mathcal I^-}$ \eqref{antipodalNA}, which confirms the minus sign in this equality.

\section{Decoupling of pure supertranslations at $i^\pm$ and $i^0$}
\label{sec:invBMS}

In non-radiative regions such as around $i^+$, $i^-$ or $i^0$, the canonical definition of the Lorentz charge $Q_Y$ transforms under supertranslations since it obeys the BMS algebra
\begin{align}
 \{ Q_{T_1},Q_{T_2}\} &= 0\,, \\
 \{Q_{T_1},Q_{Y_2}\} &= Q_{Y_2(T_1)}\,, \label{BMSalgebraAgain}\\
  \{Q_{Y_1},Q_{Y_2}\} &= Q_{[Y_1,Y_2]}\,,
\end{align}
where the scalar function over the sphere $Y(\cdot)$ was defined in \eqref{YT}. Moreover, around $i^+$, $i^-$ or $i^0$, the asymptotic metric components allow one to read-off the field $C(\theta,\phi)$ defined with only $\ell \geq 2$ harmonics that transforms under a BMS supertranslation as 
\begin{equation}\label{deltaCApp}
\delta_{Y,T} C =(Y(C) + T)\vert_{\ell \geq 2} =  \left.\left( Y^A \partial_A C - \frac{1}{2}D_A Y^A C \right)\right|_{\ell \geq 2}+\sum_{\ell \geq 2,m} T_{\ell m}Y_{\ell m}. 
\end{equation}
(Only $\ell \geq 2$ harmonics appear in the right-hand side of \eqref{deltaCApp} consistently with the absence of $\ell=0,1$ harmonics in $C$.) This transformation matches with Eq. (3.22) of \cite{Compere:2018ylh}\footnote{$\,\,$ Thanks to the recent work \cite{Fuentealba:2022xsz}, the field $C(x^A)$ can be recognized as the canonical charge associated with logarithmic supertranslation asymptotic symmetries,  which are symplectic conjugated to supertranslations. We thank M. Henneaux for helping spotting a convention clash in that equation in an earlier version.}. We do not consider adding $\ell=0,1$ harmonics in $C$ since they do not appear in any physical quantity (\emph{i.e.}, neither in the metric nor in the definition of invariant Lorentz charges as defined below).

Using the BMS algebra and the $C$ field, it is straightforward to prove that the BMS algebra can be decoupled into a direct sum of the pure supertranslation subalgebra and the Poincar\'e algebra after performing a non-linear redefinition of the Lorentz charges. In the following we present an algebraic derivation of this decoupling property mainly following \cite{Compere:2019gft} (see also the derivations  \cite{Javadinezhad:2018urv,Chen:2021szm,Chen:2021kug,Compere:2021inq,Javadinezhad:2022hhl} and earlier work \cite{Gallo:2014jda}).

We first promote the supertranslations $T$ to be field-dependent\footnote{$\,\,$ The consideration of field-dependent symmetry generators dates at least back to \cite{Barnich:2004uw}.}. More particularly, we will now allow $T$ to depend upon $C$. We can now subtract from the Lorentz charge a supermomentum charge with parameter $T=Y(C)$,
\begin{equation}
Q_Y^{\text{inv}} \equiv Q_Y-Q_{T=Y(C)}\, . 
\end{equation}
Note that $Y(C)$ contains both $\ell=1$ and $\ell >1$ harmonics when $Y$ corresponds to a boost, see Eq. \eqref{BMSexplicit2}.
In particular, in the asymptotic regions $\vert u \vert \rightarrow \infty$ of $\scri^\pm$ where the fields are non-radiative, the supertranslation-invariant Lorentz charges are defined as
\begin{align}
    Q^{\text{inv}}_Y\vert_{\scri^\pm} & = \frac{1}{8\pi}\int_{S^2}d\Omega\, Y^A (N_A - 3 m \partial_A C-C \partial_A m ) d\Omega\, . \label{QYscri+inv}
\end{align}

Let us consider a pure supertranslation $T$. By construction, using \eqref{deltaCApp} we have the bracket 
\begin{equation}
\{Q_T,Q_Y^{\text{inv}}\} = Q_{Y(T)}-Q_{Y(T)}=0\, . 
\end{equation}
The Lorentz charge $Q_Y^{\text{inv}}$ is therefore supertranslation-invariant. The commutator between a Lorentz transformation and a translation $T_\mu$ is unaffected because $\delta_{T_\mu} C=0$ and therefore
\begin{equation}
\{Q_{T_\mu},Q_Y^{\text{inv}}\} = Q_{Y(T_\mu)}\, . 
\end{equation}
This is the standard Poincar\'e commutator. After some algebra, one obtains that the commutator between two Lorentz transformations is unchanged
\begin{align}
\{Q_{Y_1}^{\text{inv}},Q_{Y_2}^{\text{inv}}\} & = Q_{[Y_1,Y_2]} -\{Q_{Y_1(C)},Q_{Y_2}\}-\{Q_{Y_1},Q_{Y_2(C)}\}\\
& = Q_{[Y_1,Y_2]}-Q_{Y_2(Y_1(C))}+Q_{Y_1(Y_2(C))}\\ 
&= Q^{\text{inv}}_{[Y_1,Y_2]} \,,
\end{align}
after using $Y_1(Y_2(C))-Y_2(Y_1(C))=[Y_1,Y_2](C)$. 

We have therefore completely decoupled the pure supertranslations with the Poincar\'e algebra. After this field-redefinition of the generators, the asymptotic symmetry algebra is the direct sum of the (commuting) pure supertranslations and the Poincar\'e algebra:
\begin{align}
 \{ Q_{T_\mu},Q_{T_\nu}\} &= 0\,, \\
 \{Q_{T_\mu},Q^{\text{inv}}_{Y}\} &= Q_{Y(T_\mu)}\,, \label{invBMSalgebra}\\
  \{Q^{\text{inv}}_{Y_1},Q^{\text{inv}}_{Y_2}\} &= Q^{\text{inv}}_{[Y_1,Y_2]}\,.
\end{align}

\section*{References}

\bibliographystyle{utphys}
\bibliography{Bibliography}
\end{document}